\documentclass[apj]{emulateapj}
\usepackage{epstopdf}
\usepackage{longtable}
\usepackage{fancyvrb}
\usepackage{multirow}
\usepackage{verbatim}
\usepackage{longtable}
\usepackage[caption=false]{subfig}
\usepackage{float}
\usepackage{color}
\usepackage[T1]{fontenc}

\newcommand{\CSCJfSigThresh}{$3$}
\newcommand{\majoraxisbmax}{$1$}
\newcommand{\CSCsigthresh}{$3$}
\newcommand{\NEDxradarcsec}{$2\farcs5$}
\newcommand{\SexDetectThresh}{$3$}
\newcommand{\AimMaxArcMin}{$5$}
\newcommand{\ChandraPSFRatioMax}{$1$}
\newcommand{\FWHMThresh}{$2$}
\newcommand{\AxisRatioThresh}{$2$}

\newcommand{\petrothreshfacmax}{$2$}

\newcommand{\petroMagrlim}{$22.5$}
\newcommand{\offsetsigthresh}{$5$}
\newcommand{\supXRBISMsigthresh}{$5$}
\newcommand{\MBHLOHITHRESH}{$10^{6}$}

\newcommand{\NEDHLXradfac}{$3$}
\newcommand{\FXLim}{$ 10^{-16}$}
\newcommand{\SDSSFWHM}{$1\farcs6$}
\newcommand{\SDSSPixScale}{$0.39''$ pix$^{-1}$}
\newcommand{\PNSTRSPixScale}{$0.26''$ pix$^{-1}$}\newcommand{\XmatchSZ}{3,515}
\newcommand{\OffsetSZ}{650}
\newcommand{\ManFlagSZ}{310}

\newcommand{\ManFlagSatSZ}{42}
\newcommand{\ManFlagNonNucSZ}{298}
\newcommand{\HLXCandSZ}{169}
\newcommand{\DetCol}{magenta}

\newcommand{\ncontSymb}{squares}
\newcommand{\ncontindSymb}{circles}
\newcommand{\SMBHCol}{green}
\newcommand{\IMBHCol}{purple}
\newcommand{\AGNCol}{red}
\newcommand{\zMax}{$0.87$}
\newcommand{\zMaxRound}{$0.9$}
\newcommand{\photozerrmean}{$0.078$}
\newcommand{\photozerrstd}{$0.047$}

\newcommand{\zAgreeDiffAvg}{$0.048$}

\newcommand{\ncontmeanmin}{$7$}
\newcommand{\ncontmeanmax}{$8$}

\newcommand{\LHunabsMEDIAN}{$4.6\times 10^{42}$}
\newcommand{\LHunabsWAMEDIAN}{$6.1\times 10^{42}$}
\newcommand{\LXSnorm}{$ 10^{41}$}
\newcommand{\LXHnorm}{$ 10^{42}$}
\newcommand{\LXFnorm}{$ 10^{42}$}
\newcommand{\LXXRBnorm}{$ 10^{39}$}
\newcommand{\LXISMnorm}{$ 10^{37}$}
\newcommand{\nHNorm}{$ 10^{22}$}
\newcommand{\HRMEDIAN}{$0.14$}

\newcommand{\NHDEF}{$0$}
\newcommand{\GAMMADEF}{1.7}
\newcommand{\GAMMAMEDIAN}{$1.78$}

\newcommand{\NHMEDIANfix}{$2.0\times 10^{22}$ cm$^{-2}$}
\newcommand{\NHMEDIANfree}{$4.8\times 10^{21}$ cm$^{-2}$}

\newcommand{\NHfreeObscFrac}{$35$}
\newcommand{\FEddMedian}{$0.24$}

\newcommand{\GammaLowerThresh}{$0.5$}
\newcommand{\GammaUpperThresh}{$3$}
\newcommand{\ErrIterSZ}{1,000}
\newcommand{\rejsig}{$1.5$}
\newcommand{\sepmax}{$1$}
\newcommand{\offseterrMean}{$0\farcs6$}
\newcommand{\DeltaThetaMin}{$0\farcs8$}
\newcommand{\DeltaThetaMean}{$3\farcs9$}
\newcommand{\DeltaSMin}{$2.5$}
\newcommand{\DeltaSMean}{$16.5$}

\newcommand{\detthresh}{$5$}
\newcommand{\ErrPosMatchThreshFac}{$5$}
\newcommand{\iOptDetFrac}{$28$}

\newcommand{\iOptDetSZ}{$49$}

\newcommand{\iOptDetLoMassSZ}{$16$}
\newcommand{\iOptDetLoMassFrac}{$32$}

\newcommand{\XtoODetpONEtoTENFrac}{$71$}

\newcommand{\supLXXRBISM}{$144$}
\newcommand{\supLXXRBISMCorrMean}{$130$}

\newcommand{\supiOptDetLoMassFrac}{$35$}

\newcommand{\WISEAGNcninetySZ}{$40$}
\newcommand{\WISEAGNcninetyFRAC}{$27$}
\newcommand{\XtoMIR}{$0.3$}

\newcommand{\FEDDlo}{$5\times 10^{-2}$}
\newcommand{\MBHLOMASSSZlo}{$22$}

\newcommand{\MBHLOMASSSZCorrlo}{$20$}

\newcommand{\MBHMINMed}{$10^{4.6}$}
\newcommand{\MBHMAXMed}{$10^{7.8}$}
\newcommand{\MBHMEDMed}{$10^{6.2}$}

\newcommand{\MBHLOMASSSZMed}{$56$}

\newcommand{\MBHLOMASSSZCorrMed}{$50$}

\newcommand{\FiltExtSZ}{$261$}

\newcommand{\AddFiltSZ}{$291$}
\newcommand{\NEDrejSZ}{$15$}
\newcommand{\RadiorejSZ}{$3$}
\newcommand{\LensrejSZ}{$1$}

\newcommand{\BrightSTD}{$0.4$}
\newcommand{\FaintSTD}{$1.0$}

\newcommand{\uFlux}{erg s$^{-1}$ cm$^{-2}$}
\newcommand{\uLum}{erg s$^{-1}$}
\newcommand{\uSFR}{\MSun~yr$^{-1}$}
\newcommand{\uColDens}{cm$^{-2}$}

\newcommand{\uXRAYE}{keV}
\newcommand{\uARCSEC}{$''$}
\newcommand{\uVEL}{km s$^{-1}$}
\newcommand{\uAngSep}{$''$}
\newcommand{\uPhySep}{kpc}

\newcommand{\Mstar}{$M_{\star}$}
\newcommand{\MstarHLX}{$M_{\star,HLX}$}
\newcommand{\MstarHost}{$M_{\star,Host}$}
\newcommand{\MSun}{$M_{\odot}$}
\newcommand{\MBulge}{$M_{\rm{Bulge}}$}
\newcommand{\MBH}{$M_{\bullet}$}
\newcommand{\LBOL}{$L_{Bol}$}
\newcommand{\LEDD}{$L_{\rm{Edd}}$}
\newcommand{\FEDD}{$f_{\rm{Edd}}$}
\newcommand{\photoz}{$z_{\rm{phot}}$}

\newcommand{\meanphotozerr}{$<E_{\rm{z,phot}}>$}
\newcommand{\stdphotozerr}{$\Delta E_{\rm{z,phot}}$}
\newcommand{\specz}{$z_{\rm{spec}}$}
\newcommand{\z}{$z$}

\newcommand{\DeltaTheta}{$\Delta \Theta$}
\newcommand{\DeltaS}{$\Delta S$}

\newcommand{\ch}{\emph{Chandra}}
\newcommand{\chcosmosleg}{\ch~COSMOS-Legacy}
\newcommand{\sdss}{SDSS}
\newcommand{\sdssname}{Sloan Digital Sky Survey}
\newcommand{\pnstrsname}{Panoramic Survey Telescope and Rapid Response System}
\newcommand{\pnstrs}{Pan-STARRS}

\newcommand{\wise}{\emph{WISE}}
\newcommand{\rosat}{\emph{ROSAT}}
\newcommand{\xmm}{\emph{XMM-Newton}}
\newcommand{\xmmtwo}{\emph{2XMM}}
\newcommand{\xmmthree}{\emph{3XMM}}
\newcommand{\wisetitle}{\emph{Wide-field Infrared Survey Explorer}}
\newcommand{\first}{FIRST}
\newcommand{\firsttitle}{Faint Images of the Radio Sky at Twenty-Centimeters}

\newcommand{\CSCname}{\ch~Source Catalog}
\newcommand{\CSC}{CSCv2}
\newcommand{\RC}{RC3}
\newcommand{\nedname}{NASA Extragalactic Database}
\newcommand{\ned}{NED}
\newcommand{\allwise}{\emph{ALLWISE}}

\newcommand{\Astropy}{\texttt{Astropy}}
\newcommand{\AstropyModeling}{\texttt{modeling}}
\newcommand{\ciao}{\texttt{CIAO}}
\newcommand{\sherpa}{\texttt{Sherpa}}
\newcommand{\mergeobs}{\texttt{merge\_obs}}
\newcommand{\colden}{\texttt{colden}}
\newcommand{\getdraws}{\texttt{get\_draws}}
\newcommand{\sampflux}{\texttt{sample\_flux}}
\newcommand{\kcorr}{\texttt{calc\_kcorr}}
\newcommand{\wvd}{\texttt{wavdetect}}
\newcommand{\se}{\texttt{Source Extractor}}

\newcommand{\uband}{\emph{u}}
\newcommand{\gband}{\emph{g}}
\newcommand{\rband}{\emph{r}}
\newcommand{\iband}{\emph{i}}
\newcommand{\zband}{\emph{z}}
\newcommand{\Vband}{\emph{V}}
\newcommand{\ugCol}{\uband$-$\gband}
\newcommand{\grCol}{\gband$-$\rband}
\newcommand{\riCol}{\rband$-$\iband}
\newcommand{\izCol}{\iband$-$\zband}
\newcommand{\sixmicron}{$6\micron$}
\newcommand{\wiseone}{W1}
\newcommand{\wisetwo}{W2}

\newcommand{\PSFsig}{$\sigma_{\rm{PSF}}$}

\newcommand{\FXH}{$F_{\rm{2-10keV}}$}

\newcommand{\fxtwokev}{$F_{\rm{2keV}}$}
\newcommand{\fx}{$F_{X}$}
\newcommand{\PhotIndx}{$\Gamma$}
\newcommand{\nHexgal}{$n_{\rm{H,exgal}}$}
\newcommand{\HR}{$HR$}
\newcommand{\LX}{$L_{X}$}
\newcommand{\LXS}{$L_{\rm{0.5-2keV,unabs}}$}
\newcommand{\LXH}{$L_{\rm{2-10keV,unabs}}$}
\newcommand{\LXF}{$L_{\rm{0.5-10keV,unabs}}$}
\newcommand{\LXHXRB}{$L_{\rm{2-10keV,XRB}}$}
\newcommand{\LXHISM}{$L_{\rm{2-10keV,ISM}}$}
\newcommand{\LXHXRBISM}{$L_{\rm{2-10keV,XRB+ISM}}$}
\newcommand{\Esoft}{$0.5-2$~\uXRAYE}
\newcommand{\Ehard}{$2-10$~\uXRAYE}
\newcommand{\Efull}{$0.5-10$~\uXRAYE}
\newcommand{\EXtoO}{$0.3-3.5$~\uXRAYE}

\newcommand{\Fsixmicron}{$F_{\rm{6\micron}}$}
\newcommand{\Lsixmicron}{$L_{\rm{6\micron}}$}

\newcommand{\FXHtoFsixmicron}{\FXH~$/$~\Fsixmicron}
\newcommand{\LXHtoLsixmicron}{\LXH~$/$~\Lsixmicron}

\newcommand{\rpetro}{$r_{P}$}

\newcommand{\fv}{$F_{V}$}
\newcommand{\fuv}{$F_{UV}$}
\newcommand{\LehmerEqn}{$=(9.05\pm0.37)\times 10^{28}$\Mstar[\MSun]$+(1.62\pm0.22)\times 10^{39}\times$\SFR[\uSFR]}

\newcommand{\XtoOname}{X-ray-to-optical}

\newcommand{\XtoO}{\fx/\fv}
\newcommand{\alphaox}{$\alpha_{ox}$}
\newcommand{\alphaoxdef}{$-0.384$~$\times$~log[\fuv/\fxtwokev]}

\newcommand{\SFR}{SFR}
\newcommand{\Kcorrection}{K-correction}

\newcommand{\fcont}{$f_{\rm{cont}}$}
\newcommand{\fcontind}{$f_{\rm{cont,LG06}}$}

\newcommand{\PA}{P.A.}

\shorttitle{A Catalog of HLXs and IMBH Candidates}
\shortauthors{Barrows et al.}

\begin{document}

\submitted{Accepted for publication in ApJ}

\title{A Catalog of Hyper-luminous X-ray Sources and Intermediate-Mass Black Hole Candidates out to High Redshifts}
\author{R. Scott Barrows$^{1}$, Mar Mezcua$^{2,3}$, and Julia M. Comerford$^{1}$}
\affiliation{$^{1}$Department of Astrophysical and Planetary Sciences, University of Colorado Boulder, Boulder, CO 80309, USA; Robert.Barrows@Colorado.edu}
\affiliation{$^{2}$Institute of Space Sciences (ICE, CSIC), Campus UAB, Carrer de Magrans, 08193 Barcelona, Spain}
\affiliation{$^3$ Institut d'Estudis Espacials de Catalunya (IEEC), Carrer Gran Capit\`{a}, 08034 Barcelona, Spain}

\bibliographystyle{apj}

\begin{abstract}

Hyper-luminous X-ray sources (HLXs; \LX~$>10^{41}$ \uLum) are off-nuclear X-ray sources in galaxies and strong candidates for intermediate-mass black holes (IMBHs). We have constructed a sample of \HLXCandSZ~HLX candidates by combining X-ray detections from the \CSCname~(Version 2) with galaxies from the \sdssname~and registering individual images for improved relative astrometric accuracy. The spatial resolution of \ch~allows for the sample to extend out to $z\sim$~\zMaxRound. Optical counterparts are detected among one-fourth of the sample, one-third of which are consistent with dwarf galaxy stellar masses. The average intrinsic X-ray spectral slope indicates efficient accretion, potentially driven by galaxy mergers, and the column densities suggest one-third of the sample has significant X-ray absorption. We find that \supLXXRBISM~of the HLX candidates have X-ray emission that is significantly in excess of the expected contribution from star formation and hot gas, strongly suggesting that they are produced by accretion onto black holes more massive than stars. After correcting for an average background or foreground contamination rate of \ncontmeanmax$\%$, we estimate that at least $\sim$\MBHLOMASSSZCorrlo~of the HLX candidates are consistent with IMBH masses, and this estimate is potentially several times higher assuming more efficient accretion. This catalog currently represents the largest sample of uniformly-selected, off-nuclear IMBH candidates. These sources may represent scenarios in which a low-mass galaxy hosting an IMBH has merged with a more massive galaxy and provide an excellent sample for testing models of low-mass BH formation and merger-driven growth.  

\end{abstract}

\section{Introduction}
\label{sec:intro}

A subset of galaxies host ultra-luminous X-ray sources (ULXs), which are defined as off-nuclear X-ray sources with luminosities \LX~$>10^{39}$ \uLum. ULXs are an intriguing class of objects as they represent potential challenges to our understanding of black hole (BH) accretion and formation \citep[see][and references therein]{Kaaret:2017}. In particular, the intrinsically luminous nature of ULXs exceeds the theoretical limit for radiative energy production in the accretion disks of stellar-mass BHs \citep{Remillard:McClintock:2006}. Environmental studies of ULXs reveal that they show a preference for star-forming regions \citep{Madau:1998,Ghosh:White:2001,Swartz:2009} and possibly indicate they may be powered by either sub-Eddington accretion \citep{Colbert:Mushotzky:1999} onto BHs more massive than typical stellar remnants ($>10$ \MSun) or super-Eddington accretion onto stellar-mass BHs \citep{Begelman:2002} and neutron stars \citep{Bachetti:2014,Israel:2016} in X-ray binaries (XRBs). 

The X-ray luminosity function of ULXs in star-forming galaxies features a break at luminosities of \LX~$>1-2\times10^{40}$ \uLum~that suggests the more luminous off-nuclear X-ray sources may be a different sub-population \citep{Swartz:2011}. Sources above this break are observationally much rarer, and the term hyper-luminous X-ray source (HLX) is attributed to those ULXs with luminosities \LX~$>10^{41}$ \uLum~\citep{Gao:2003}. These luminosities can only be produced by stellar-mass compact objects in certain short-lived conditions of efficient accretion that are unlikely to be observed \citep[see][and references therein]{Miller:Colbert:2004}. However, HLX luminosities can realistically be produced by accretion onto more massive BHs. Indeed, the X-ray spectra of bright ULXs and HLXs provide some evidence for relatively low blackbody temperatures that correspond to BH masses of \MBH~$>100$ \MSun~\citep{Miller:2003,Davis:2011:HLX1}. 

Interestingly, the corresponding stellar counterparts of HLXs are typically faint and suggest small masses that are not consistent with the typical stellar bulges of galaxies where supermassive BHs reside \citep{Matsumoto:2001,Farrell:2009,Mezcua:2015,Comerford:2015}. Therefore, HLXs may be associated with intermediate-mass BHs (IMBHs) with masses of \MBH~$=10^{2}-10^{6}$\MSun~\citep{King:2005}. Identifying IMBHs is of significant cosmological interest as they represent primordial seed masses that can theoretically grow to supermassive BH (SMBH) masses in reasonable timescales \citep{Volonteri:2009b,Volonteri:2010,Mezcua:2017}. The mass of an IMBH may depend strongly on its origin \citep[for reviews see][]{Volonteri:2012,Woods:2018}, with the most likely scenarios being the direct collapse of pre-galactic gas disks \citep[\MBH~$=10^{4}-10^{5}$~\MSun;][]{Lodato:2006}, the end-stage of massive Population III stars \citep[\MBH~$=10^{2}-10^{3}$~\MSun;][]{Volonteri:2010} or through the collapse of dense stellar clusters \citep[\MBH~$=10^{2}-10^{4}$~\MSun;][]{Zwart:2002}.  These scenarios predict that IMBHs should exist in the nuclei of low-mass galaxies formed in the early Universe \citep{Volonteri:2009b,Van_Wassenhove:2010,Greene:2012}. 

After their formation, IMBHs could then grow through accretion to produce today's observed SMBH population, though the mechanisms that drove this accretion are poorly constrained \citep[for reviews see][]{Volonteri:2010,Mezcua:2017}. IMBHs that have not yet evolved to SMBH masses represent our best opportunity to observe how they grow. Much work has focused on identifying IMBHs as active galactic nuclei (AGN) in low-mass galaxies \citep{Greene:2007,Reines:2013,Lemons:2015,Baldassare:2016,Mezcua:2016,Chilingarian:2018,Mezcua:2018}. These samples can provide important constraints on the dominant formation mechanisms of IMBHs and their subsequent growth. For example, theoretical models predict that, at low galaxy bulge masses (\MBulge~$<10^{9.5}$ \MSun), merger-driven BH growth is inefficient due to significant supernova and stellar feedback in the nucleus \citep{Angles-Alcazar:2017}. This picture is supported by \citet{Martin-Navarro:2018}, who find that low-mass galaxies are dominated by supernova feedback, and by \citet{Mezcua:2018}, who find a low fraction of active IMBHs in dwarf galaxies out to high redshift. This observation could be explained by inefficient BH accretion in low-mass galaxies \citep{Pacucci:2018} and suggests that IMBHs may only settle onto the typical \MBH~$-$~\MBulge~relation once their host galaxy bulge masses become sufficiently massive through merger-driven growth \citep{Pacucci:2018}. 

Indeed, mergers may be important phases in the lifetimes of IMBHs and play a significant role in their growth histories starting from primordial seed masses \citep{Mezcua:2019}. This evolutionary route may be occurring among HLXs since they potentially represent IMBH growth within a dwarf galaxy that is triggered by an encounter with a more massive galaxy \citep{Mapelli:2012}. The merger-induced torques will remove angular momentum from the gas and dust to trigger accretion onto the IMBH so that it appears as a bright, off-nuclear X-ray source \citep{Springel:Hernquist:2005,Oser:2012}. Perhaps the best-studied off-nuclear IMBH candidate is HLX-1 \citep{Farrell:2009} and evidence suggests it is consistent with a minor merger \citep{Mapelli:2012,Webb:2017}. Several other nearby ULXs/HLXs present varied evidence for association with IMBHs \citep{Matsumoto:2001,Jonker:2010,Mezcua:2013a,Mezcua:2013c,Mezcua:2015,Kim:2015,Mezcua:2018b} though their origins are often unclear.
 
Several groups have performed systematic archival searches for ULXs/HLXs as off-nuclear X-ray sources within galaxy light profiles. Some of the earliest works utilized \rosat~\citep{Liu:2005a,Liu:2005b} while most catalogs have relied primarily on the \xmm~\citep{Walton:2011,Zolotukhin:2016,Earnshaw:2018} or \ch~\citep{Swartz:2004,Gong:2016} X-ray observatories. These catalogs have focused on nearby galaxies primarily from the Third Reference Catalogue of Bright Galaxies \citep[\RC;][]{deVaucouleurs:1991}. Each of these works include a significant focus on the identification and attempted removal of potential sources of contamination (i.e foreground stars or background AGN). Indeed, follow-up spectroscopy has revealed numerous examples in which ULXs/HLXs are instead higher redshift AGN \citep{Dadina:2013,Sutton:2015}.

Since minor mergers are common in the Universe \citep{Lotz:2011}, significant numbers of IMBHs may be in off-nuclear regions of more massive galaxies such as HLX-1 \citep{Soria:2013}. Therefore, measuring the fraction of HLXs in mergers can help to constrain their possible formation mechanisms. Moreover, understanding how HLX properties and environments evolve with redshift is necessary for constraints on the conditions of triggering IMBH growth. However, the off-nuclear X-ray sources in previous catalogs are all at redshifts of $z\lesssim0.06$ and hence can not be used to study the redshift evolution of their host galaxies or merger histories. Furthermore, previous systematic catalogs of off-nuclear X-ray sources contain few sources with luminosities exceeding $10^{41}$ \uLum. While \citet{Zolotukhin:2016} discovered nearly $100$ HLX candidates above this luminosity threshold, the large angular sizes of their host galaxies result in a $\sim70\%$ frequency of contamination from background or foreground sources.

In this paper we describe a new HLX catalog that complements previous catalogs by significantly increasing the number of known HLX candidates. The genesis of this project was motivated by a procedure we developed for identifying candidate spatially offset AGN based on \ch~X-ray signatures that are displaced from the nuclei of \sdssname~(\sdss) galaxies \citep{Barrows:2016}. The premise of that project relied on the registration of individual pairs of \ch-\sdss~images and the subsequent relative astrometric uncertainties. Here we apply that same procedure to the search for HLX and IMBH candidates. Hence, one of the chief advantages provided by the catalog we present in this work is the use of \ch~combined with the individual registration and astrometry for each pair of images, thus allowing for off-nuclear X-ray sources to be identified down to small angular offsets. Consequently, the host galaxy redshifts in the catalog presented here extend to $z\sim$\zMaxRound, thereby significantly augmenting the redshift range of previous HLX samples and increasing the number of HLX candidates by a factor of more than three. 

This paper is structured as follows: in Section \ref{sec:sample} we describe the selection of HLX candidates; in Section \ref{sec:analysis_results} we analyze the sample and present our results concerning the nature of the X-ray sources, and in Section \ref{sec:conc} we present our conclusions. Throughout we use the cosmological parameters from the Nine-year Wilkinson Microwave Anisotropy Probe Observations \citep{Hinshaw:2013}: $H_{0}=69.3$ km s$^{-1}$ Mpc$^{-1}$, $\Omega_{M}=0.287$, and $\Omega_{\Lambda}=0.713$.

\section{Building the Sample}
\label{sec:sample}

In this section we describe the process for building our sample of HLX candidates: spatially cross-matching galaxies and X-ray sources (Section \ref{sec:XMatch}), estimating astrometric uncertainties and selection of spatially offset X-ray sources within galaxy profiles (Section \ref{sec:SpatialOffset}), filtering out bad photometry and contaminating or extended X-ray sources (Section \ref{sec:filters}), and selection of X-ray sources with luminosities indicating accretion onto massive BHs (Section \ref{sec:MassiveBH}). 

\subsection{Spatial Match of Galaxies and X-Ray Sources}
\label{sec:XMatch}

In this section we describe the samples and procedure used for cross-matching galaxies and X-ray sources. The galaxy selection is discussed in Section \ref{sec:galselect}, and the spatial match to X-ray sources is discussed in Section \ref{sec:initmatch}.

\subsubsection{Galaxy Selection}
\label{sec:galselect}

Our initial galaxy sample is drawn from the catalog of \sdss~detections that are classified as galaxies\footnotemark[1] and with photometric redshifts\footnotemark[2] (\photoz) derived by the \sdss~pipeline. The \photoz~estimates and associated uncertainties for each galaxy are based on matches in multi-dimensional color space (\ugCol, \grCol, \riCol, and \izCol) to a training set with spectroscopic redshifts \citep{Cunha:2009}. The accuracy of \photoz~estimates can vary significantly among the overall \sdss~galaxy sample, though the mean \photoz~error in our final sample (Section \ref{sec:finalsamp}) is \meanphotozerr~$=$~\photozerrmean~with a standard deviation of \stdphotozerr~$=$~\photozerrstd. We use the galaxy photometric redshift catalog from Data Release 7 \citep[DR7;][]{Abazajian09} but also include any additional galaxies detected in subsequent data releases through DR14 \citep{Abolfathi:2018}. 

\footnotetext[1]{https://www.sdss.org/dr14/algorithms/classify/\#photo\_class}
\footnotetext[2]{https://www.sdss.org/dr14/algorithms/photo-z/}

If a spectroscopic redshift (\specz) from the \sdss~is available, then we adopt it as the final redshift value (\z). Otherwise, we search for any available spectroscopic redshifts of the galaxies from external sources by querying the \nedname~(\ned). We consider a \ned~source to be associated with the galaxy if its world coordinates (RA and DEC) place it within \NEDxradarcsec~(radius determined from a visual examination of \ned~matches to a subset of our sample) of the galaxy centroid. If multiple \ned~sources with redshifts are found within this search region then we adopt the redshift of the closest \ned~source. Furthermore, if a \ned~source has multiple published redshifts then we adopt the redshift with the smallest uncertainty. If a spectroscopic redshift from \ned~is found using the above search criteria, then it is adopted as the final \z~value. For reference, we find that the mean absolute magnitude of \photoz~$-$~\specz~among our final sample is \zAgreeDiffAvg~and similar to the \photoz~errors quoted above. 

\begin{figure}[t!]
\hspace*{0.05in} \includegraphics[width=0.45\textwidth]{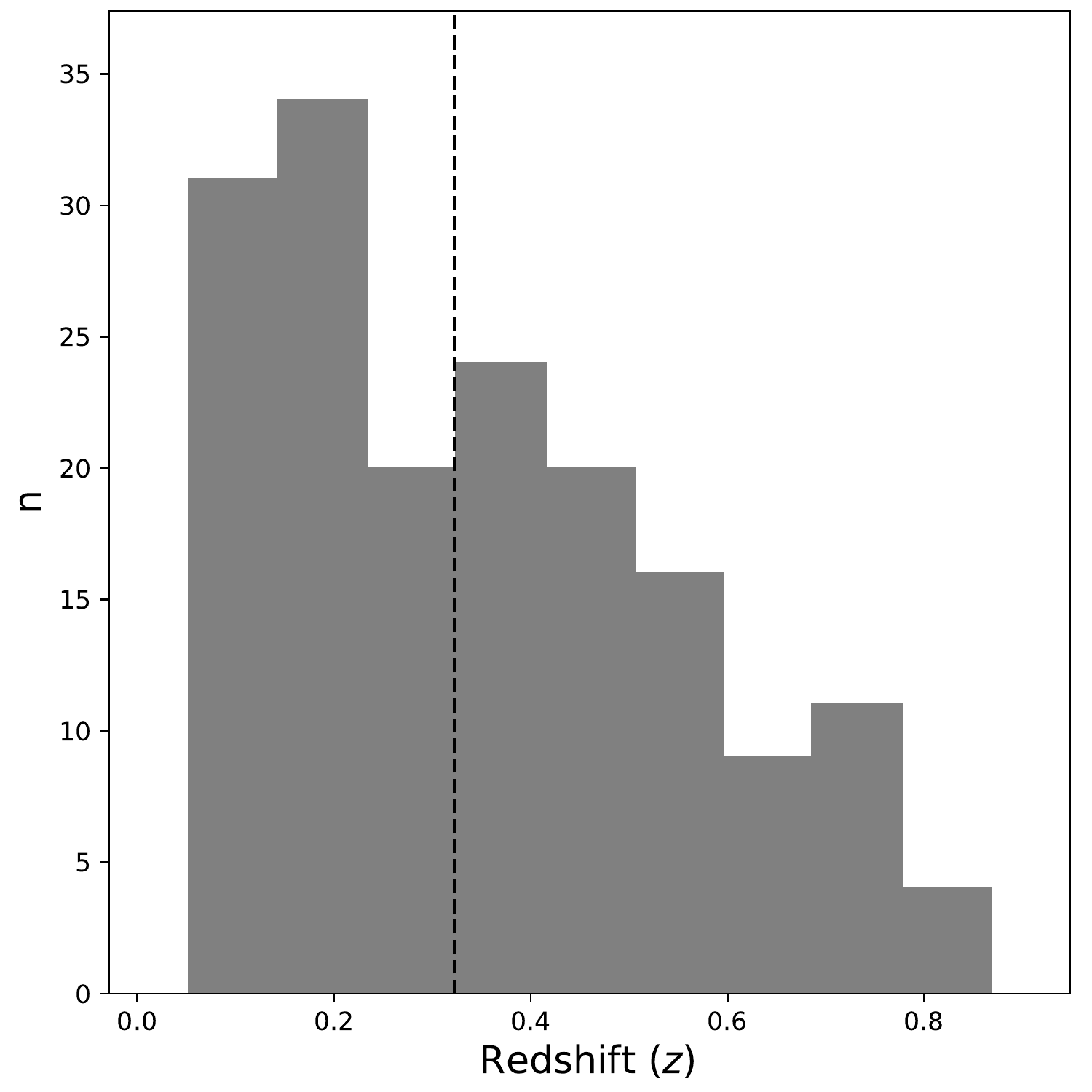}
\caption{\footnotesize{Distribution of redshifts (\z) for the final sample of HLX candidates (Section \ref{sec:finalsamp}). The values of \z~are determined from either photometry or spectra (Section \ref{sec:galselect}). The median value is denoted by the vertical dashed line. The HLX candidate redshifts extend out to \z~$=$~\zMax.}}
\label{fig:Z}
\end{figure}

If no spectroscopic redshift is available, then the \photoz~value is used for the final \z~value. We do not impose a maximum redshift limit on the galaxy selection since one of our ultimate aims with this sample is to address redshift evolution of the HLX population. However, we acknowledge that the sample will therefore likely be affected by a luminosity bias (see Section \ref{sec:comparison} for further discussion of this effect). The distribution of \z~values for the final sample of HLX candidates (Section \ref{sec:finalsamp}) is shown in Figure \ref{fig:Z} and extends out to \z~$=$~\zMax.

\subsubsection{Match to X-Ray Sources}
\label{sec:initmatch}

The \ch~X-ray observatory is ideal for identifying off-nuclear X-ray sources since it has the best spatial resolution of current X-ray telescopes. To obtain the most comprehensive list of robust source detections from \ch, we use the \ch~Source Catalog \citep[][]{Evans:2010} Version 2 (\CSC)\footnotemark[3] Master Source Catalog as our parent sample of X-ray sources. 
\footnotetext[3]{http://cxc.harvard.edu/csc2}
Before matching X-ray detections to the SDSS galaxy sample, we require that all X-ray sources are detected at  $>$\CSCsigthresh$\sigma$ significance to remove spurious detections. Furthermore, since HLXs are expected to be point sources (i.e. associated with accretion onto BHs), we also implement a strict cut that rejects sources with $1\sigma$ major axis spatial extents (from the \CSC~Master Source Catalog) greater than \majoraxisbmax\uARCSEC, corresponding to $\sim2$ times the Full Width at Half Maximum (FWHM) of on-axis \ch/ACIS spatial resolution. A specific rejection of spatially resolved X-ray sources is also implemented and described in Section \ref{sec:resolved}.

\begin{deluxetable*}{cccccccc}
\tabletypesize{\footnotesize}
\tablecolumns{8}
\tablecaption{Astrometric Properties of the HLX Candidates.}
\tablehead{
\colhead{I.D. \vspace*{0.05in}} &
\colhead{\z} &
\colhead{Galaxy Name} &
\colhead{HLX Name} &
\colhead{$<n>$} &
\colhead{$\Delta \Theta$} &
\colhead{$\Delta S$} &
\colhead{\PA} \\
\colhead{($-$) \vspace*{0.05in}} &
\colhead{($-$)} &
\colhead{($-$)} &
\colhead{($-$)} &
\colhead{($-$)} &
\colhead{($''$)} &
\colhead{(kpc)} &
\colhead{($^{\circ}$)} \\
\colhead{1} &
\colhead{2} &
\colhead{3} &
\colhead{4} &
\colhead{5} &
\colhead{6} &
\colhead{7} &
\colhead{8}
}
\startdata
1 & $0.17$ & SDSS J000150.94$+$232957.1 & 2CXO J000151.2$+$232959 & $3$ & $5.3\pm0.4$ & $15.2\pm1.1$ & $64.4$\\ 
2 & $0.22$ & SDSS J001320.07$-$192746.4 & 2CXO J001320.0$-$192748 & $1$ & $2.8\pm0.5$ & $10.0\pm1.8$ & $187.3$\\ 
3 & $0.38$ & SDSS J001837.98$+$163319.9 & 2CXO J001838.1$+$163319 & $5$ & $2.0\pm0.3$ & $10.5\pm1.4$ & $99.3$\\ 
4 & $0.11$ & SDSS J002234.45$+$001603.1 & 2CXO J002234.1$+$001609 & $4$ & $7.2\pm0.6$ & $14.3\pm1.2$ & $327.1$\\ 
5 & $0.64$ & SDSS J003608.60$+$182105.4 & 2CXO J003608.8$+$182100 & $0$ & $5.8\pm1.3$ & $40.4\pm8.8$ & $149.4$\\ 
6 & $0.54$ & SDSS J003933.79$+$031830.4 & 2CXO J003933.6$+$031828 & $2$ & $2.9\pm0.2$ & $18.8\pm1.3$ & $216.1$\\ 
7 & $0.27$ & SDSS J004114.58$-$210737.6 & 2CXO J004114.3$-$210737 & $0$ & $3.7\pm0.3$ & $15.5\pm1.1$ & $276.9$\\ 
8 & $0.54$ & SDSS J004201.96$-$092426.3 & 2CXO J004201.7$-$092427 & $1.4$ & $3.7\pm0.1$ & $23.7\pm0.9$ & $247.2$\\ 
9 & $0.69$ & SDSS J004709.94$-$080802.0 & 2CXO J004709.7$-$080802 & $4$ & $3.0\pm0.4$ & $21.6\pm3.2$ & $262.3$\\ 
10 & $0.30$ & SDSS J005550.41$+$262337.2 & 2CXO J005550.1$+$262341 & $1.8$ & $5.8\pm0.1$ & $26.3\pm0.5$ & $315.8$  
\enddata
\tablecomments{Column $1$: unique identifier for the galaxy$-$HLX candidate pair; Column $2$: host galaxy redshift; Column $3$: name of the host galaxy; Column $4$: name of the HLX candidate; Column $5$: average number of matches between the \sdss~\rband-band image and all \ch~OBSID(s) in which the HLX candidate is detected; Columns $6-7$: HLX candidate angular and projected physical offset from the host galaxy center; and Column $8$: position angle (East of North) of the HLX candidate offset. (This table is available in its entirety in a machine-readable form in the online journal. A portion is shown here for guidance regarding its form and content.)}
\label{tab:Astrometry}
\end{deluxetable*}

Matches between the \sdss~galaxy sample and the \CSC~are based on their world coordinates. The RA and DEC of the \sdss~galaxies represent the galaxy \rband-band photometric centroids since those are the images in which the galaxy positions are measured and that have the highest sensitivity \citep{York:2000}. The RA and DEC of each master \CSC~source is based on the solution from \wvd~and the point spread function (PSF) from the co-added observations. To create a list of X-ray sources that are within the light profile of a galaxy, we require that each \CSC~source centroid is within two Petrosian radii (\rpetro; measured by the \sdss~pipeline) of an \sdss~galaxy centroid. An aperture of \petrothreshfacmax~$\times$~\rpetro~includes most of a galaxy's flux \citep{Graham:2005b}, and \sdss~photometry indicates that it provides the optimal combination of maximizing the integrated galaxy flux while minimizing sky noise\footnotemark[4]. 
\footnotetext[4]{https://www.sdss.org/dr12/algorithms/magnitudes/}
While adopting a smaller offset threshold will reduce the probability of the X-ray source being an unrelated chance projection, doing so will also exclude many truly related sources (the probability of chance projections is discussed in Section \ref{sec:ContEst}). Note that multiple X-ray sources may satisfy this criterion for a single galaxy.

Finally, while all objects in the \sdss~photometric catalog are detected at $>5\sigma$ significance, to further mitigate uncertainties on association with the candidate host galaxy we also filter out galaxies with Petrosian \rband-band magnitudes $>$~\petroMagrlim. This threshold is chosen because we find that the Petrosian radius errors are relatively stable for \rband~$<22.5$ (standard deviation of \BrightSTD) but can increase significantly for fainter magnitudes (standard deviation of \FaintSTD). This step results in \XmatchSZ~distinct pairs of \sdss~galaxies and \CSC~X-ray sources.

\begin{figure*}[t!]
\hspace*{0.1in} \includegraphics[width=0.96\textwidth]{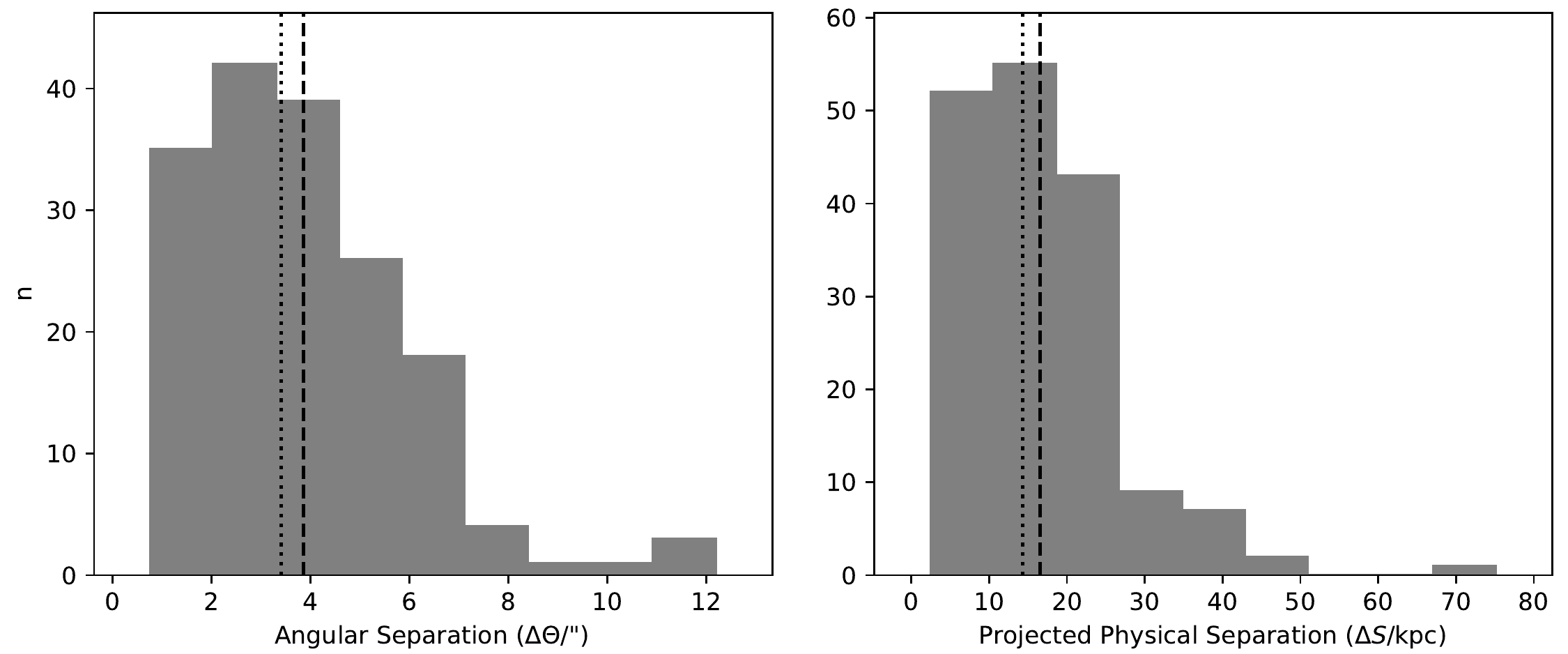}
\caption{\footnotesize{Distribution of angular offsets (\DeltaTheta; left) and projected physical offsets (\DeltaS; right) from the host galaxy nucleus for the final sample of HLX candidates (Section \ref{sec:finalsamp}). The angular offsets are determined after applying the relative astrometric corrections (Section \ref{sec:regastr}) and the physical offsets are determined from \DeltaTheta~and the cosmology stated in Section \ref{sec:intro}. In each panel, the vertical, dashed lines denote the mean values and the vertical, dotted lines denote the mean Petrosian radius}. Note that the mean offset is similar to the mean Petrosian radius.}
\label{fig:OFFSET_ANG_PHYS}
\end{figure*}

\subsection{Spatially Offset X-Ray Sources}
\label{sec:SpatialOffset}

In this section we describe the procedure used to identify X-ray sources that are spatially offset from the centers of galaxies from the spatial cross-match (Section \ref{sec:initmatch}). We first register the individual images containing the X-ray sources and the galaxies to obtain the relative astrometric corrections and uncertainties (Section \ref{sec:regastr}). Then we find those with statistically significant spatial offsets (Section \ref{sec:finalastr}).

\subsubsection{Registration and Astrometry}
\label{sec:regastr} 

We use the SDSS \rband-band images in the registration procedure since they are the images in which the galaxy positions (used for cross-matching in Section \ref{sec:initmatch}) are measured. Full details of the astrometry and registration procedure are described in \citet{Barrows:2016}, though here we reiterate the basic steps. 

We detect sources in the SDSS \rband-band images using \se~\citep{Bertin:Arnouts:1996} and a detection threshold of \SexDetectThresh$\sigma$. We detect sources in the \ch~images using \wvd~with a detection probability threshold of $10^{-8}$ that corresponds to less than one false detection per \AimMaxArcMin$'$ search region. We then apply several filters to each pair of source lists to optimize accuracy of the relative astrometric solutions. First, we filter out unreliable detections from both source lists. From the \se~output, unreliable detections are defined as flagged sources. From the \wvd~output, unreliable detections are defined as follows: detections with a significance $<$~\CSCJfSigThresh$\sigma$ (potentially spurious) and detections that are more than \AimMaxArcMin$'$ from the detector aimpoint \citep[the PSF becomes large and asymmetric beyond this radius;][]{Gaetz:2004}. We then filter out extended sources in both lists. From the \se~output, extended sources are defined as having a FWHM that is larger than \FWHMThresh~times the typical \sdss~image resolution FWHM \citep[\SDSSFWHM;][]{York:2000} and with a major-to-minor axis ratio $>$~\AxisRatioThresh. From the \wvd~output, extended sources are defined as those having a source extent-to-PSF size ratio (generated by \wvd) greater than \ChandraPSFRatioMax. Finally, the host galaxy and candidate off-nuclear X-ray source(s) are excluded from the lists of matched pairs to produce astrometric solutions that are independent of the spatial offsets being tested. 

Matched pairs of sources between the \se~and \wvd~lists are identified within a \sepmax\uAngSep~threshold radius. This threshold corresponds to slightly more than the $90\%$ absolute astrometry of \ch~and our results do not change when a larger (up to $2''$) threshold radius is used. Linear transformations along the Cartesian X and Y axes of the \sdss~image coordinates are computed as the mean offset between the final matched lists after iteratively rejecting matched pairs that are outliers by more than \rejsig$\sigma$. The relative astrometric uncertainties are computed from the errors on the source centroids in the final matched list. If no matches are found between a pair of images, then the linear transformations are set to $0$, and the astrometric uncertainties are set to the quadrature sum of the absolute astrometric errors from the \sdss~($0\farcs035$) and \ch~($0\farcs8$).

The uncertainty of the X-ray source position relative to the galaxy centroid is then the quadrature sum of the relative astrometric uncertainties, the \CSC~master source centroid uncertainty, and the galaxy centroid uncertainty. These uncertainties correspond to $1\sigma$ confidence intervals and are computed separately for both the RA and DEC dimensions.

\subsubsection{Selection of Spatially Offset X-Ray Sources}
\label{sec:finalastr}

Since each \CSC~Master Source may be based on detections from multiple individual observations (OBSIDs)\footnotemark[5], 
\footnotetext[5]{http://cxc.harvard.edu/csc2/organization.html}
for each galaxy-X-ray source pair we register the \sdss~\rband-band image with each \ch~OBSID in which the \CSC~source is detected. The final transformations between an \sdss~galaxy and \CSC~source are then the error-weighted averages of the astrometric corrections between each of the individual image pairs. The uncertainties on those transformations are the standard error of the weighted mean. 

We consider spatially offset X-ray sources to be those that are offset from the host galaxy center by $\geq$~\offsetsigthresh~times the uncertainty of the X-ray source position relative to the galaxy centroid (i.e. the final uncertainty computed in Section \ref{sec:regastr}). This step results in \OffsetSZ~spatially offset X-ray sources. The redshifts, host galaxy names, HLX candidate names, and astrometric properties are listed in Table \ref{tab:Astrometry}. The distributions of the angular offsets (\DeltaTheta) and projected physical offsets (\DeltaS; computed using the host galaxy redshifts and the cosmology stated in Section \ref{sec:intro}) between the X-ray sources and galaxy centroids for the final HLX candidate sample (Section \ref{sec:finalsamp}) are shown in Figure \ref{fig:OFFSET_ANG_PHYS}. The mean value of \DeltaTheta~is \DeltaThetaMean~(approximately one Petrosian radius), reflecting the average offset error of \offseterrMean~and the \offsetsigthresh$\sigma$~offset criterion, and the mean value of \DeltaS~is \DeltaSMean~\uPhySep.

\begin{deluxetable*}{cccccccccc}
\tabletypesize{\footnotesize}
\tablecolumns{10}
\tablecaption{X-Ray Properties of the HLX Candidates.}
\tablehead{
\colhead{I.D. \vspace*{0.05in}} &
\colhead{\PSFsig} &
\colhead{\LXS} &
\colhead{\LXH} &
\colhead{\LXF} &
\colhead{\PhotIndx} &
\colhead{\nHexgal} &
\colhead{$S$} &
\colhead{$H$} &
\colhead{\HR} \\
\colhead{($-$) \vspace*{0.05in}} &
\colhead{(\uAngSep)} &
\colhead{(\LXSnorm~\uLum)} &
\colhead{(\LXHnorm~\uLum)} &
\colhead{(\LXFnorm~\uLum)} &
\colhead{($-$)} &
\colhead{(\nHNorm~\uColDens)} &
\colhead{($-$)} &
\colhead{($-$)} &
\colhead{($-$)} \\
\colhead{1} &
\colhead{2} &
\colhead{3} &
\colhead{4} &
\colhead{5} &
\colhead{6} &
\colhead{7} &
\colhead{8} &
\colhead{9} &
\colhead{10}
}
\startdata
1 \vspace*{0.05in} & $0.90$ & $5.96_{-2.46}^{+3.51}$ & $1.31_{-0.54}^{+0.77}$ & $1.90_{-0.78}^{+1.12}$ & $1.78_{-0.32}^{+0.28}$ & $0.28_{-0.17}^{+0.14}$ & $43.0_{-6.9}^{+5.9}$ & $25.8_{-5.8}^{+4.5}$ & $-0.25_{-0.12}^{+0.11}$\\ 
2 \vspace*{0.05in} & $1.11$ & $0.14_{-0.02}^{+0.02}$ & $3.92_{-0.43}^{+0.43}$ & $3.93_{-0.44}^{+0.43}$ & $1.7\tablenotemark{a}$ & $11.66_{-0.50}^{+0.54}$ & $31.8_{-6.2}^{+4.9}$ & $52.4_{-7.9}^{+6.1}$ & $\phantom{+}0.25_{-0.10}^{+0.11}$\\ 
3 \vspace*{0.05in} & $1.46$ & $28.1_{-15.3}^{+20.0}$ & $4.42_{-2.40}^{+3.15}$ & $7.22_{-3.92}^{+5.13}$ & $2.32_{-0.45}^{+0.36}$ & $0.52_{-0.32}^{+0.25}$ & $18.9_{-4.8}^{+3.6}$ & $24.9_{-5.6}^{+4.2}$ & $\phantom{+}0.14_{-0.15}^{+0.15}$\\ 
4 \vspace*{0.05in} & $2.09$ & $0.46_{-0.24}^{+0.54}$ & $0.14_{-0.08}^{+0.17}$ & $0.19_{-0.10}^{+0.22}$ & $1.77_{-0.44}^{+0.31}$ & $0.19_{-0.11}^{+0.11}$ & $14.6_{-4.4}^{+3.1}$ & $\phantom{1}9.4_{-3.7}^{+2.3}$ & $-0.22_{-0.21}^{+0.19}$\\ 
5 \vspace*{0.05in} & $1.59$ & $67.3_{-15.0}^{+14.2}$ & $12.2_{-2.7\phantom{0}}^{+2.6\phantom{0}}$ & $18.9_{-4.2\phantom{0}}^{+4.0\phantom{0}}$ & $1.7\tablenotemark{a}$ & $0\tablenotemark{a}$ & $16.9_{-4.5}^{+3.4}$ & $12.9_{-4.1}^{+2.7}$ & $-0.13_{-0.19}^{+0.17}$\\ 
6 \vspace*{0.05in} & $2.01$ & $49.1_{-21.3}^{+22.6}$ & $17.6_{-7.6\phantom{0}}^{+8.1\phantom{0}}$ & $22.5_{-9.8}^{+10.3}$ & $1.31_{-0.12}^{+0.14}$ & $0.04_{-0.02}^{+0.03}$ & $113_{-12}^{+10}$ & $106_{-12}^{+10}$ & $-0.03_{-0.06}^{+0.07}$\\ 
7 \vspace*{0.05in} & $0.50$ & $46.6_{-4.7\phantom{0}}^{+4.7\phantom{0}}$ & $8.47_{-0.85}^{+0.85}$ & $13.1_{-1.3\phantom{0}}^{+1.3\phantom{0}}$ & $1.7\tablenotemark{a}$ & $0\tablenotemark{a}$ & $22.0_{-5.2}^{+3.9}$ & $21.9_{-5.2}^{+3.9}$ & $-0.00_{-0.15}^{+0.15}$\\ 
8 \vspace*{0.05in} & $0.86$ & $41.0_{-4.9\phantom{0}}^{+4.9\phantom{0}}$ & $7.45_{-0.89}^{+0.89}$ & $11.5_{-1.4\phantom{0}}^{+1.4\phantom{0}}$ & $1.7\tablenotemark{a}$ & $0\tablenotemark{a}$ & $15.6_{-4.9}^{+3.6}$ & $75.8_{-9.4}^{+8.2}$ & $\phantom{+}0.66_{-0.07}^{+0.09}$\\ 
9 \vspace*{0.05in} & $2.70$ & $349_{-151}^{+158}$ & $90.9_{-39.3}^{+41.0}$ & $126_{-55\phantom{0}}^{+57\phantom{0}}$ & $1.65_{-0.22}^{+0.18}$ & $0.70_{-0.39}^{+0.39}$ & $18.9_{-4.9}^{+3.6}$ & $29.9_{-6.1}^{+4.7}$ & $\phantom{+}0.22_{-0.13}^{+0.14}$\\ 
10 \vspace*{0.05in} & $0.51$ & $1.87_{-0.38}^{+0.47}$ & $1.37_{-0.28}^{+0.34}$ & $1.56_{-0.32}^{+0.39}$ & $0.79_{-0.44}^{+0.15}$ & $0.13_{-0.02}^{+0.02}$ & $45.0_{-7.1}^{+6.2}$ & $57.4_{-8.3}^{+7.0}$ & $\phantom{+}0.12_{-0.10}^{+0.10}$                     
\enddata
\tablecomments{Column $1$: unique identifier for the galaxy$-$HLX candidate pair; Column 2: $1\sigma$ PSF size at the HLX position; Columns $3-5$: unabsorbed, rest-frame luminosities in the soft, hard, and full bands; Column $6$: intrinsic photon index; Column $7$: intrinsic column density; Columns $8-9$: soft and hard X-ray photon counts; and Column $10$: hardness ratio. (This table is available in its entirety in a machine-readable form in the online journal. A portion is shown here for guidance
regarding its form and content.)\\$^{a}$fixed (Section \ref{sec:xrayspec}).}
\label{tab:Xray}
\end{deluxetable*}

\subsection{Additional Filters}
\label{sec:filters}

In this section we apply additional filters meant to remove erroneous spatially offset X-ray sources. First, we remove all host galaxies for which the photometry may be inaccurate (Section \ref{sec:filter_bad_phot}). We then cross-match the X-ray sources with catalogs of previously classified sources to identify known contaminants, i.e. spatially offset X-ray sources that are spectroscopically confirmed to not be HLXs (Section \ref{sec:filter_known_cont}). Finally, we remove any X-ray sources that are spatially resolved (Section \ref{sec:resolved}).

\subsubsection{Bad Photometry}
\label{sec:filter_bad_phot}

We remove all detections which the SDSS pipeline flagged as \texttt{saturated} since the nuclear positions may be compromised (\ManFlagSatSZ~sources). A visual inspection also reveals that, while all of the SDSS photometric detections in the galaxy catalog are associated with a galaxy, they are not always coincident with the galaxy's nucleus. These cases are identified by eye and removed (\ManFlagNonNucSZ~sources). Categorically, they can be grouped into the following classes:

\begin{enumerate}

\item The detection is an off-nuclear knot of star formation.

\item The detection is in a galaxy with no clearly defined nucleus (irregular galaxies and galaxy mergers).

\item The detection is in a galaxy with obvious dust lanes that may affect the detection of the \rband-band nucleus. These galaxies are typically nearby and with angular sizes sufficient to resolve dust lanes.

\item The detection photometry may be contaminated by bright neighboring sources. These galaxies are typically blended with foreground galaxies, are within clusters or brightest cluster galaxies, or are contaminated by diffraction spikes. 

\end{enumerate}

Removing the above sources with bad photometry results in \ManFlagSZ~spatially offset X-ray sources.

\subsubsection{Known Contaminants}
\label{sec:filter_known_cont}

We cross-match each off-nuclear X-ray source with \ned~to identify any that might be associated with known sources that have spectroscopic redshifts marking them as background or foreground sources. The cross-match radius we use is \NEDHLXradfac~times the quadrature sum of the RA and DEC errors in the X-ray source offset from the host galaxy nucleus. For a redshift to be uniquely associated with the spatially offset X-ray source, the spectroscopic observation must not overlap with the cross-match radius used to assign host galaxy redshifts (\NEDxradarcsec~from the host galaxy centroid; Section \ref{sec:galselect}). This step identifies \NEDrejSZ~spatially offset X-ray sources that have unique redshifts, and they are all larger than the host galaxy redshift by $>700,000$ \uVEL. Therefore, they are considered to be background sources and removed.

We also search the results of the \ned~cross-match for host galaxies associated with astrophysical phenomena that could mimic the offset signatures for which we select. First, we search \ned~for HLX candidates near known extended radio jets containing X-ray hotspots that are offset from the galaxy nucleus. We also cross-match the X-ray source positions with the \firsttitle~(\first) survey detection catalog\footnotemark[6] to search for any matches that display extended radio morphologies. 
\footnotetext[6]{http://sundog.stsci.edu/}
Since we are looking for jets, we consider a relatively large ($10''$) cross-match radius. We find that \RadiorejSZ~of the offset X-ray sources are coincident with extended radio detections from \first~(they are also known radio jet sources with central AGN). Visual inspection reveals that the X-ray sources are likely hotspots in the radio jets. Since these X-ray sources are associated with extended radio jets, we remove them from the list of spatially offset X-ray sources. Second, we search for HLX candidates near known gravitational lenses that can make an X-ray AGN appear offset from the galaxy nucleus. This search identifies \LensrejSZ~gravitationally lensed X-ray source, which is a broad emission line quasi-stellar object (QSO) where the spatially offset X-ray source is coincident with one of the lensed components. This QSO has been removed from the catalog because the X-ray source is not truly offset from the galaxy center. Removing the above known contaminants (background sources, jet hotspots, and lenses) results in \AddFiltSZ~spatially offset X-ray sources.

\subsubsection{Spatially Resolved X-ray Sources}
\label{sec:resolved}

As stated in Section \ref{sec:initmatch}, our focus is on X-ray point sources. Therefore, in addition to the removal of sources with $1\sigma$ major axis extents $>$~\majoraxisbmax\uARCSEC~(Section \ref{sec:initmatch}), we also omit sources with $1\sigma$ major axis extents greater than the PSF $1\sigma$ radius at the source position\footnotemark[7]. 
\footnotetext[7]{http://cxc.harvard.edu/csc2/columns/srcextent.html}
This restriction ensures that no X-ray sources are spatially resolved. For each source, we determine the PSF sizes from PSF maps generated by \mergeobs. The OBSIDs combined within \mergeobs~are those within the `best' Bayesian block of observations, where each block is determined by identifying the set of observations considered to have a constant photon flux \citep{Scargle:2013}. The block with the largest combined exposure time is the `best' block\footnotemark[8] from which source properties in the \CSC~Master Catalog are generated.
\footnotetext[8]{http://cxc.harvard.edu/csc2/data\_products/master/blocks3.html}
This step results in \FiltExtSZ~spatially offset X-ray sources. The $1\sigma$ PSF sizes are listed in Table \ref{tab:Xray}.

\subsection{Selecting HLX Candidates}
\label{sec:MassiveBH}

In this section we describe our procedure for selecting spatially offset X-ray sources that are consistent with accretion onto massive BHs (\MBH~$>10^{2}$ \MSun). This procedure involves measuring the intrinsic X-ray fluxes (Section \ref{sec:xrayspec}) and then computing the intrinsic luminosities to select those above a target threshold (Section \ref{sec:finalsamp}). 

\subsubsection{X-Ray Spectral Modeling}
\label{sec:xrayspec}

If the spatially offset X-ray sources are legitimately associated with the host galaxies identified from the cross-match (Section \ref{sec:initmatch}), then their redshifts can be assumed the same as the host galaxy. Therefore, we use \sherpa, the modeling and fitting package within the \ch~Interactive Analysis of Observations (\ciao) software, to extract and fit to each of the X-ray sources a spectral model that includes rest-frame components so that we can measure intrinsic luminosities (for the final HLX candidate selection) and additional physical properties (for subsequent analysis of the X-ray source nature). The extraction regions for the source and background are purposefully made to be identical to those used in the measurements from the \CSC. Specifically, the source region is the $90\%$ PSF enclosed energy flux ellipse, and the background region is an annulus with inner radius directly surrounding the source. The intrinsic source flux is represented by a power-law component ($F\sim E^{-\Gamma}$, where $E$ is the rest-frame energy and \PhotIndx~is the photon index). The flux is attenuated by photoelectric absorption from a column of neutral hydrogen that is set to the host galaxy redshift (\nHexgal). The models also include a non-redshifted column of neutral hydrogen that is fixed to the Galactic value at each source's Galactic coordinates (estimated from the \colden~function within \ciao).

The components \PhotIndx, \nHexgal, and the power-law normalization are allowed to vary freely during the fits. To maximize the signal-to-noise ratio and significance of the parameter estimates, we apply the same model simultaneously\footnotemark[9] to extracted spectra from each of the OBSIDs in the `best' Bayesian block (Section \ref{sec:resolved}) since they are consistent with a constant photon flux. 
\footnotetext[9]{http://cxc.harvard.edu/sherpa/threads/sourceandbg/}
Within \ciao, the \PhotIndx~and \nHexgal~errors are computed numerically using \getdraws, and the absorbed and unabsorbed fluxes are computed numerically using \sampflux. All errors are based on \ErrIterSZ~samples. In some fits the sample distribution does not converge on a value for \nHexgal, and in these cases we set \nHexgal~$=$~\NHDEF~and re-run the fit. If the final \PhotIndx~solution is not within the typical range for BH accretion \citep[\PhotIndx~$=$~\GammaLowerThresh~$-$~\GammaUpperThresh;][]{Nandra:1994,Reeves:2000,Piconcelli:2005,Ishibashi:2010} then we rerun the fit with the photon index fixed at \PhotIndx~$=$~\GAMMADEF~\citep[a commonly-used value for AGN;][]{Middleton:2008}. The fluxes are computed over the observed energy range of $0.5-7$ keV, and \Kcorrection s (calculated using the \ciao~function \kcorr) are applied to determine the rest-frame fluxes over the soft (\Esoft), hard (\Ehard), and full (\Efull) energy ranges.

\begin{figure}[t!]
\hspace*{0.05in} \includegraphics[width=0.45\textwidth]{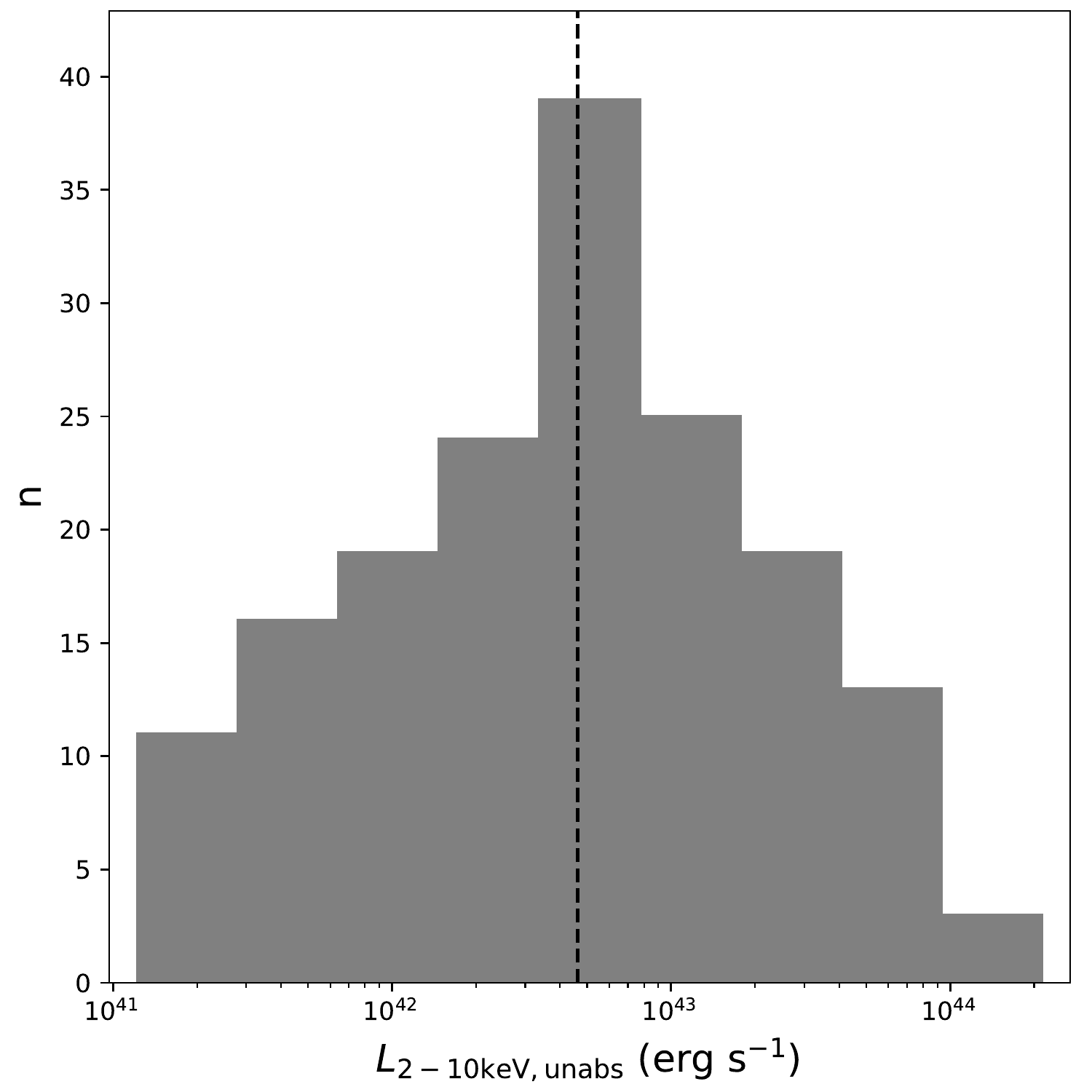}
\caption{\footnotesize{Distribution of unabsorbed, hard X-ray luminosities (\LXH) for the final sample of \HLXCandSZ~HLX candidates. The values of \LXH~are determined from X-ray spectral fitting (Section \ref{sec:xrayspec}). The median value is denoted by the vertical dashed line and is typical of SMBHs but also efficiently accreting IMBHs.}}
\label{fig:LXHUNABS}
\end{figure}

\subsubsection{Final Sample of HLX Candidates}
\label{sec:finalsamp}

The rest-frame X-ray fluxes are converted to luminosities using the host galaxy redshifts and the cosmology stated in Section \ref{sec:intro}. Errors in the redshifts and in the flux measurements are propagated into the luminosity calculations. We note that the X-ray spectral models in this project will be a useful bi-product of our study as they provide intrinsic X-ray source properties for each of these \CSC~sources (assuming they are not chance projections; see Section \ref{sec:ContEst} for a further discussion).

To select sources classified as HLXs, we restrict the rest frame, unabsorbed, hard X-ray luminosities (\LXH) to be \LXH~$\geq 10^{41}$ \uLum. The X-ray sources that pass this criterion represent the sample of \HLXCandSZ~HLX candidates. The X-ray luminosities, spectral photon indices, and extragalactic column densities are listed in Table \ref{tab:Xray}. The distribution of \LXH~is shown in Figure \ref{fig:LXHUNABS}. The median value of \LXH~$=$~\LHunabsMEDIAN~\uLum~is typical for SMBHs but is also consistent with some IMBHs observed in dwarf galaxies \citep{Chilingarian:2018,Mezcua:2018} and in off-nuclear regions of galaxies \citep{Farrell:2009}.

\section{Analysis and Results}
\label{sec:analysis_results}

In this section we analyze our sample of HLX candidates to put constraints on their physical origin. In Section \ref{sec:ContEst} we estimate the fraction of unknown contaminants, in Section \ref{sec:optcntrpart} we compute values or upper limits on the stellar counterpart fluxes and masses, in Section \ref{sec:XRayProps} we analyze the X-ray properties, in Section \ref{sec:AGNprop} we analyze the multi-wavelength properties for AGN signatures, in Section \ref{sec:BHmass} we estimate BH masses and identify potential IMBH candidates, and in Section \ref{sec:comparison} we compare our sample with previous catalogs.

\begin{figure*}[t!]
\hspace*{0.1in} \includegraphics[width=0.95\textwidth]{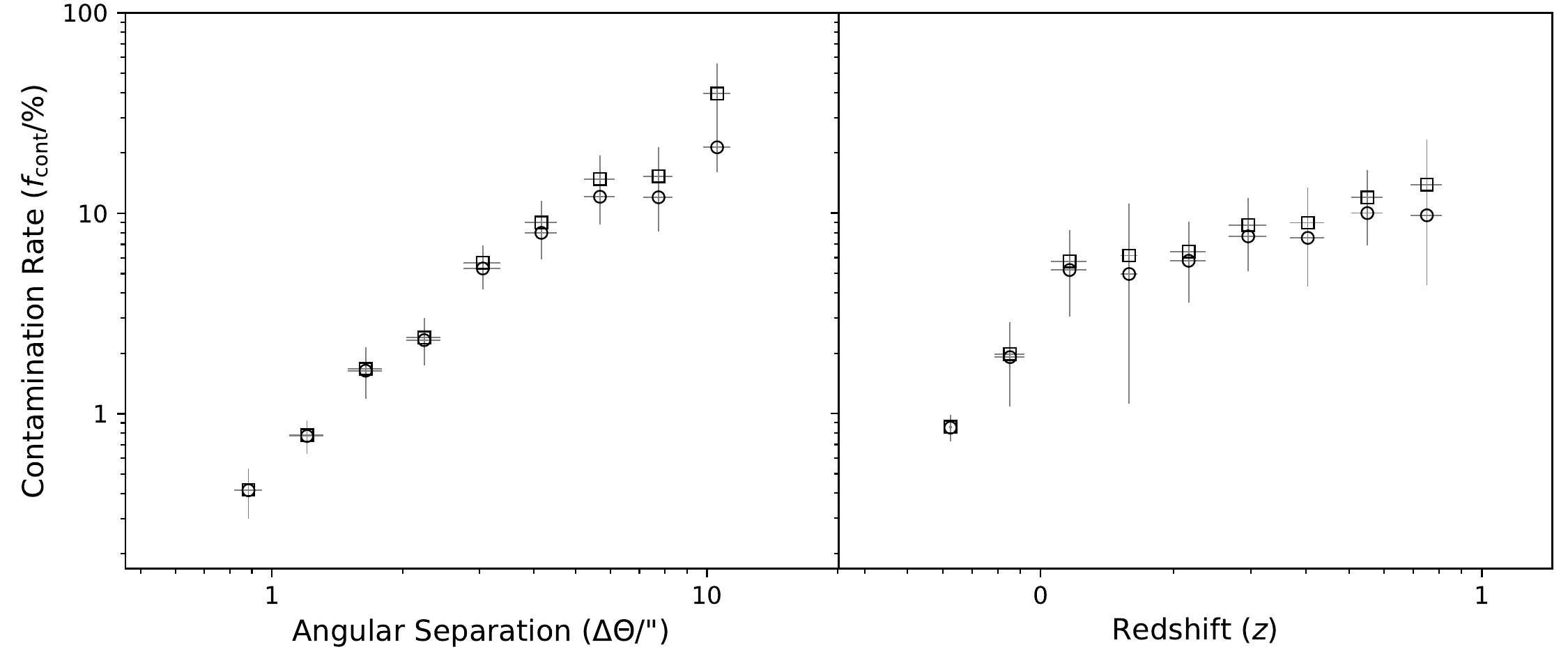}
\caption{\footnotesize{Contamination rate (\fcont) as a function of angular separation (\DeltaTheta; left) and redshift (\z; right). The values of \fcont~are determined from the X-ray cumulative flux distribution (Section \ref{sec:ContEst}). The ordinates reflect average values in evenly-spaced bins along the abscissa, and the error bars represent the two-dimensional scatter in each bin. The \ncontSymb~represent estimates using our original method, and the \ncontindSymb~represent estimates using the modification from \citet{Corredoira:2006}. Note that \fcont~has a stronger (positive) dependence on \DeltaTheta~than on \z.}}
\label{fig:FCONT_SEP_Z}
\end{figure*}

\subsection{Unknown Contaminants}
\label{sec:ContEst}

The HLX candidates in our sample do not have known redshifts, and some of them may be contaminants (background or foreground sources not physically interacting with the host galaxy). In this section we use X-ray source surface densities to estimate the number of unknown X-ray contaminants for each host galaxy. Our approach closely follows that of \citet{Walton:2011} and \citet{Zolotukhin:2016}, which utilizes the cumulative flux distribution of hard X-ray sources \citep{Moretti:2003} to compute the number of X-ray sources per solid angle above a given limiting sensitivity. For each galaxy, the limiting sensitivity is computed individually and is set to the smallest of the following two quantities: the \CSC~limiting flux of our sample (faintest X-ray source in the original matched sample: \FXLim~\uFlux) and the minimum flux to detect an X-ray source with \LXH~$>10^{41}$ \uLum~at the host galaxy redshift. 

For each galaxy, we then use the surface density to compute the expected number of X-ray sources within a solid angle defined by the angular offset between the HLX candidate and galaxy centroid. Finally, normalizing by the number of observed HLX candidates for each galaxy yields the contamination fraction. As in \citet{Zolotukhin:2016}, we also use the correction value of $0.7$ to account for the ratio between the broad-band \ch~flux used in our X-ray source detections and the hard X-ray sources that determine the cumulative flux distribution from \citet{Moretti:2003}.

Using this method, there are two observables that control the predicted number of random sources falling within a specified solid angle around a galaxy: the angular offset of the X-ray source, \DeltaTheta~(used to compute the solid angle), and the galaxy redshift, \z~(used to compute the limiting sensitivity from the luminosity threshold of \LXH~$>10^{41}$ \uLum). Figure \ref{fig:FCONT_SEP_Z} shows that the contamination rates are positively dependent on both \DeltaTheta~and \z~(reaching $\sim20-40\%$ for the highest angular offsets and redshifts). For comparison, we have also computed contamination rates following the approach in \citet{Corredoira:2006}. This method estimates the probability that all X-ray sources associated with a galaxy are contaminants from a Poisson distribution that is described by the maximum number of contaminating sources. This number is further modified by a term that accounts for true HLXs being brighter and closer to galaxy nuclei than randomly expected. Figure \ref{fig:FCONT_SEP_Z} shows that the estimates from these two methods are consistent to within the uncertainties over the full range of parameter space explored. The contamination rates are listed in Table \ref{tab:Add}.

The two approaches indicate that the average contamination rate among the full sample is \ncontmeanmin~$-$~\ncontmeanmax~$\%$. This predicted frequency of background or foreground sources is smaller than average estimates from previous catalogs \citep[$\sim20-30\%$;][]{Walton:2011,Zolotukhin:2016,Gong:2016}, and this difference is discussed in Section \ref{sec:comparison}. Ultimately, spectroscopic redshifts of the HLX candidates are necessary to prove or disprove association with the host galaxy.

\begin{figure*}[t!]
\includegraphics[width=\textwidth]{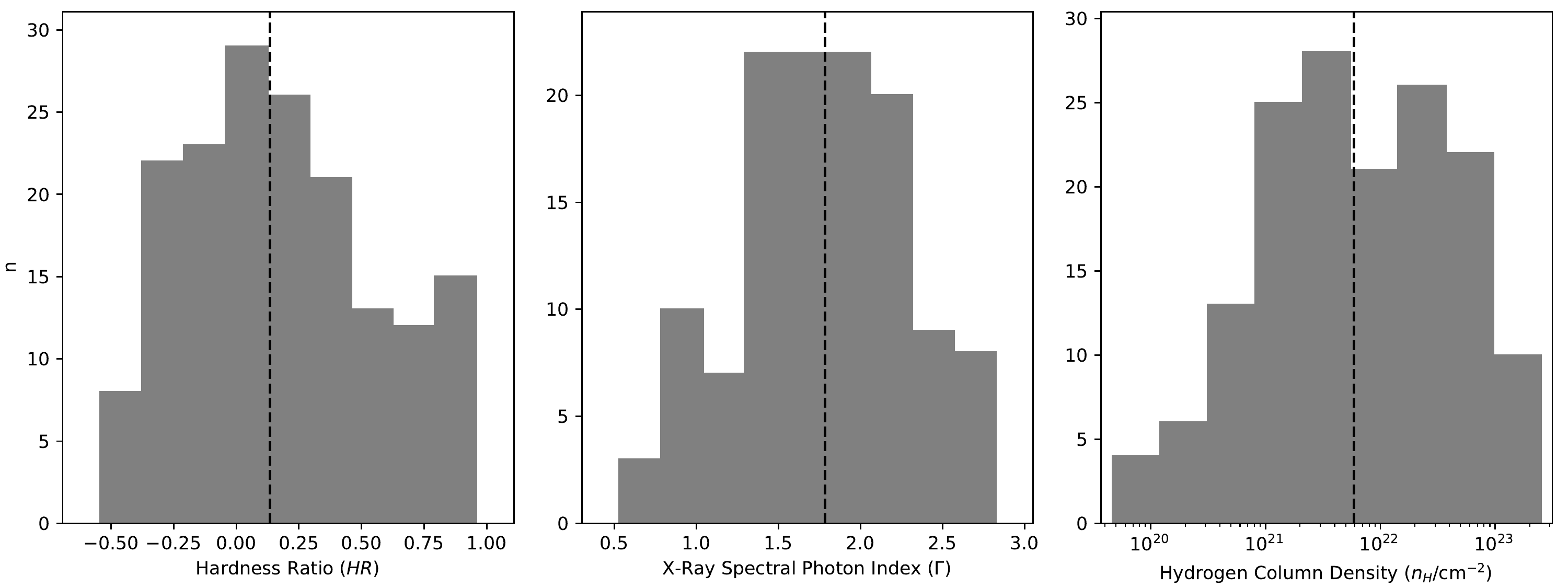}
\caption{\footnotesize{The distribution of hardness ratios (\HR) for the full HLX candidate sample is shown in the left panel. The middle panel shows the distribution of intrinsic X-ray spectral photon indices (\PhotIndx; middle) for the subset with freely varying values of \PhotIndx, and the right panel shows the extragalactic hydrogen column densities (\nHexgal; right) for the subset with freely varying values of \nHexgal. \HR~is described in Section \ref{sec:SpecShape}, and the values of \PhotIndx~and \nHexgal~are determined in Section \ref{sec:xrayspec}. In all panels the solid vertical lines denote the median values of the each distribution. Note that the distribution of photon indices indicates relatively efficient accretion, while the extragalactic hydrogen column densities indicate modest to significant intrinsic absorption (\nHexgal~$>10^{22}$ \uColDens) in \NHfreeObscFrac$\%$ of the sample.}}
\label{fig:HR_GAMMA_NH}
\end{figure*}

\subsection{Optical Counterparts and Stellar Masses}
\label{sec:optcntrpart}

In this section we measure or place upper limits on the magnitudes and stellar masses of optical counterparts that are spatially coincident with the HLX candidates. To detect optical counterparts, we use stacked images from Data Release 1 of the \pnstrsname~(\pnstrs) since they have a finer pixel scale (\PNSTRSPixScale) compared to that of the \sdss~(\SDSSPixScale), and the sky coverage provides uniform imaging for all of our HLX candidates and host galaxies.

We do this analysis on the \pnstrs~\rband- and \iband-band images \citep[filters with the highest overall responses;][]{Chambers:2016} to compute mass-to-light ratios and estimate stellar masses. After registering the \pnstrs~\rband- and \iband-band Stack images with the \ch~images following the same procedure as described in Section \ref{sec:regastr}, we search for detections associated with the HLX candidates by modeling the images with a combination of multiple Sersic profiles (intended to represent stellar bulges) and a uniform background component. The procedure is based on \citet{Barrows:2017b}, though here we summarize the basic steps: the model initially consists of two Sersic components, with one set to the primary galaxy position and the other set to the HLX candidate position (in the \pnstrs~reference frame). Additional Sersic components are added to the model to fit other sources in the field that are detected by \se. The models are fit using the implementation of the Levenberg-Marquardt algorithm \citep{Levenberg:1944,Marquardt:1963} within the \AstropyModeling~package in \Astropy~\citep{Astropy:2013,Astropy:2018}, and all parameters of each Sersic component and the background amplitude are allowed to vary freely. An HLX candidate is considered to have an optical counterpart within a given band if its model Sersic component position is within the \ErrPosMatchThreshFac$\sigma$ error ellipse of the X-ray source position and has an amplitude that is $\geq$~\detthresh$\sigma$ above the average local background determined from an annulus directly surrounding the source (inner radius of two effective radii) and a width of two times the primary host galaxy effective radius.

To compute stellar masses (\Mstar) of the optical detections associated with the primary galaxy and the HLX, we use the \iband-band mass-to-light ratio from \citet{Bell:2003} that is based on the \rband~$-$~\iband~color. The \iband-band luminosities were computed from the Sersic component \iband-band flux and the \iband-band \Kcorrection~generated from galaxy template fits through the \sdss~pipeline. Stellar masses are only computed for optical counterparts detected in both the \iband- and \rband-bands. We find that \iOptDetSZ~of the HLX candidates (\iOptDetFrac$\%$) have stellar counterpart detections, \iOptDetLoMassSZ~of which (\iOptDetLoMassFrac$\%$) are consistent with being dwarf galaxies (\Mstar~$<10^{9.5}$\MSun). The host galaxy stellar masses and the HLX stellar counterpart masses (if detected) are listed in Table \ref{tab:Add}. For non-detections, the host galaxy stellar mass should represent a hard upper limit to the HLX stellar counterpart mass.

\begin{deluxetable*}{ccccccccc}
\tabletypesize{\footnotesize}
\tablecolumns{9}
\tablecaption{Galaxy Properties and Multi-Wavelength Detections of the HLX Candidates.}
\tablehead{
\colhead{I.D. \vspace*{0.05in}} &
\colhead{\fcont} &
\colhead{\fcontind} &
\colhead{\MstarHost} &
\colhead{\MstarHLX} &
\colhead{log[\XtoO]} &
\colhead{\LXHXRB} &
\colhead{\LXHISM} &
\colhead{MIR AGN} \\
\colhead{($-$) \vspace*{0.05in}} &
\colhead{($\%$)} &
\colhead{($\%$)} &
\colhead{(log[\MSun])} &
\colhead{(log[\MSun])} &
\colhead{($-$)} &
\colhead{(\LXXRBnorm~\uLum)} &
\colhead{(\LXISMnorm~\uLum)} &
\colhead{($-$)} \\
\colhead{1} &
\colhead{2} &
\colhead{3} &
\colhead{4} &
\colhead{5} &
\colhead{6} &
\colhead{7} &
\colhead{8} &
\colhead{9}
}
\startdata
1 \vspace*{0.05in} & $7.2$ & $6.7$ & $10.0\pm7.7$ & $9.6\pm8.7$ & $-1.01$\tablenotemark{~} & $0.88_{-0.48}^{+1.05}$ & $0.49_{-0.27}^{+0.59}$ & yes\\ 
2 \vspace*{0.05in} & $2.8$ & $2.7$ & $10.5\pm7.9$ & $$-$$ & $-1.71\tablenotemark{a}$ & $6.07_{-3.34}^{+7.23}$ & $148_{-80}^{+175}$ & no\\ 
3 \vspace*{0.05in} & $2.6$ & $2.5$ & $11.2\pm8.1$ & $$-$$ & $-0.81\tablenotemark{a}$ & $16.3_{-8.9}^{+19.4}$ & $38.5_{-20.9}^{+45.7}$ & no\\ 
4 \vspace*{0.05in} & $6.8$ & $6.3$ & $11.1\pm7.5$ & $$-$$ & $-3.05\tablenotemark{a}$ & $14.0_{-7.7}^{+16.7}$ & $147_{-80}^{+174}$ & no\\ 
5 \vspace*{0.05in} & $29.8$ & $22.1$ & $10.8\pm8.0$ & $10.0\pm9.0$ & $\phantom{+}0.07$\tablenotemark{~} & $26.0_{-14.4}^{+31.1}$ & $1030_{-560}^{+1220}$ & no\\ 
6 \vspace*{0.05in} & $8.2$ & $7.5$ & $11.5\pm8.7$ & $$-$$ & $-0.79\tablenotemark{a}$ & $51.1_{-28.1}^{+60.9}$ & $1120_{-610}^{+1330}$ & no\\ 
7 \vspace*{0.05in} & $6.2$ & $5.8$ & $10.9\pm7.5$ & $$-$$ & $-0.70\tablenotemark{a}$ & $6.81_{-3.71}^{+8.10}$ & $0.18_{-0.10}^{+0.22}$ & no\\ 
8 \vspace*{0.05in} & $12.0$ & $10.7$ & $10.2\pm8.3$ & $$-$$ & $-0.43\tablenotemark{a}$ & $3.67_{-2.03}^{+4.37}$ & $119_{-65}^{+142}$ & no\\ 
9 \vspace*{0.05in} & $7.6$ & $7.0$ & $12.0\pm9.7$ & $$-$$ & $-0.05\tablenotemark{a}$ & $106_{-58}^{+126}$ & $272_{-148}^{+323}$ & no\\ 
10 \vspace*{0.05in} & $16.9$ & $14.3$ & $11.1\pm7.1$ & $$-$$ & $-2.12\tablenotemark{a}$ & $11.7_{-6.4}^{+13.9}$ & $53.5_{-29.1}^{+63.5}$ & no     
\enddata
\tablecomments{Column $1$: unique identifier for the galaxy$-$HLX candidate pair; Columns $2-3$: contamination rates; Column $4$: host galaxy stellar mass; Column $5$: HLX stellar counterpart mass (if detected; otherwise, \MstarHost~is considered to be the upper limit); Column $6$: \XtoOname~flux ratio; Columns $7-8$: estimated hard X-ray luminosities from XRBs and hot ISM gas; and Column $9$: whether or not the HLX candidate is detected as a MIR AGN. (This table is available in its entirety in a machine-readable form in the online journal. A portion is shown here for guidance
regarding its form and content.)\\$^{a}$lower limit (Section \ref{sec:XtoO}).}
\label{tab:Add}
\end{deluxetable*}

\subsection{X-Ray Properties}
\label{sec:XRayProps}

In this section we analyze the HLX candidate X-ray properties to constrain their nature. Specifically, in Section \ref{sec:SpecShape} we describe the X-ray spectral parameters, in Section \ref{sec:XtoO} we compare the X-ray fluxes to the optical counterparts, and in Section \ref{sec:Excess} we determine which HLX candidates have X-ray luminosities requiring the presence of an accreting massive BH (i.e. IMBH or SMBH).

\subsubsection{Spectral Shape}
\label{sec:SpecShape}

The X-ray spectral shapes of the HLX candidates encode information about the emission and absorption processes near the X-ray source. The X-ray spectral hardness ratio - commonly defined as \HR~$=(H-S)/(H+S)$, where $H$ and $S$ are the numbers of rest-frame hard and soft counts, respectively - is a commonly used parameter to describe the spectral shape of X-ray sources. We compute hardness ratios from the merged observations (see Section \ref{sec:resolved} for a description of merging OBSIDs) using the \texttt{Bayesian Estimation of Hardness Ratios} program \citep{Park:2006}. Values of $S$, $H$, and \HR~are listed in Table \ref{tab:Xray}. The left panel of Figure \ref{fig:HR_GAMMA_NH} shows the distribution of $HR$ for the full HLX candidate sample. The median value (\HR~$=$~\HRMEDIAN) is relatively hard \citep[compared to e.g. the \chcosmosleg~Survey median of \HR~$=-0.2$;][]{Civano:2016} and suggests either intrinsically hard X-ray spectra or significant absorption. To disentangle the components of emission and absorption, we use the results of the X-ray spectral modeling (Section \ref{sec:xrayspec}) to put constraints on the physical nature of the HLX candidates. 

In our models the intrinsic shape of the energy flux emitted by the accreting source is controlled by the photon index parameter (\PhotIndx), which can provide information about the accretion efficiency. For instance, Eddington ratios (\FEDD~$=$~\LBOL$/$\LEDD, where \LBOL~is the bolometric luminosity and \LEDD~is the Eddington luminosity) of luminous (\LX~$>10^{41}$ \uLum) SMBHs and IMBHs are observed to positively correlate with \PhotIndx~such that more efficiently accreting sources have softer X-ray spectra (larger values of \PhotIndx), corresponding to a higher fraction of soft X-ray and far-UV photons produced in the inner portion of the accretion disk \citep{Bian:2005,Shemmer:2006,Greene:2007}.

The middle panel of Figure \ref{fig:HR_GAMMA_NH} shows the distribution of photon indices for the subset of our HLX candidates for which \PhotIndx~is not fixed. The median value is \PhotIndx~$=$~\GAMMAMEDIAN. For comparison, among a sample of $40$ dwarf galaxies \citet{Mezcua:2018}~find hardness ratios that are consistent with intrinsic photon indices of \PhotIndx~$=1-1.4$ for X-ray AGN with luminosities comparable to our sample. The relatively softer slopes among our sample may indicate larger values of \FEDD~compared to nuclear IMBHs in isolated dwarf galaxies. The enhanced accretion rates may be due to our selection of potential galaxy mergers that can trigger BH accretion, or alternatively to the presence of luminous QSOs (physically interacting or background AGN) that are known to have high accretion rates. 

In our models the X-ray photon attenuation is parameterized by the extragalactic column density (\nHexgal). In addition to measuring the level of photo-electric absorption, \nHexgal~is also a probe of the relative abundance of gas and dust near the X-ray source. The right panel of Figure \ref{fig:HR_GAMMA_NH} shows the distribution of \nHexgal~for the subset of our HLX candidates for which \nHexgal~is not fixed. The median value is \nHexgal~$=$~\NHMEDIANfree~and \NHfreeObscFrac$\%$ of the HLX candidates have \nHexgal~$>10^{22}$ \uColDens, which is a canonical threshold for significant absorption \citep{Civano:2012}. Large values of intrinsic obscuration may suggest inflows of gas and dust during a phase of BH accretion. Such events are thought to be likely in galaxy mergers, and this would be consistent with the IMBH or SMBH interpretation of the HLX candidates. In this minor merger scenario the BH is located in the nucleus of the less massive galaxy and may experience the most significant enhancement in accretion efficiency relative to the primary galaxy SMBH \citep{Capelo:2015}. However, significant extragalactic column densities may also exist for a distant X-ray source (e.g. a background AGN) that is seen through intervening clouds of gas and dust.

We note that values of \PhotIndx~and \nHexgal~have a level of mutual degeneracy such that large extragalactic column densities can instead be modeled by small photon indices and vice versa. Indeed, when restricted to the subset for which \PhotIndx~is fixed at $1.7$, the average value (\nHexgal~$=$~\NHMEDIANfix) is about four times larger than for the subset with freely-varying \PhotIndx. This difference may indicate that some of the X-ray sources with \PhotIndx~$<1.7$ may have underestimated values of \nHexgal.

\subsubsection{X-Ray-to-Optical Flux Ratios}
\label{sec:XtoO}

The relative dominance of X-ray to optical flux (parameterized by the \XtoOname~flux ratio) can put important constraints on the nature of X-ray sources. The \XtoOname~flux ratio is formally defined as \XtoO, where \fx~is the \EXtoO~observed flux and \fv~is the \Vband-band observed flux \citep{Maccacaro:1988}. Empirically, \XtoO~can generally distinguish between stars ($0.3<$~\XtoO~$<0.9$) and AGN ($0.1<$~\XtoO~$<16$) \citep{Maccacaro:1988,Stocke:1991,Lin:2012}. The \XtoOname~flux ratios for HLXs sometimes approach higher values (\XtoO~$>10$) since they are often identified as bright X-ray sources with faint optical counterparts in nearby galaxies \citep{Tao:2011,Zolotukhin:2016}. \XtoO~is qualitatively similar, though inverted relative, to the \alphaox~spectroscopic parameter (\alphaox~$=$~\alphaoxdef~\citep{Tananbaum:1979}, and indeed SMBHs show larger \alphaox~values \citep{Yuan:1998} compared to IMBHs \citep{Plotkin:2016}. Assuming HLXs are powered by lower-mass BHs, the larger values of \XtoO~can be explained by the higher accretion disk temperatures (and correspondingly higher X-ray production) compared to SMBHs \citep{RDong:2012}. However, stellar continuum contributions from the host galaxies of AGN can also affect the optical flux. To compute \XtoO~for the HLX candidates in our sample, we use the X-ray models to calculate the observed \EXtoO~flux, and we use the \pnstrs~\rband- and \iband-band magnitudes to convert the optical counterpart detections or upper limits (Section \ref{sec:optcntrpart}) to \Vband-band magnitudes following the conversion from \citet{Jester05}. The \XtoOname~flux ratios are listed in Table \ref{tab:Add}.

\begin{figure}[t!]
\hspace{-0.025in} \includegraphics[width=0.48\textwidth]{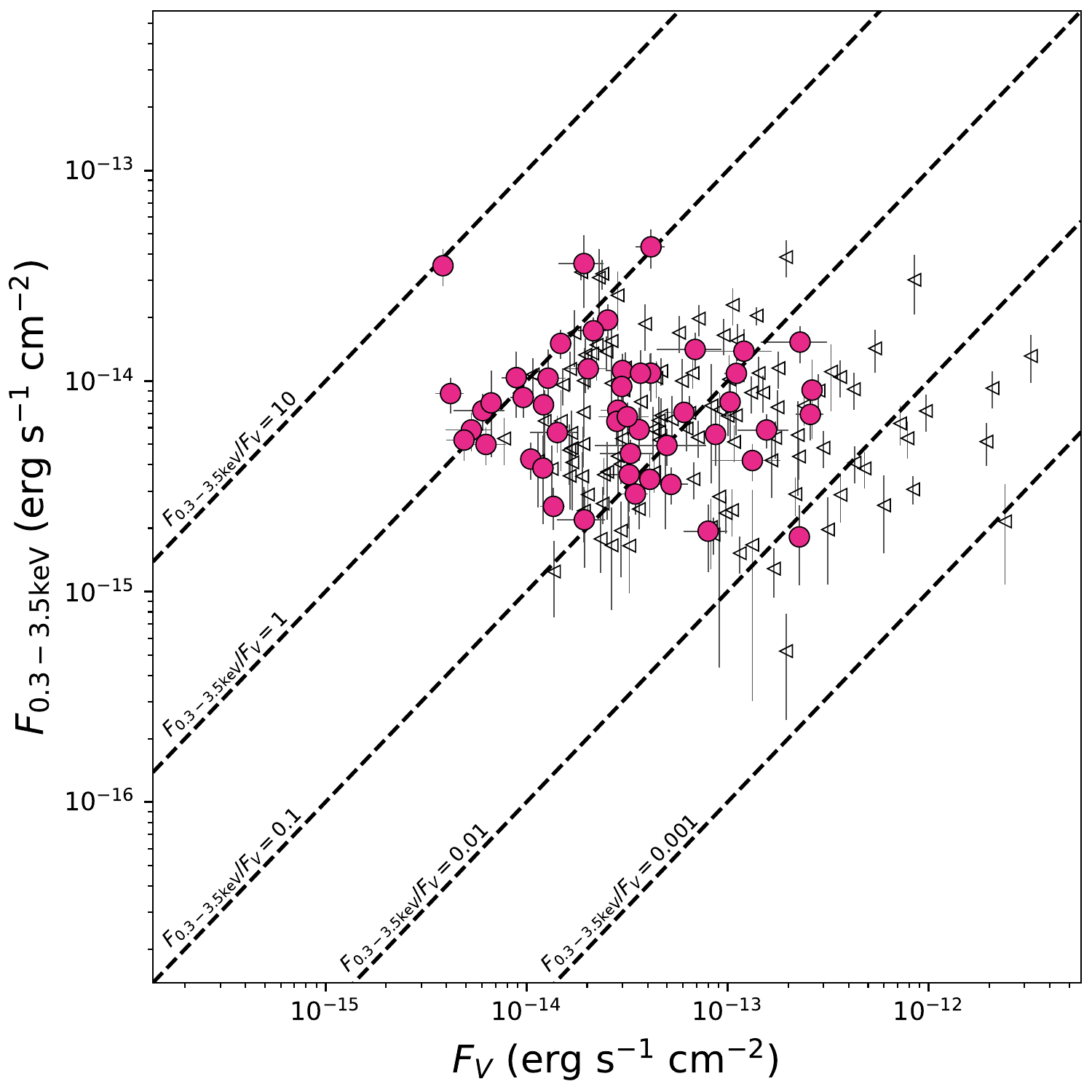} 
\caption{\footnotesize{Observed $0.3-3.5$~\uXRAYE~flux (\fx) as a function of $V$-band flux (\fv). Optical detections are shown as filled, \DetCol~circles while upper limits are shown as open, leftward pointing triangles. The five dashed lines indicate ratios of \fx~$/$~\fv~$=10, 1, 0.1, 0.01, \rm{and}~0.001$ (labeled). Note that most HLX candidates with optical detections are consistent with expectations for AGN (\XtoO~$=0.1-10$).}}
\label{fig:FX_FR}
\end{figure}

Figure \ref{fig:FX_FR} plots the hard X-ray fluxes (\fx) versus the \Vband-band fluxes or upper limits (\fv) for the HLX candidate sample. Figure \ref{fig:FX_FR} reveals that \XtoODetpONEtoTENFrac$\%$ of the optical counterpart detections have \XtoO~values between $0.1$ and $10$ (consistent with expectations for AGN). The \XtoO~values suggest that most of the HLX candidates may be associated with AGN (SMBHs or potentially IMBHs in typical galaxy stellar bulges). Assuming that non-detections of optical counterparts are due to intrinsic faintness, those HLXs without detected optical counterparts are likely to have \XtoO~values generally larger than those with detections, possibly exceeding \XtoO~$=10$ and consistent with the high values observed for some HLXs.

\subsubsection{X-Ray Signatures of AGN}
\label{sec:Excess}
 
HLXs are typically found in spiral galaxies or star-forming galaxies with significant star formation and a large supply of gas, both of which may contribute to the observed hard X-ray emission that is used to select HLXs. Therefore, in this section we determine the expected X-ray contributions from stars (XRBs) and hot ISM gas. We then use these estimates to select the HLX candidates with a significant excess above the XRB and hot ISM contributions since they must be associated with AGN (i.e. accreting IMBHs or SMBHs). 

To estimate the hard X-ray contribution expected from XRBs (\LXHXRB), we employ the relation from \citet{Lehmer:2010} that describes \LXHXRB~as a function of both galaxy stellar mass (\Mstar) and star-formation rate (\SFR): \LXHXRB[\uLum]~\LehmerEqn. While we only need to consider the star formation associated with the X-ray source optical counterpart, in most cases their magnitudes are only measured as upper limits (Section \ref{sec:optcntrpart}). Therefore, we estimate \Mstar~and \SFR~for the entire host galaxy in each case and acknowledge that these values are upper limits. Values of \Mstar~for the host galaxies are computed as described in Section \ref{sec:BHmass} but using the host galaxy photometry instead. Values of \SFR~for the host galaxies are estimated from the relation of \citet{Hopkins:2003}: \SFR$_{u}$[\uSFR]~$=(L_{u}/1.81\times 10^{21} [\rm{W~Hz}^{-1}])^{1.186}$. We calculate $L_{u}$ from the (dust-corrected) \sdss~apparent \uband-band magnitudes of the host galaxy and the photometric \Kcorrection s derived by the \sdss~pipeline. If the galaxy has an \sdss~spectrum then it has values of \Mstar~and \SFR~derived from the \sdss~pipeline, and we use those values instead. The uncertainties on \LXHXRB~are derived from the scatter (0.34 dex) and coefficient uncertainties of the \citet{Lehmer:2010} relation, and by further propagating the uncertainties of \Mstar~and \SFR~through the calculations. We note that, while \citet{Lehmer:2016} present an updated form of this relation, the \citet{Lehmer:2016} AGN threshold of \LX~$>3\times 10^{42}$ \uLum~would miss a significant fraction of X-ray sources in our sample and is therefore not appropriate for distinguishing between AGN and star formation among HLXs. To estimate the contribution expected from hot ISM gas (\LXHISM), we employ the relation in \citet{Mineo:2012} describing the 0.5-2 keV luminosity from ISM gas as a function of SFR and extrapolating to rest-frame $2-10$ keV using a thermal power-law model of spectral index \PhotIndx~$=3$ as in \citet{Mezcua:2018}. The hard X-ray luminosities from XRBs and hot ISM gas are listed in Table \ref{tab:Add}.

\begin{figure}
\includegraphics[width=0.47\textwidth]{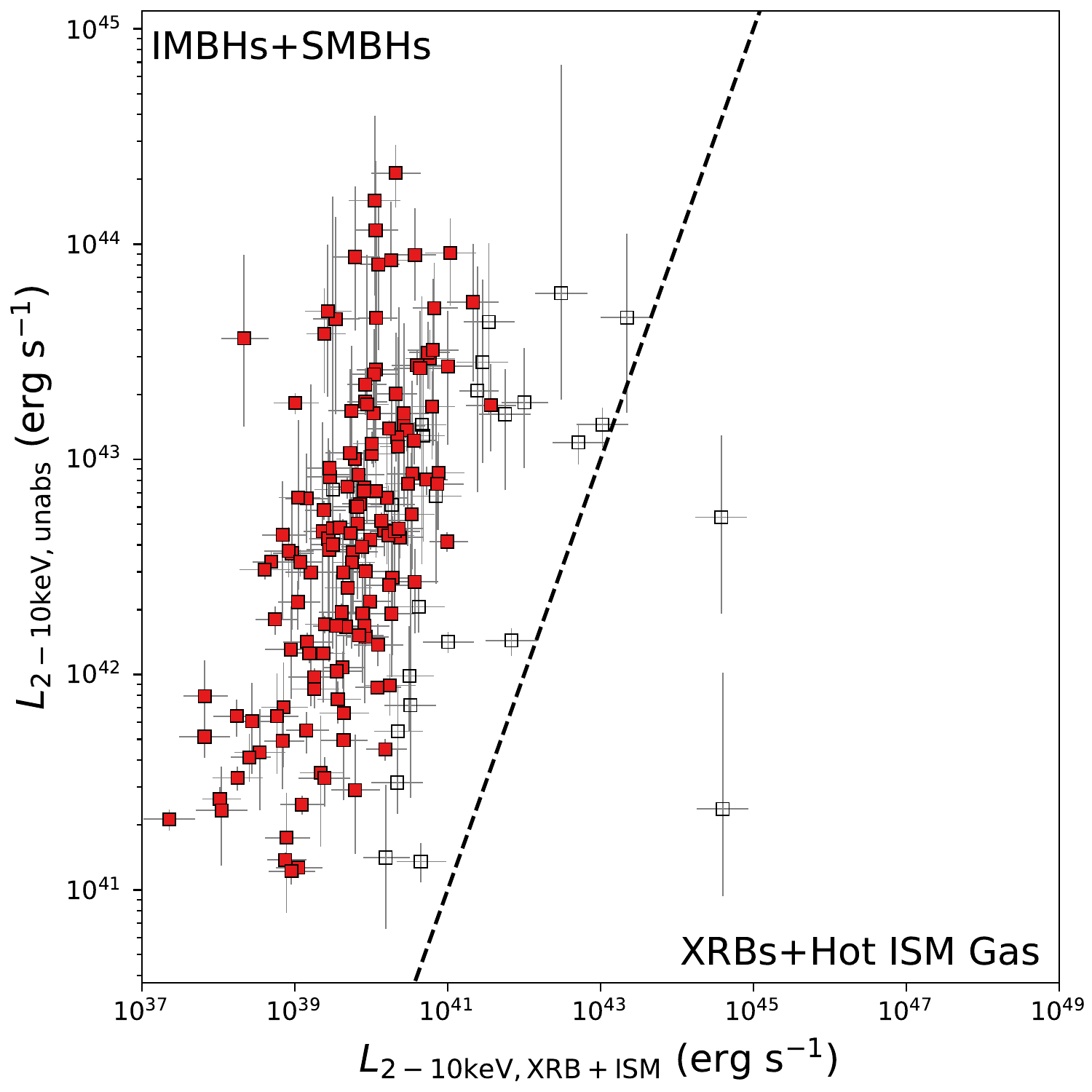} 
\caption{\footnotesize{Rest-frame, unabsorbed, hard X-ray luminosity (\LXH) as a function of expected contribution from XRBs and hot ISM gas (\LXHXRBISM) for the HLX candidates. The dashed line denotes the one-to-one relation (\LXH~$=$~\LXHXRBISM). AGN (i.e. accreting IMBHs and SMBHs) are those HLX candidates where \LXH~$>$~\LXHXRBISM~by more than $5\sigma$ in both axes (\AGNCol-filled symbols).}}
\label{fig:LXH_LXHXRBISM}
\end{figure}

If the \LXH~value of an HLX is more than \supXRBISMsigthresh$\sigma$ greater than \LXHXRB~$+$~\LXHISM~(\LXHXRBISM) then it is considered to have excess emission beyond star formation and hot ISM gas. By implication, the source of the HLX X-ray power must include accretion onto a massive BH (\MBH~$>10^{2}$ \MSun). Figure \ref{fig:LXH_LXHXRBISM} shows \LXH~against \LXHXRBISM~for the full HLX candidate sample, illustrating the selection of sources that pass the above criteria (\supLXXRBISM). We refer to these sources as our sample of HLX AGN candidates. In general, the most luminous of these sources are likely produced by SMBHs, while the best IMBH candidates will be associated with the least luminous ones. However, this distinction is predicated on the assumption of a common Eddington ratio distribution throughout the full sample (this is discussed further in Section \ref{sec:BHmass}). 

\subsection{Multi-Wavelength AGN Detections}
\label{sec:AGNprop}

In the previous section (Section \ref{sec:Excess}) we identified a subset of HLX candidates that must be associated with AGN (accreting IMBHs or SMBHs) based on their X-ray signatures. In this section we search for additional signatures of BH accretion in these sources from multi-wavelength AGN detections to further constrain the nature of the HLX candidates. 

First, we consider radio AGN detections. We perform a cross-match of the HLX AGN candidates with \first~using a search radius equal to the quadrature sum of the RA and DEC errors for the X-ray source offset from the host galaxy nucleus. Following the procedure from \citet{Lofthouse:2018}, we select radio AGN from the \first~cross-match (Section \ref{sec:filter_known_cont}) as sources with $1.4$ GHz luminosities that can not be accounted for by star formation alone  based on the empirical division between purely star-forming galaxies and AGN from \citet{Lofthouse:2018}. This match returns no sources.

Second, we consider mid-infrared (MIR) AGN detections. We perform a cross-match of the HLX AGN candidates with the \wisetitle~\citep[\wise;][]{Wright:2010} \allwise~Source Catalog\footnotemark[10]. 
\footnotetext[10]{http://wise2.ipac.caltech.edu/docs/release/allwise/}
MIR AGN are selected according to the $90\%$ completeness criterion originally defined in \cite{Assef:2013} and further refined in \citet{Assef:2018}: \wiseone~$-$~\wisetwo~$>0.5$. While this criterion has a relatively high contamination rate of non-AGN, we use it to be conservatively inclusive since MIR AGN detections may be preferentially associated with background AGN (see the discussion below). For this selection, we mimic several pertinent cuts to the \allwise~catalog that were used in \citet{Assef:2018}: we require that the \wisetwo~signal-to-noise ratio is $>5$, and the \wiseone~and \wisetwo~magnitudes are fainter than $8$ and $7$ respectively (the saturation limits). This selection yields \WISEAGNcninetySZ~HLX AGN candidates that are also consistent with \wise~AGN detections (\WISEAGNcninetyFRAC$\%$). The MIR AGN status of each HLX candidate is listed in Table \ref{tab:Add}. We note that the $6''$ FWHM spatial resolution of the \wise~\wiseone~and \wisetwo~photometry is such that the galaxy nucleus is also consistent with the \wise~detections in the majority of cases. Thus, the MIR AGN may also be in the galaxy nucleus and not the X-ray source. However, this scenario implies that the putative nuclear MIR AGN would need to be heavily obscured if a nuclear X-ray detection is not present.

\newcommand{\IMBHbriefCaptionText}{Ten examples of IMBH candidates (selected assuming \FEDD~$=$~\FEddMedian; Section \ref{sec:BHmass}) that represent the full redshift range of the sample. Redshift increases left to right, top to bottom. Left: SDSS $g+r+i$ color composite image; middle:  grayscale \pnstrs~\iband-band image; and right: \pnstrs~\iband-band image contours with X-ray (0.5-10kev) counts (binned to the \pnstrs~pixel scale) from the merged \ch~images (Section \ref{sec:resolved}) overlaid. In all panels North is up and East is to the left, the red `+' denotes the galaxy centroid, and the cyan ellipse denotes the HLX candidate position and $5\sigma$ offset significance from the galaxy centroid. Dotted ellipses denote optical counterpart upper limits while solid ellipses denote detections (Section \ref{sec:optcntrpart})}.

\begin{figure*} $
\begin{array}{c c}
\includegraphics[width=0.49\textwidth]{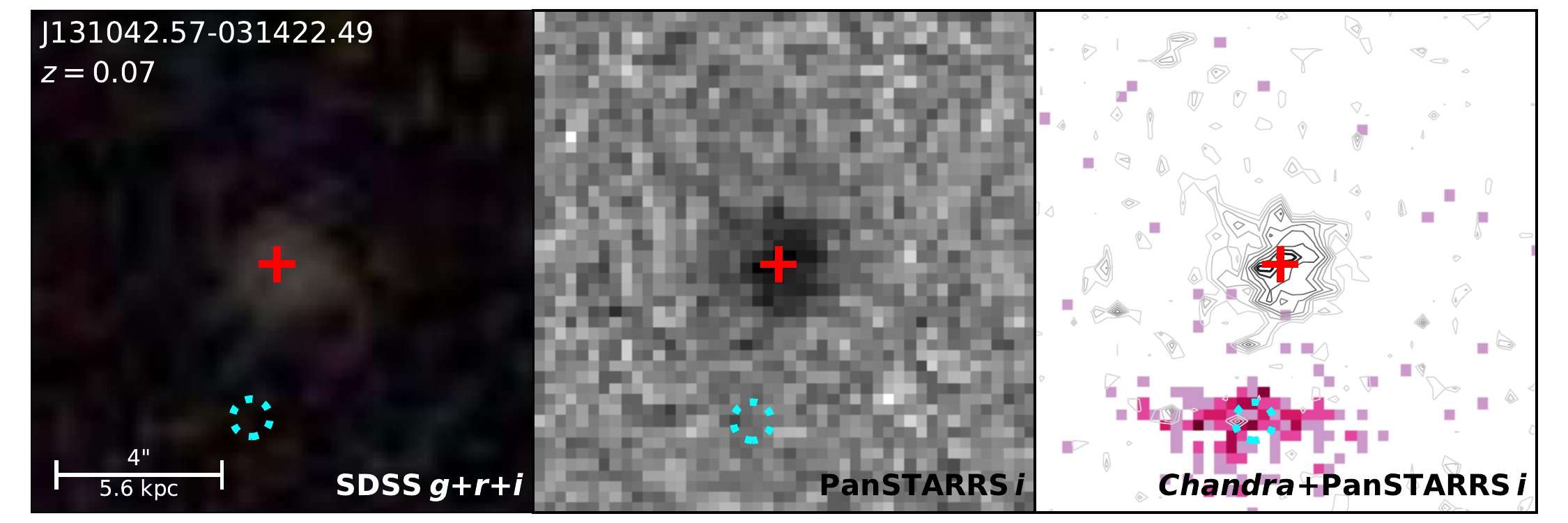} &
\includegraphics[width=0.49\textwidth]{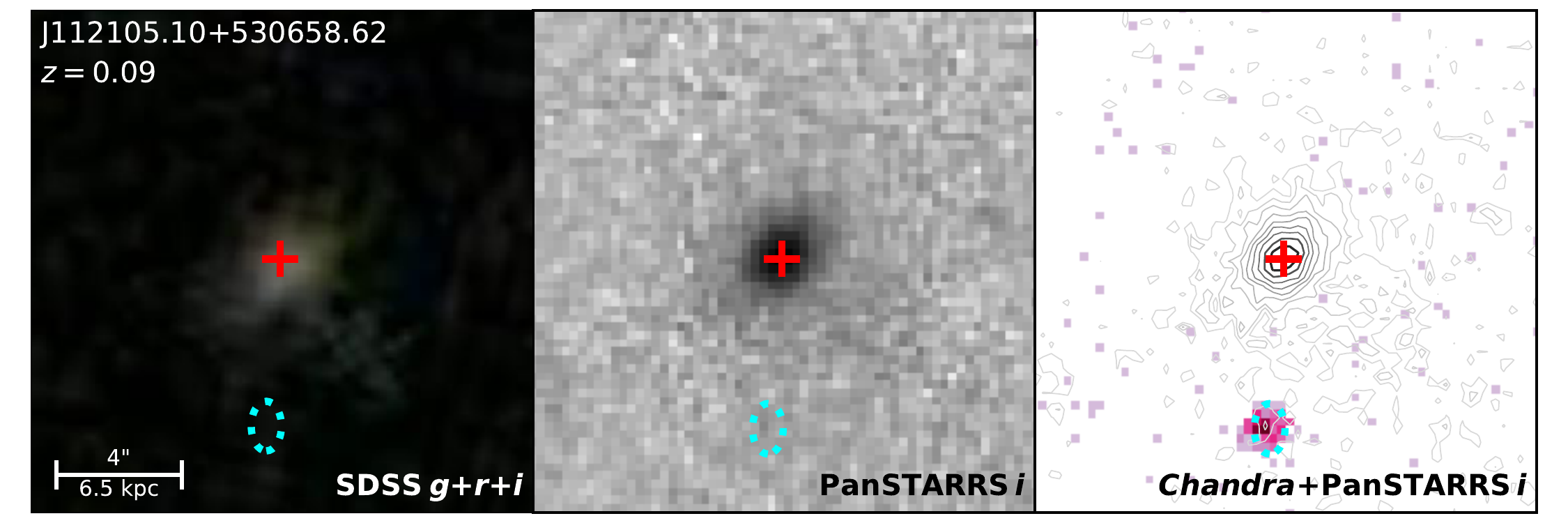} \\
\includegraphics[width=0.49\textwidth]{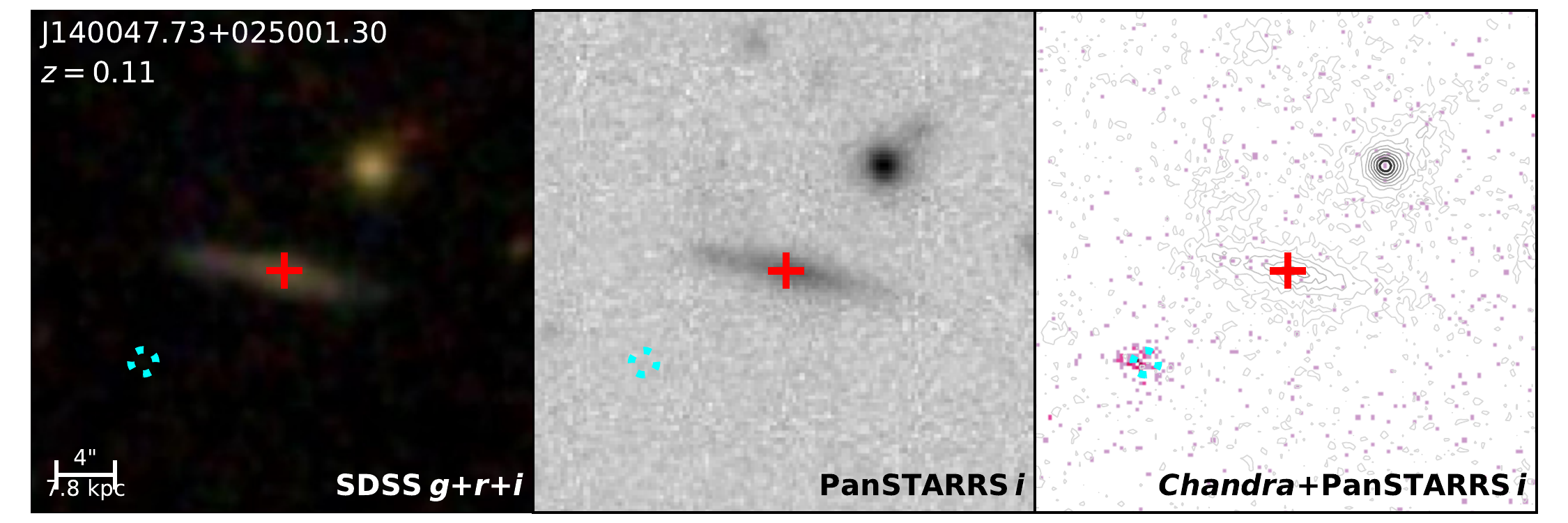} &
\includegraphics[width=0.49\textwidth]{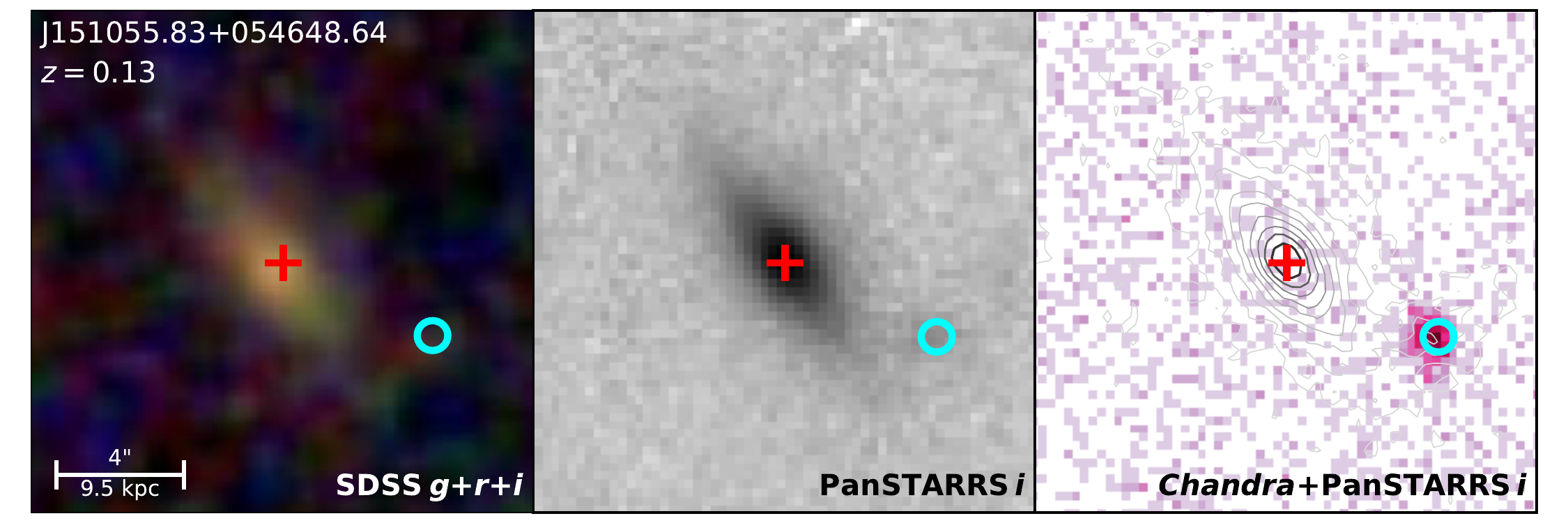} \\
\includegraphics[width=0.49\textwidth]{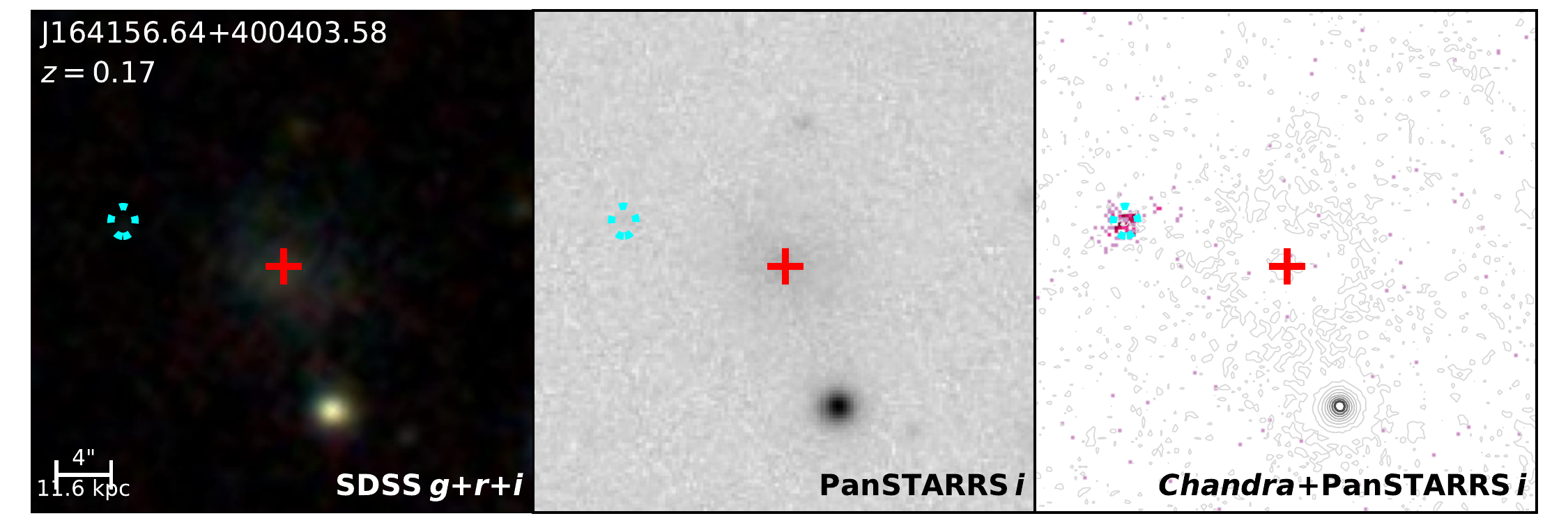} &
\includegraphics[width=0.49\textwidth]{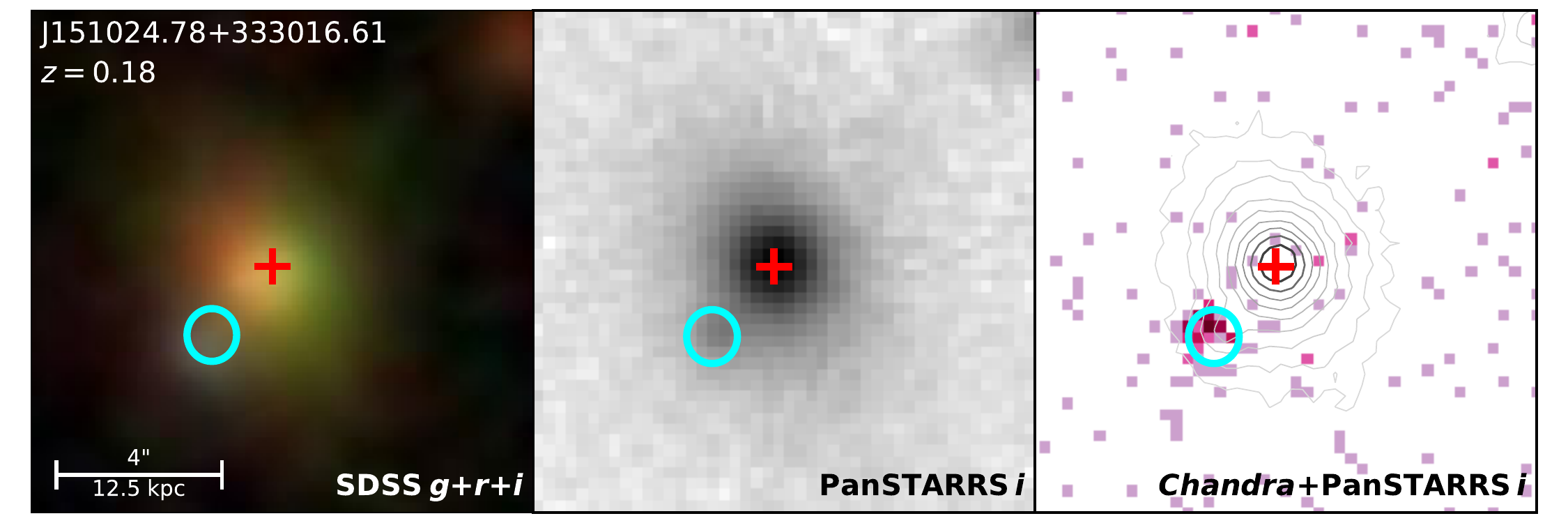} \\
\includegraphics[width=0.49\textwidth]{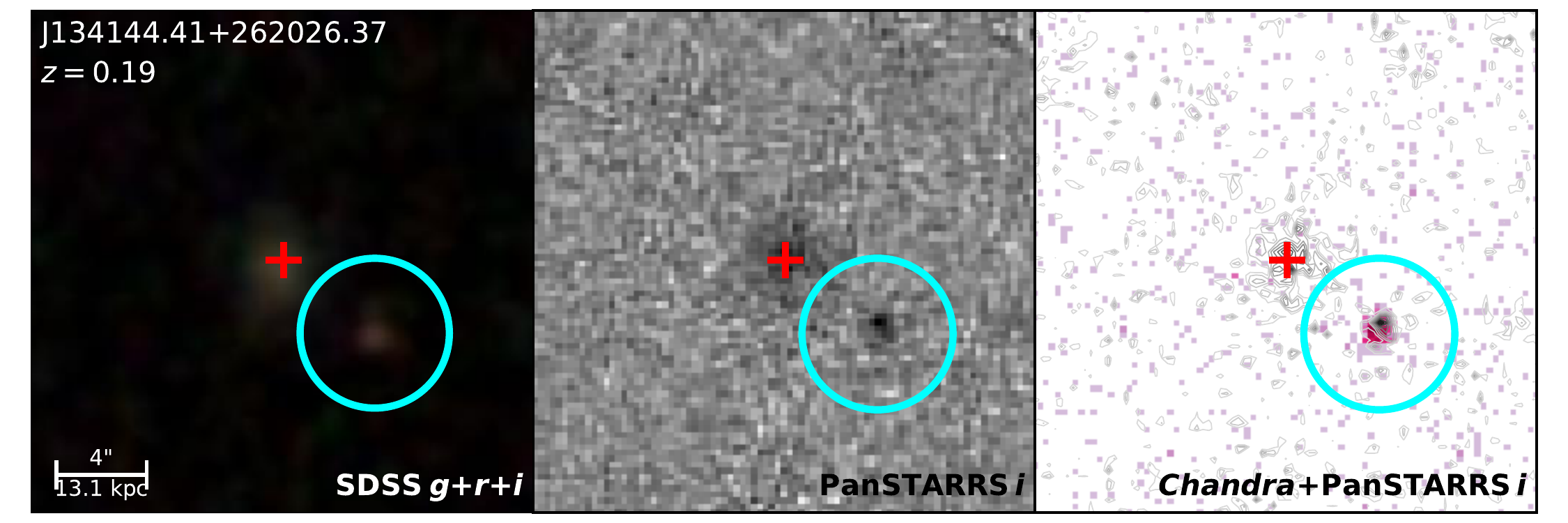} &
\includegraphics[width=0.49\textwidth]{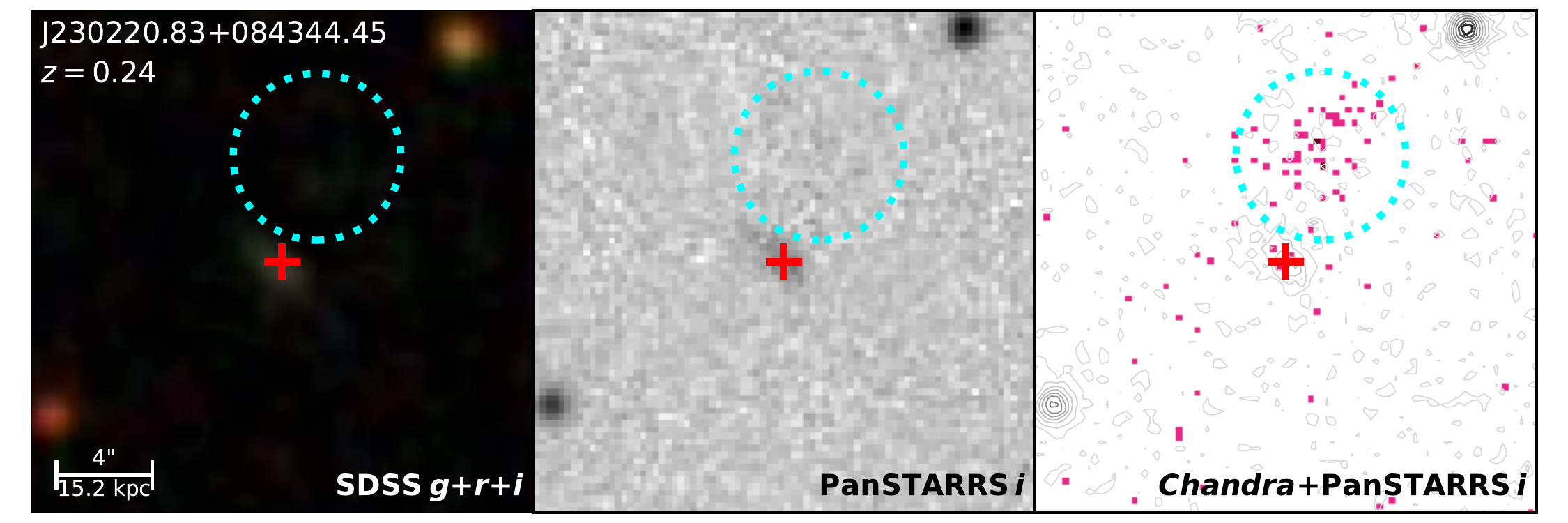} \\
\includegraphics[width=0.49\textwidth]{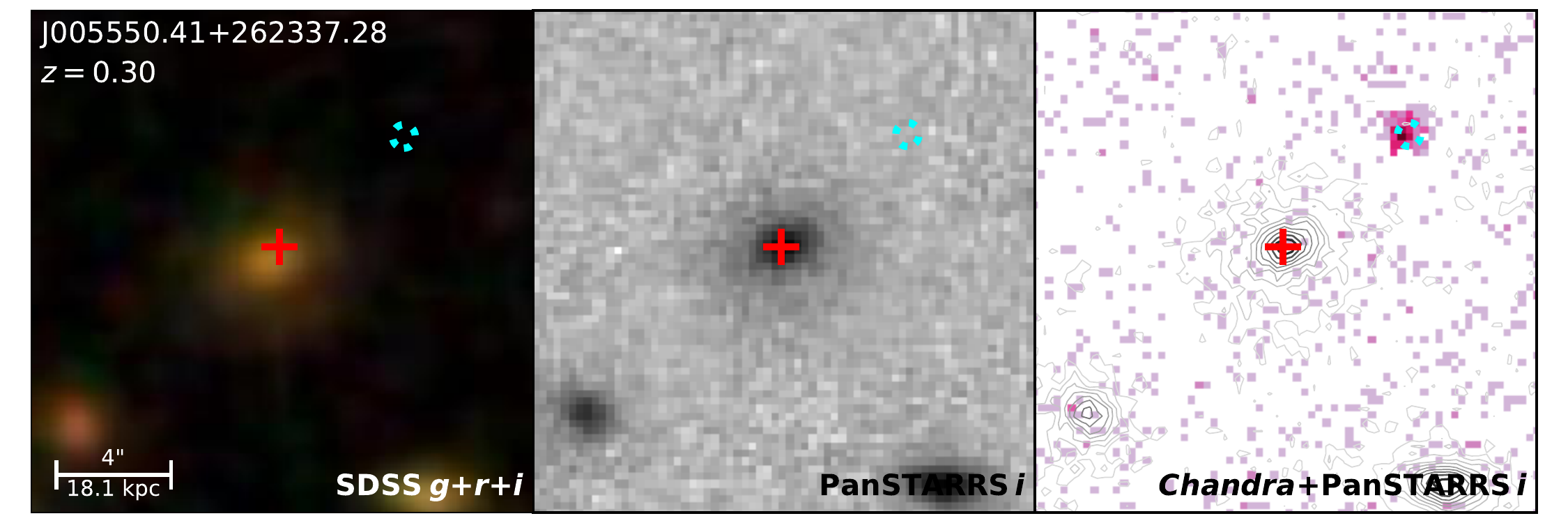} &
\includegraphics[width=0.49\textwidth]{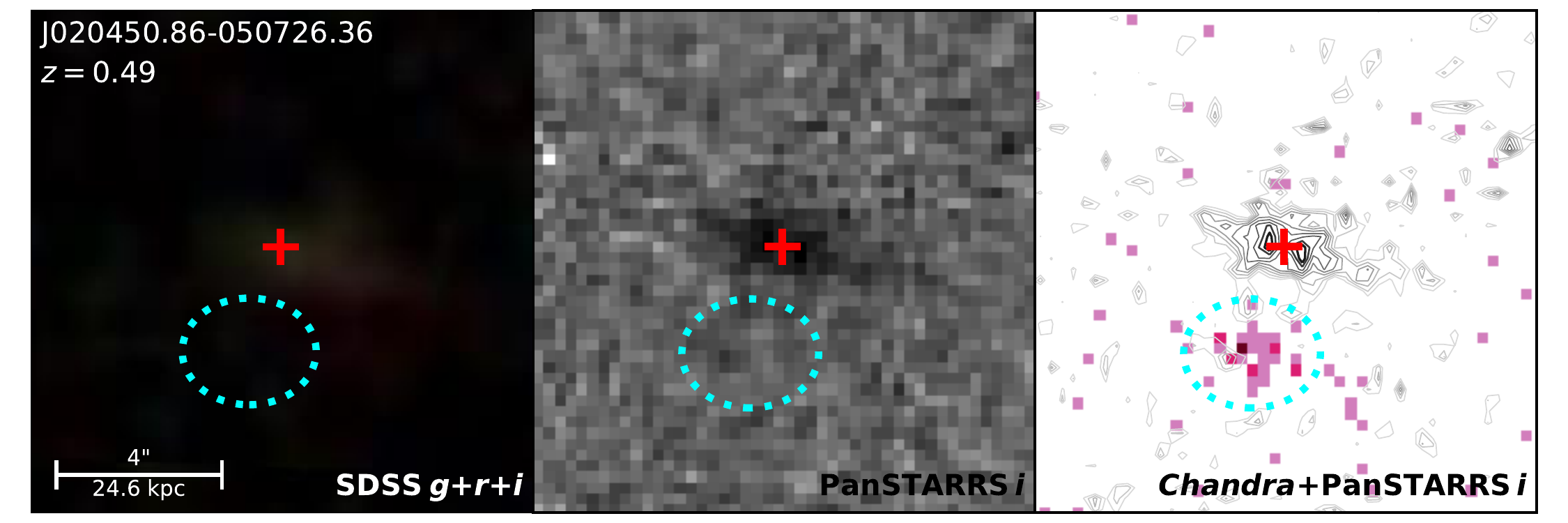} \\
\end{array} $
\caption{\footnotesize{\IMBHbriefCaptionText}}
\label{fig:IMBHbrief}
\end{figure*}

MIR AGN detections are commonly associated with SMBHs \citep[][though see below for a discussion of association with IMBHs]{Stern:2012} and thus may indicate that the X-ray source associated with an HLX candidate is in fact a SMBH. The SMBH producing the AGN signature may also be a chance projection of a physically unrelated source with a small transverse separation but large radial separation. Since the number density of luminous AGN peaks at redshifts of \z~$\sim2$ \citep{Hewitt:Burbidge:1993}, the most likely instance of this scenario is a background AGN. Indeed, MIR AGN detections exhibit a bias toward luminous AGN that outshine their host galaxy's MIR emission from star formation \citep{Hickox:2009,Assef:2011}, further increasing the likelihood of the AGN being at a high redshift. The larger median hard X-ray luminosity among the MIR AGN subsample (\LHunabsWAMEDIAN~\uLum) compared to the full sample (\LHunabsMEDIAN~\uLum) is consistent with this prediction, though the difference is not statistically significant.

Alternatively, the MIR AGN may be associated with a SMBH or IMBH that is physically interacting with the candidate host galaxy. To test if the HLXs with MIR AGN detections are consistent with SMBHs or IMBHs, we compare them with the characteristic correlation between the X-ray and MIR luminosities of accreting AGN. In particular, \citet{Stern:2015} shows that the \LXHtoLsixmicron~ratio for AGN varies from $\sim1$ down to $\sim0.03$ over a hard X-ray luminosity range of \LXH~$=10^{42}-10^{46}$\uLum. After computing \Fsixmicron~by linear extrapolation to rest-frame \sixmicron~as in \citet{Stern:2015}, we find that the average \FXHtoFsixmicron~ratio is \XtoMIR. This \LXHtoLsixmicron~ratio corresponds to luminosities of \LXH~$\sim10^{43}-10^{44}$~\uLum~which are larger than the median \LXH~value computed using the host galaxy redshifts of our sample (Figure \ref{fig:LXHUNABS}) and therefore suggests that some of the HLX candidates with MIR AGN detections may be intrinsically more luminous AGN and hence potential background sources. However, we also note that low-mass BHs are known to possess larger X-ray-to-near-IR fluxes compared to AGN \citep{Heida:2014}. If the same pattern persists for X-ray-to-MIR fluxes, then the HLXs with the largest \FXHtoFsixmicron~ratios among our sample may be potential IMBHs with MIR AGN detections.

\subsection{BH Masses and IMBH Candidates}
\label{sec:BHmass} 

The stellar masses estimated from optical counterpart detections in Section \ref{sec:optcntrpart} imply that \supiOptDetLoMassFrac$\%$ of the HLX AGN candidates are hosted by low-mass (\Mstar~$<10^{9.5}$ \MSun) galaxies. An extrapolation of scaling relationships between BH masses and the stellar masses of their host galaxies \citep{Cisternas:2011,Marleau:2013,Reines:2015} suggests that many of these HLX candidates may therefore be low-mass AGN with BH masses \MBH~$<10^{6}$ \MSun~(i.e. IMBHs). To test this hypothesis, we estimate BH masses for the HLX AGN candidates by assigning Eddington ratios to the accreting X-ray sources and using the relation \LEDD[\uLum]~$=1.3\times10^{38}$ \MBH[\MSun] along with the hard X-ray bolometric correction \LBOL~$=10\times$\LXH~\citep{Marconi04}.

We estimate Eddington ratios based on photon indices using a log-linear function relating \PhotIndx~ and \FEDD~from the  \citet{Greene:2007} sample of X-ray-detected IMBHs. Since we do not have \PhotIndx~fits for all of our sample (i.e. some of them were fixed; see Section \ref{sec:xrayspec}) we can not estimate \FEDD~for each source individually. Therefore, we instead use the median value of \PhotIndx~among all fits with a freely varying photon index (\PhotIndx~$=$~\GAMMAMEDIAN; Section \ref{sec:SpecShape}) to estimate a single Eddington ratio (\FEDD~$=$~\FEddMedian) and apply it to each HLX AGN candidate. This median Eddington ratio is consistent with Eddington ratios among other samples of similarly luminous (\LX~$>10^{41}$ \uLum) low-mass BHs \citep{Greene:Ho:2004,RDong:2012}. This assumption yields a BH mass range of \MBH~$=$~\MBHMINMed~$-$~\MBHMAXMed\MSun~with a median value of \MBH~$=$~\MBHMEDMed\MSun. Based on this range, \MBHLOMASSSZMed~of the \supLXXRBISM~HLX AGN candidates have \MBH~$<$~\MBHLOHITHRESH\MSun~and are consistent with being IMBH candidates. Examples of these IMBH candidates throughout the sample redshift range are shown in Figure \ref{fig:IMBHbrief}. The remaining IMBH candidates are shown in the appendix.

\begin{figure}
\hspace{-0.025in} \includegraphics[width=0.48\textwidth]{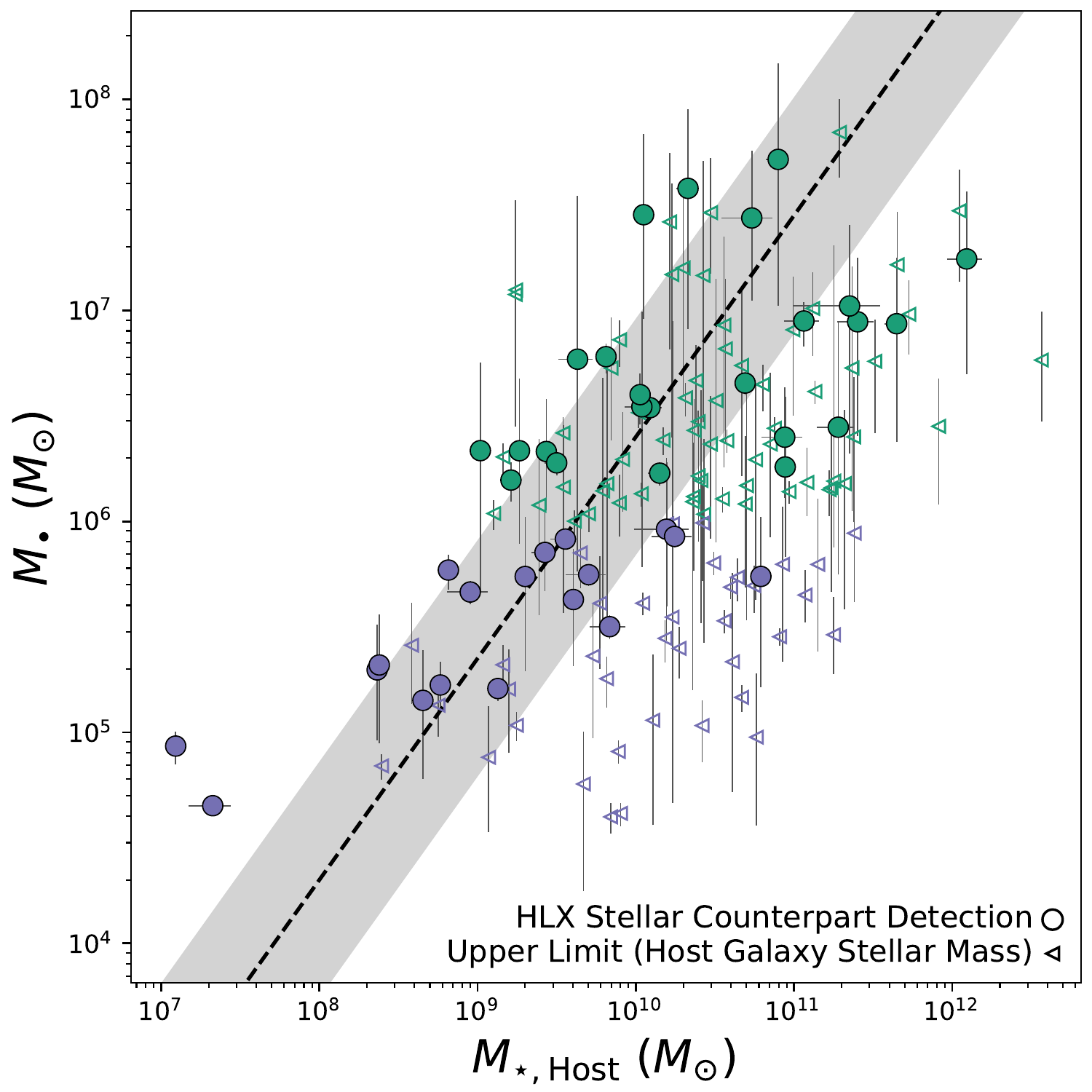} 
\caption{\footnotesize{BH mass estimates (\MBH) for the HLX AGN candidates as a function of stellar counterpart mass (\Mstar). Values of \MBH~are estimated from the X-ray luminosities and assuming an Eddington ratio of \FEDD~$=$~\FEddMedian~(Section \ref{sec:BHmass}) and values of \Mstar~are estimated from image modeling (Section \ref{sec:optcntrpart}). Detections are shown as filled circles while upper limits (set at the host galaxy stellar mass) are shown as open, leftward facing triangles. IMBH candidates (\MBH~$<10^{6}$ \MSun) are shown in \IMBHCol~while SMBH candidates (\MBH~$>10^{6}$ \MSun) are shown in \SMBHCol. The black dashed line is the relation between \MBH~and \Mstar~from \citet{Reines:2015} for AGN that includes a sample of low-mass galaxies, and the gray-shaded region indicates the scatter of their fit. The HLX AGN candidates are generally consistent with the \citet{Reines:2015} relation, suggesting that the BHs have grown with their host galaxies at a rate similar to the general population of AGN. As expected, the stellar counterpart mass upper limits are generally over-massive compared to the BH masses.}}
\label{fig:MBH_MGAL}
\end{figure}

These BH mass estimates are plotted against the masses for their stellar counterpart detections or upper limits (\Mstar~from Section \ref{sec:optcntrpart}) in Figure \ref{fig:MBH_MGAL}. Since the majority of the detected stellar counterpart masses are in the low-mass regime (\Mstar~$<10^{9.5}$ \MSun), for comparison we show the relation between BH mass and host galaxy total stellar mass from \citet{Reines:2015} for AGN since it includes a sample of low-mass galaxies. No systematic offset toward over- or under-massive BHs is detected. Therefore, under our assumption of \FEDD~$=$~\FEddMedian, the relationship between the HLX BH masses and their stellar counterparts follows the same relationship between active BHs and their host galaxies and strongly suggests that the HLX AGN candidates are generally associated with accreting BHs in galaxy nuclei.

The lack of a systematic offset above the relation indicates that significant tidal stripping of the secondary galaxy has not yet occurred. However, a few individual sources do have significantly under-massive stellar counterparts. Furthermore, if the undetected stellar counterparts are intrinsically faint, then they may also reside in stripped stellar cores. Alternatively, the over-massive BHs may have grown efficiently through stochastic mechanisms independent of merger-driven galaxy growth \citep{Martin:2018}. Additionally, for the subset of host galaxy stellar masses or upper limits that are consistent with dwarf galaxies, the X-ray sources might instead be BHs that formed in-situ but are wandering throughout the galaxy and hence appear to be off-nuclear \citep{Bellovary:2019}.

The X-ray luminosity threshold of \LXH~$\geq10^{41}$ \uLum~used to select HLX candidates likely introduces a bias toward efficiently accreting sources and will naturally miss all IMBHs with Eddington ratios \FEDD~$\lesssim10^{-2}$. Indeed, several studies have identified low-mass BHs with Eddington ratios of \FEDD~$\sim10^{-2}$ \citep{Yuan:2014,Baldassare:2015} and even as low as \FEDD~$\sim10^{-3}$ \citep{Baldassare:2017}. Therefore, to test the effect of these smaller Eddington ratios on our interpretations, we apply the median value of \FEDD~$=$~\FEDDlo~from the sample of \citet{Baldassare:2017} that results in \MBHLOMASSSZlo~of the \supLXXRBISM~HLX AGN candidates satisfying the IMBH criteria of \MBH~$<$~\MBHLOHITHRESH\MSun. Under this assumption, the BH mass estimates are systematically offset by up to $\sim1$ order-of-magnitude above the \citet{Reines:2015} relation, implying that tidal stripping or highly efficient, merger-free BH growth has occurred.

\subsection{Comparison with Other Catalogs}
\label{sec:comparison}

In this section, we compare our catalog to other searches for off-nuclear X-ray sources and IMBH candidates from the literature. For this comparison, we focus on recent catalogs (within the last decade) that feature source lists of size comparable to ours. In particular, we note similarities while also emphasizing how our sample complements results from these works.

The majority of recent ULX/HLX catalogs have used archival \xmm~detections from either \xmmtwo~\citep{Walton:2011} or \xmmthree~\citep{Zolotukhin:2016,Earnshaw:2018}. While the catalog of \citet{Gong:2016} used detections from \ch, all of these studies cross-matched X-ray sources with relatively nearby galaxies (\z$\lesssim0.06$) in the \RC~catalog. In some cases the samples were further limited in distance to yield complete samples \citep{Zolotukhin:2016,Gong:2016}. These characteristics describe the primary difference from our catalog which utilizes the \ch~archives and cross-matches those detections with galaxies from the \sdss~that generally span a much larger redshift range and have much smaller angular sizes. These differences result in a significantly larger number of HLX candidates in our catalog ($>150$) compared to previous catalogs ($\lesssim100$). As mentioned in Section \ref{sec:galselect}, the larger redshift range means our sample is biased toward intrinsically more luminous sources at higher redshifts. However, the larger luminosities in our catalog also result in the currently largest sample of uniformly-selected IMBH candidates ($\sim$\MBHLOMASSSZCorrlo$-$\MBHLOMASSSZCorrMed, after correcting for the average contamination rate of \ncontmeanmax$\%$).

Perhaps the largest drawback of our sample is the relatively poor resolution in the galaxy images. Since previous catalogs are composed of galaxies with large angular sizes, multiple off-nuclear X-ray sources can be resolved within a single galaxy. Furthermore, in some cases the stellar counterparts can be resolved. The galaxies in our catalog, on the other hand, are ubiquitously smaller in angular size and for the vast majority resolving the optical counterparts is not possible. Even with sufficient resolution, the \sdss~or \pnstrs~sensitivity is likely not sufficient to detect the faint stellar counterparts expected for many HLXs. 

The method of cross-matching X-ray sources with optical galaxy light profiles is qualitatively similar between our catalog and those using the \RC~catalog. In particular, previous catalogs have required the X-ray source to be within the $D_{\rm{25}}$ elliptical isopohote of the galaxy \citep{Walton:2011,Gong:2016,Earnshaw:2018} or two times the Petrosian radius \citep{Zolotukhin:2016} as in our criteria. However, the catalogs based on \xmm~detections also implemented a uniform cutoff in nuclear angular size offset that was empirically derived to remove a high fraction of nuclear AGN contaminants while preserving an optimal number of ULX or HLX candidates. While \citet{Zolotukhin:2016} implemented a $3\sigma$ cutoff in offset significance, they also imposed a uniform minimum offset of $5''$. In contrast, our procedure (and that of \citealt{Gong:2016} who also use \ch~data) implements a uniform cutoff in statistical significance of the angular offset from the nucleus. Our approach is based entirely on offset significance and is made possible by the relative astrometric uncertainties that we measure between each pair of \ch~and \sdss~images (Section \ref{sec:regastr}). This approach is most practical for our catalog when considering the larger redshift range of our sample over which a uniform angular size would not be appropriate.

Our catalog construction includes statistical estimates of the number of unknown contaminants following a procedure that is nearly identical to previous catalogs \citep{Walton:2011,Zolotukhin:2016,Gong:2016,Earnshaw:2018}. Furthermore, our steps to remove known contaminants are also similar to those of previous catalogs and consist of identifying and removing X-ray sources that are matched to sources with known spectroscopic redshifts that mark them as background or foreground sources. However, the cross-matching radii we use are based on the positional uncertainties for each X-ray source, as opposed to a uniform radial search used in other catalogs. These cross-match radii are always smaller than the search radii used in the other catalogs, and as stated above, this choice is more appropriate for the larger redshift range and smaller angular sizes of our candidate host galaxies. 

A final notable difference is that our contamination rate estimates are smaller than previous catalogs, a difference due primarily to the smaller angular sizes of our galaxies. In particular, Figure \ref{fig:FCONT_SEP_Z} shows that the rate is strongly dependent on angular size and can easily account for the differences in estimated contamination rates. The smaller contamination rate estimates suggest that as many as \supLXXRBISMCorrMean~of the HLX candidates may be accreting IMBHs or SMBHs that are truly interacting with their host galaxies.

\section{Conclusions}
\label{sec:conc}

We have constructed a catalog of \HLXCandSZ~HLX candidates by utilizing the spatial resolution of \ch~and a detailed procedure for registration of individual image pairs and estimates of relative astrometric uncertainties. This procedure results in selection of HLX candidates with offsets from the host galaxy centroid  down to \DeltaThetaMin~(\DeltaSMin~\uPhySep) and with redshifts extending out to \z~$\sim$~\zMaxRound. Our main conclusions are as follows:

\begin{enumerate}

\item{The estimated mean contamination rate of unrelated X-ray sources is $\sim$~\ncontmeanmax$\%$ and has a strong positive dependence on HLX angular offset from the host galaxy centroid and a positive dependence on redshift (Figure \ref{fig:FCONT_SEP_Z}). This rate is smaller than for previous catalogs ($20-30\%$) due to the smaller angular sizes of our candidate host galaxies. The contamination rate may also potentially be higher among HLXs with MIR AGN detections as they might preferentially be associated with more luminous AGN.}

\item{Optical counterparts are detected among \iOptDetFrac$\%$ of the HLX candidates. These counterparts are consistent with dwarf galaxy stellar masses for \iOptDetLoMassFrac$\%$ of these detections and therefore may harbor IMBHs based on observed correlations between BH masses and host galaxy stellar masses.}

\item{The median X-ray spectral index is consistent with relatively efficient accretion (\PhotIndx~$=$~\GAMMAMEDIAN) which could be merger-driven or associated with QSOs (middle panel of Figure \ref{fig:HR_GAMMA_NH}). A subset of HLX candidates (\NHfreeObscFrac$\%$) have X-ray spectral evidence for significant X-ray absorption (\nHexgal~$>10^{22}$ \uColDens) that could be the result of mergers but would also be consistent with obscured background AGN (right panel of Figure \ref{fig:HR_GAMMA_NH}).}

\item{The X-ray-to-optical flux ratios for the majority (\XtoODetpONEtoTENFrac$\%$) of HLX candidates with optical counterpart detections are consistent with AGN (Figure \ref{fig:FX_FR}). Furthermore, many of the non-detected stellar counterparts} may have much larger X-ray-to-optical flux ratios that are similar to other confirmed nearby HLXs.

\item{We have identified a subset of HLX candidates (\supLXXRBISM) with X-ray emission that far exceeds the contribution from XRBs and hot ISM gas, suggesting they are associated with accretion onto IMBHs or SMBHs (Figure \ref{fig:LXH_LXHXRBISM}). Of these HLX AGN candidates, \WISEAGNcninetyFRAC$\%$ are detected as AGN at MIR wavelengths.} 

\item{Assuming a conservative Eddington ratio of \FEDDlo~for all HLX candidates, \MBHLOMASSSZlo~are consistent with being IMBHs. If we instead adopt the Eddington ratio that corresponds to the median photon index, this number is significantly higher (\MBHLOMASSSZMed~IMBH candidates) and yields BH masses that are consistent with the expected relation between BH mass and host galaxy stellar mass (Figure \ref{fig:MBH_MGAL}).}

\end{enumerate}

In future works we will use the unique size and dynamic range of redshifts in this sample to study HLX environments, their redshift evolution, and the merger-driven growth of IMBHs. \\

We thank an anonymous referee for detailed and helpful comments that have improved the quality of the manuscript. Support for this work was provided by NASA through Chandra Award Number AR7-18010X issued by the Chandra X-ray Observatory Center, which is operated by the Smithsonian Astrophysical Observatory for and on behalf of NASA under contract NAS8-03060. M.M. acknowledges support from the Spanish Juan de la Cierva program (IJCI-2015-23944). This research has made use of data obtained from the Chandra Source Catalog, provided by the Chandra X-ray Center (CXC) as part of the Chandra Data Archive.

\appendix

\newcommand{\IMBHrestCaptionText}{The remaining IMBH candidates (assuming \FEDD~$=$~\FEddMedian; Section \ref{sec:BHmass}) in order of ascending redshift. Annotations are the same as in Figure \ref{fig:IMBHbrief}.}

\begin{figure*}[b!] $
\begin{array}{c c}
\includegraphics[width=0.48\textwidth]{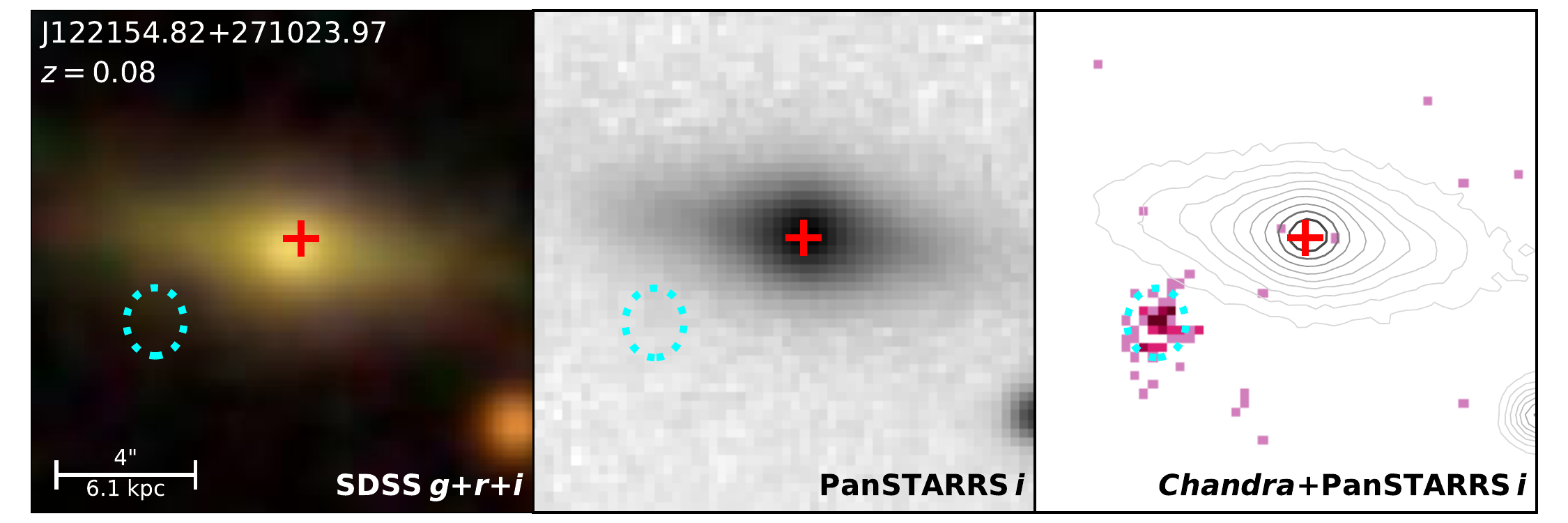} &
\includegraphics[width=0.48\textwidth]{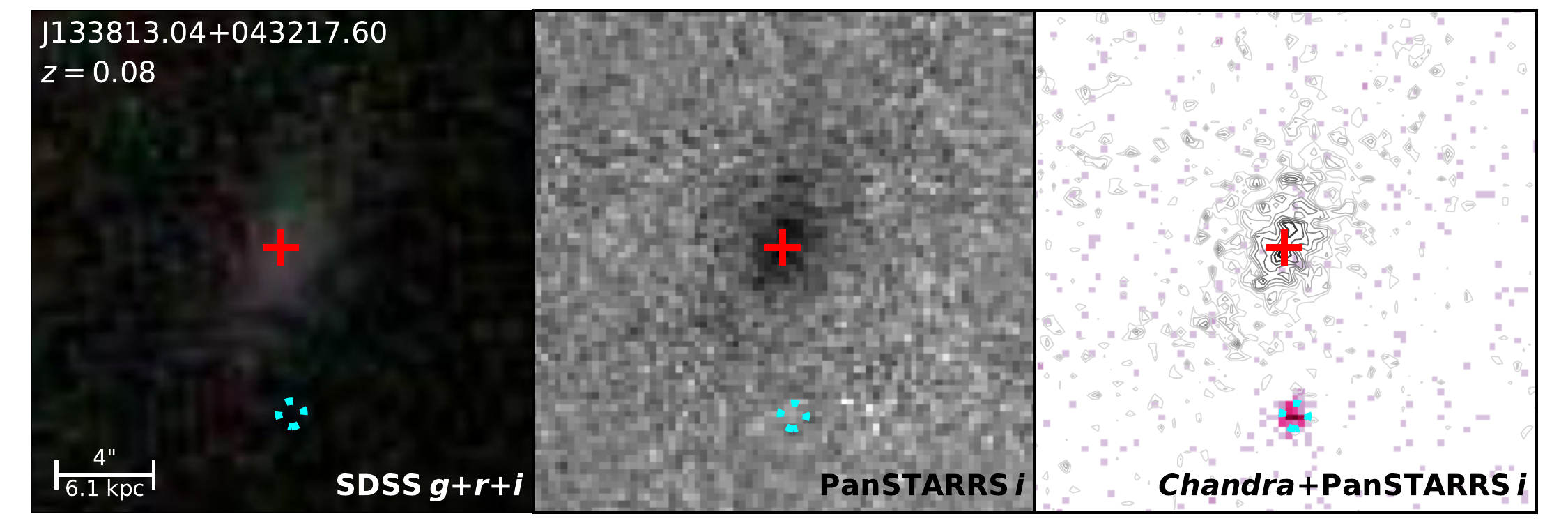} \\
\includegraphics[width=0.48\textwidth]{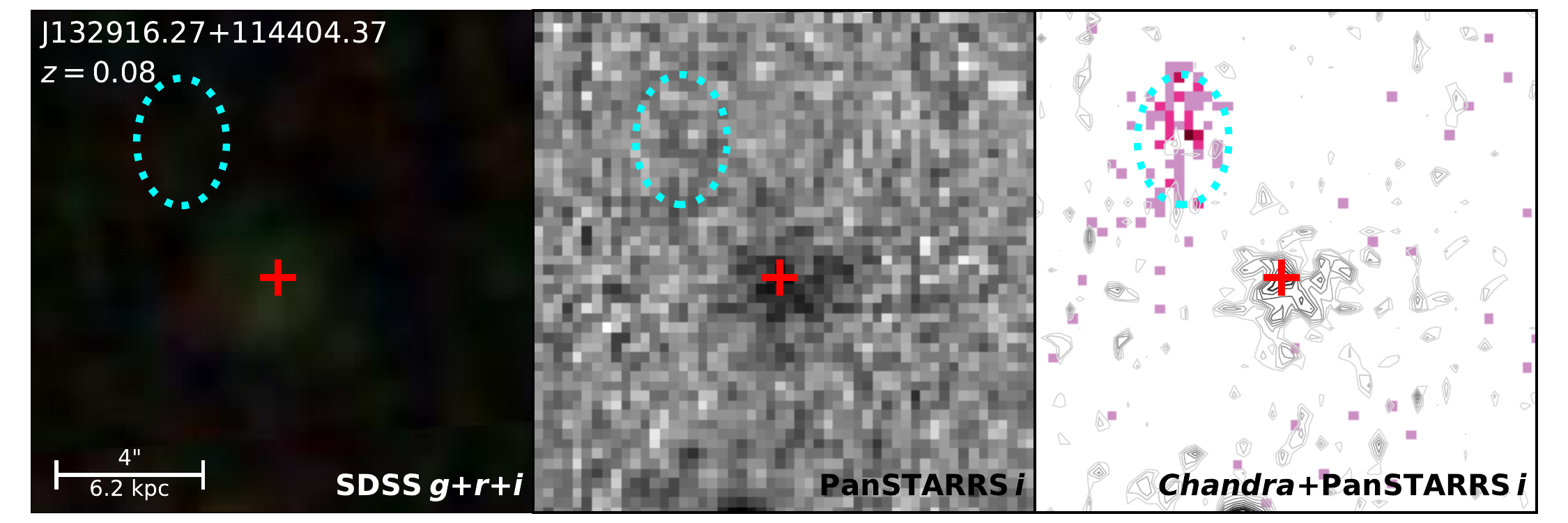} &
\includegraphics[width=0.48\textwidth]{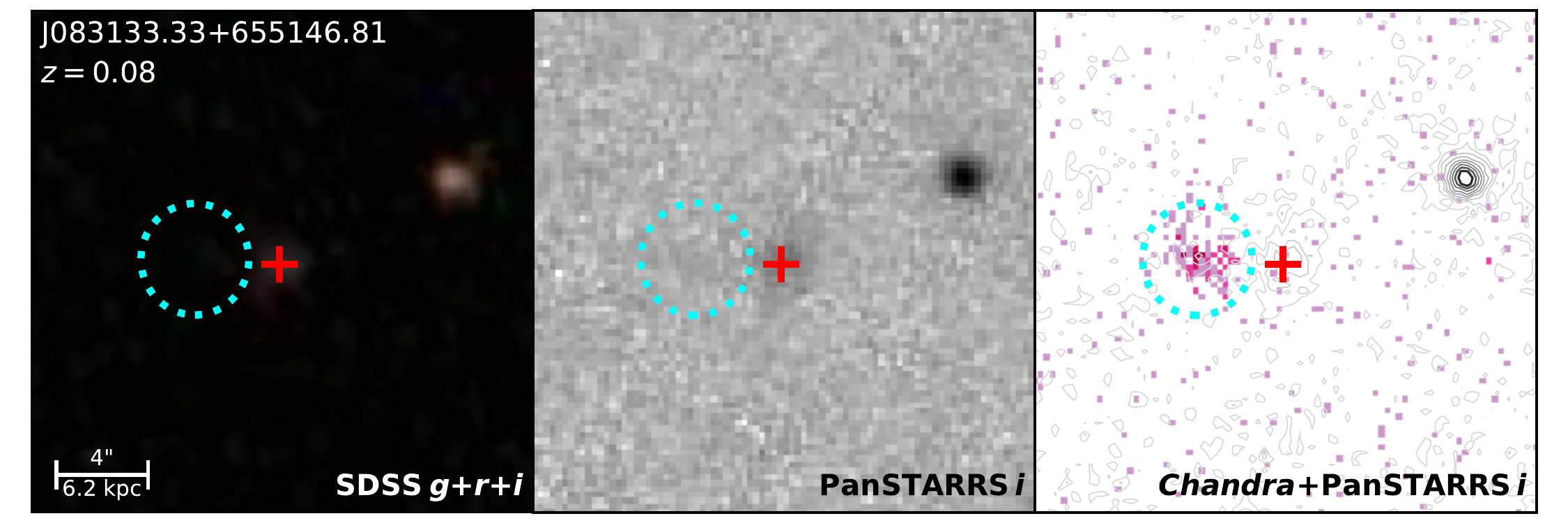} \\
\includegraphics[width=0.48\textwidth]{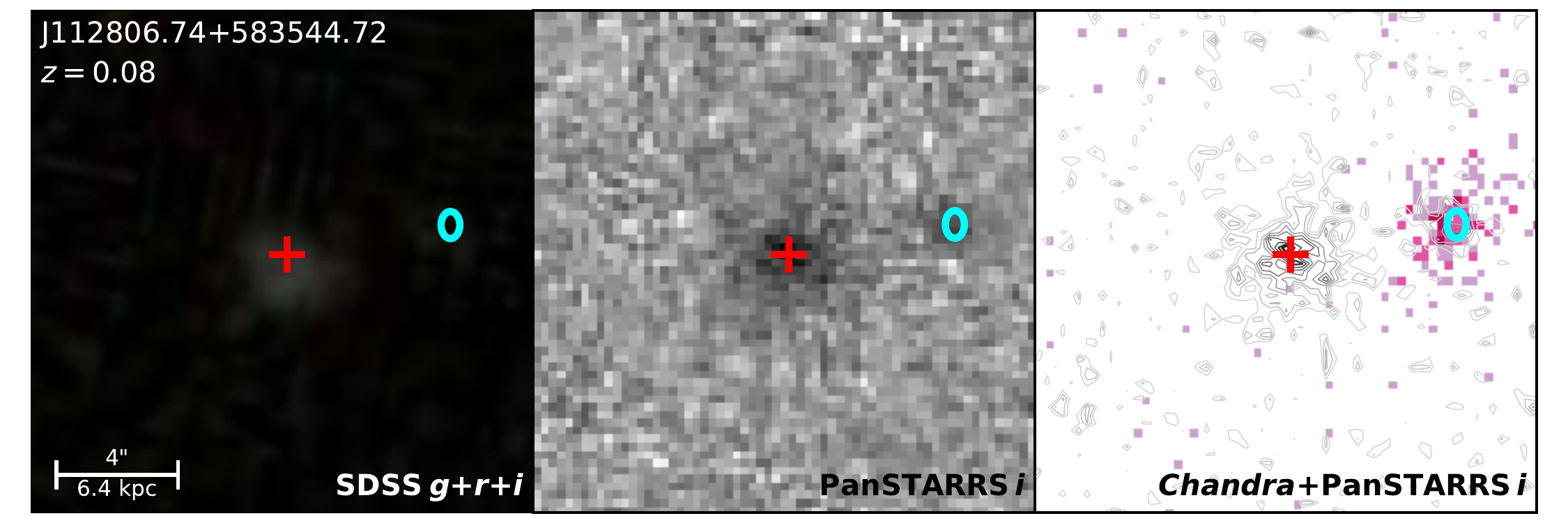} &
\includegraphics[width=0.48\textwidth]{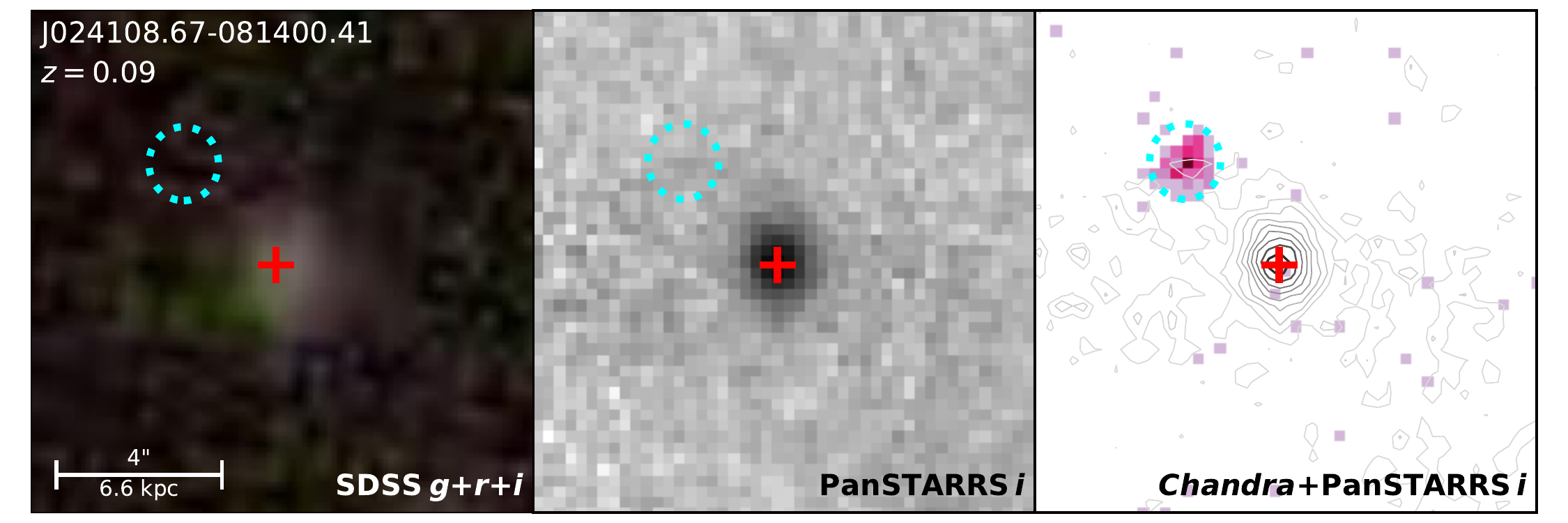} \\
\includegraphics[width=0.48\textwidth]{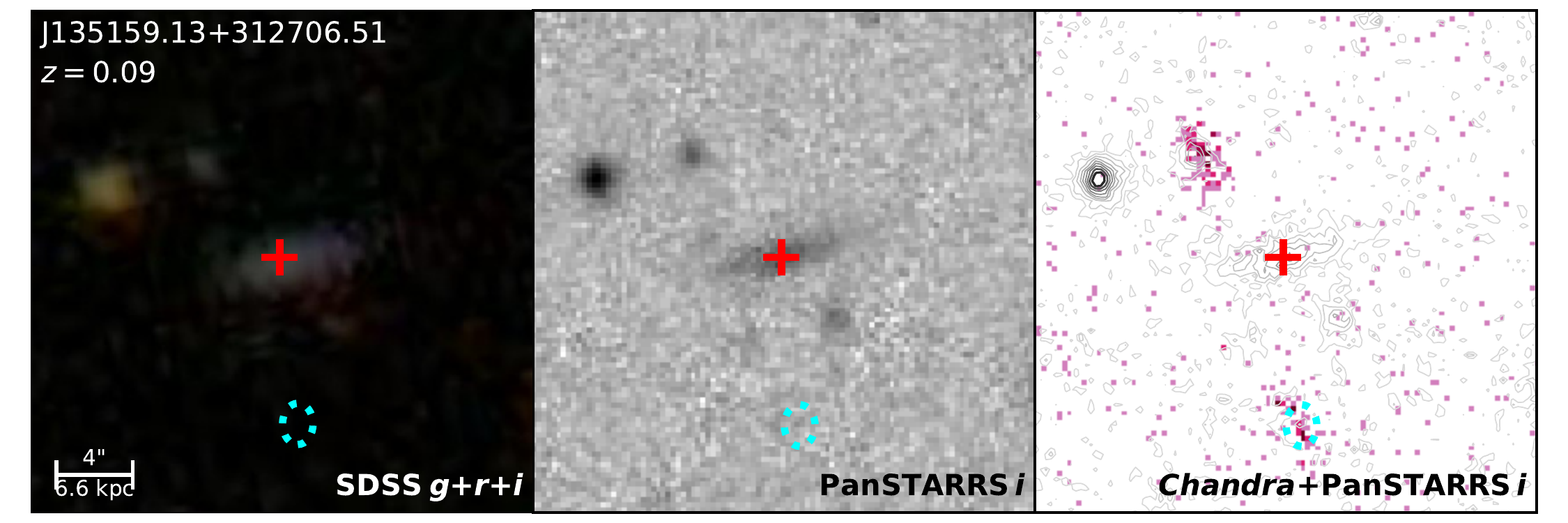} &
\includegraphics[width=0.48\textwidth]{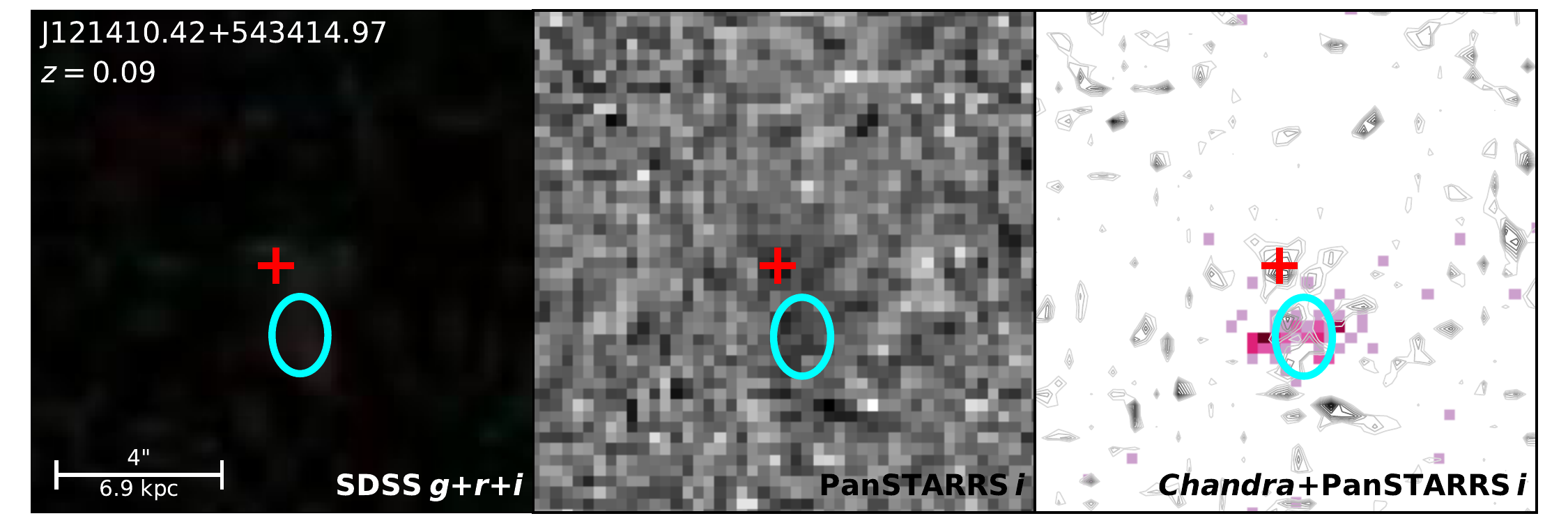} \\
\includegraphics[width=0.48\textwidth]{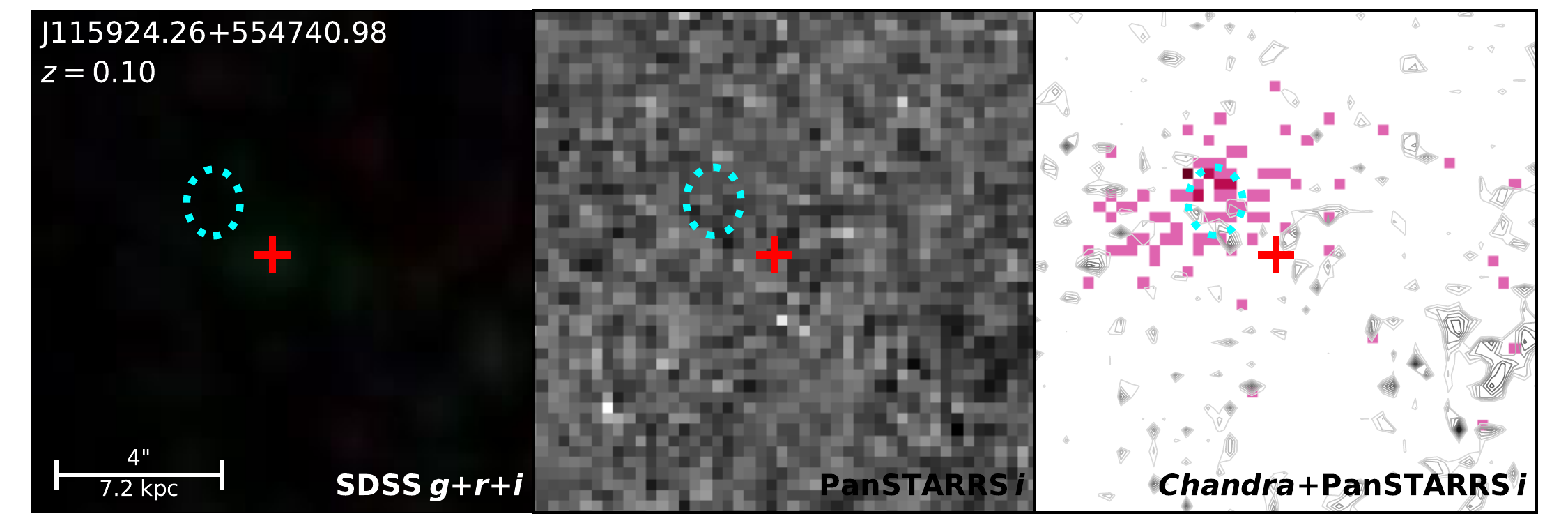} &
\includegraphics[width=0.48\textwidth]{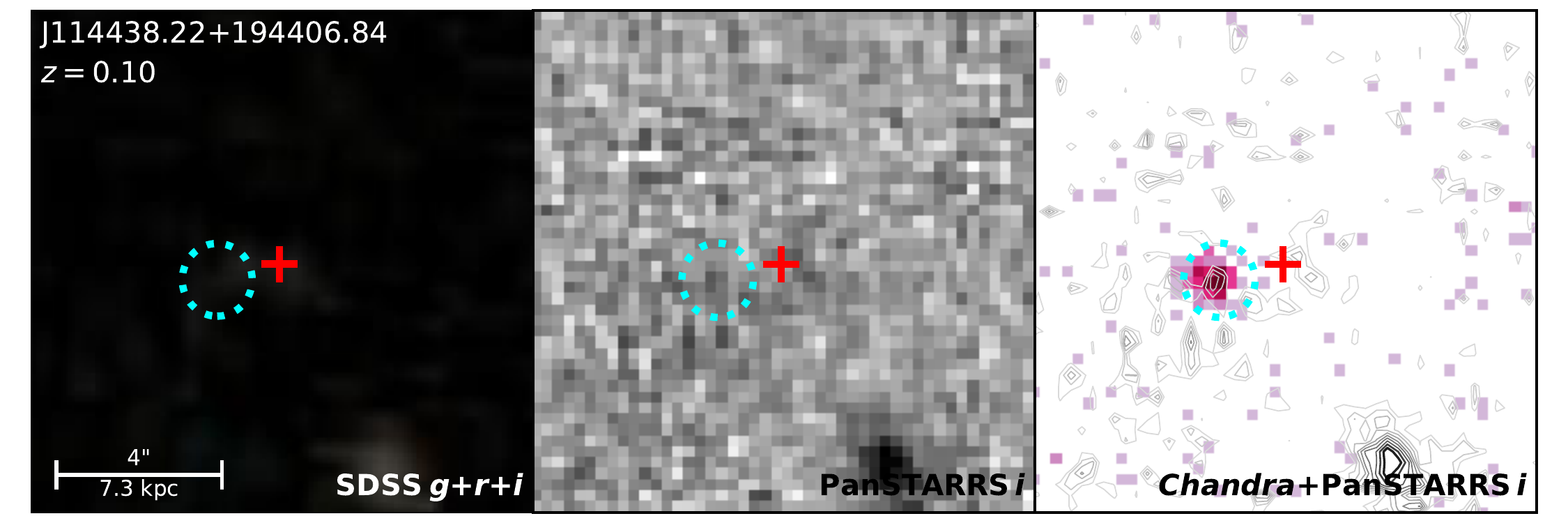} \\
\includegraphics[width=0.48\textwidth]{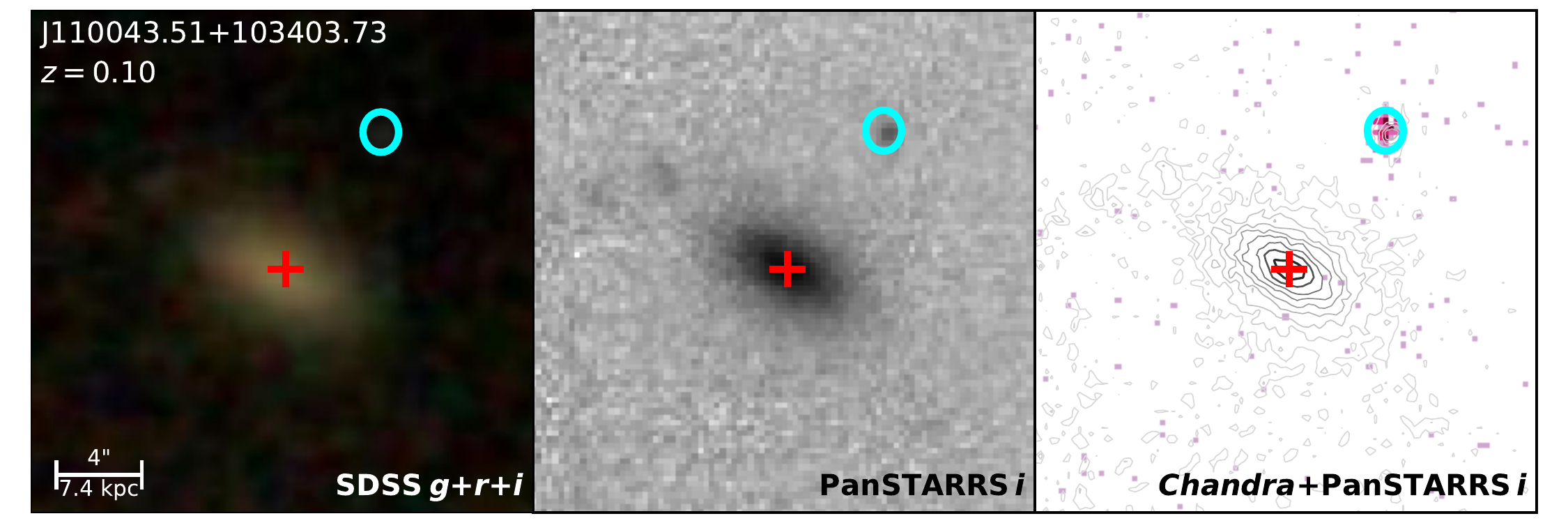} &
\includegraphics[width=0.48\textwidth]{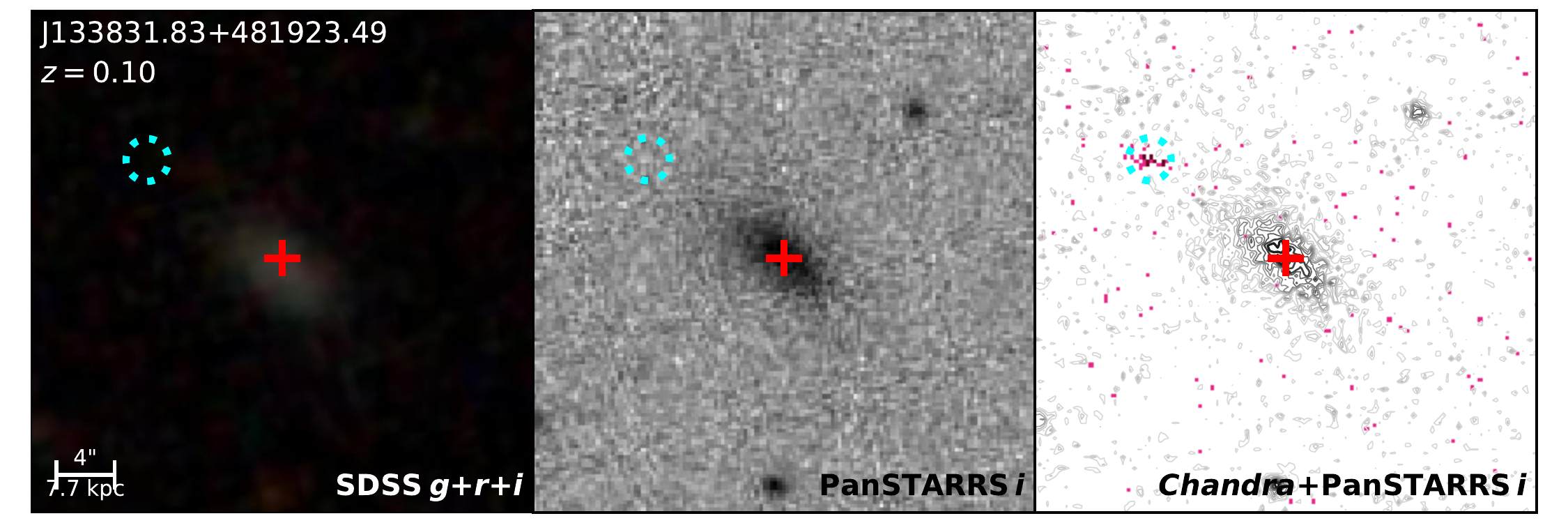} \\
\includegraphics[width=0.48\textwidth]{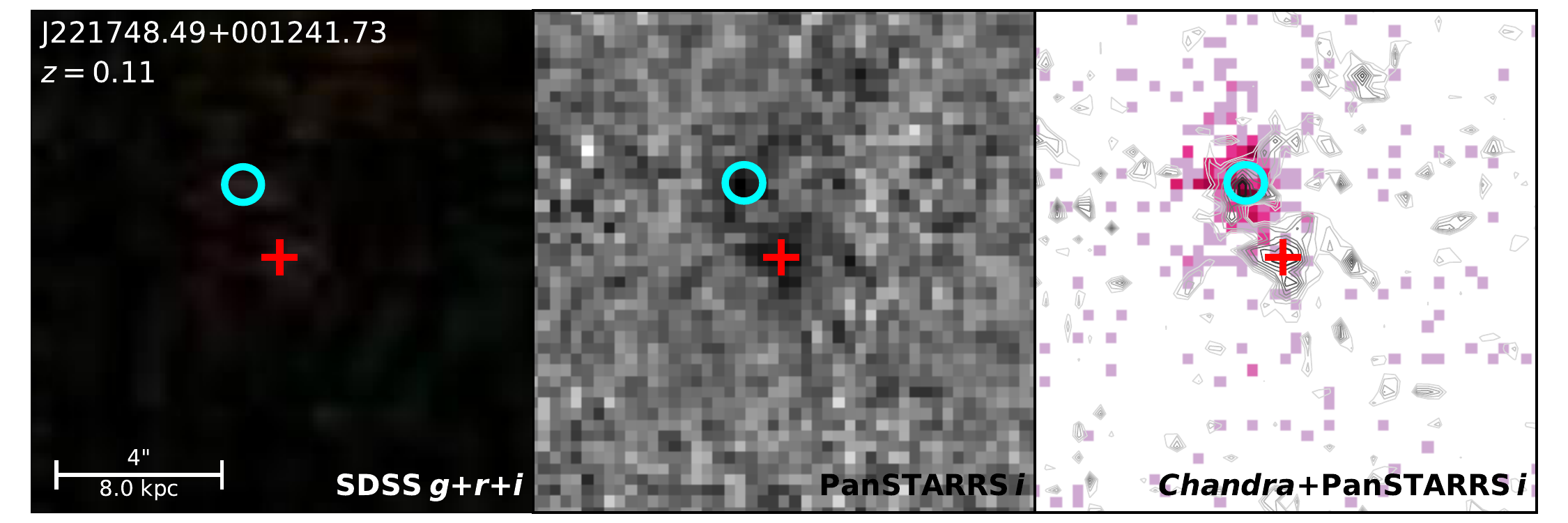} &
\includegraphics[width=0.48\textwidth]{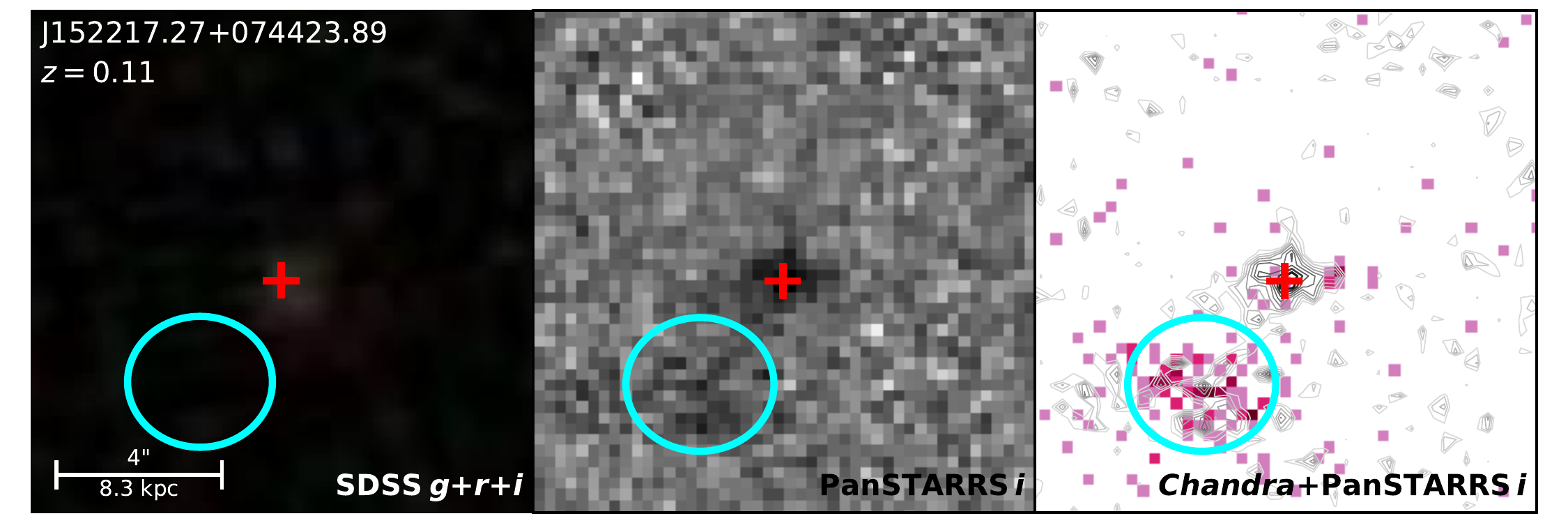} \\
\end{array} $
\caption{\footnotesize{\IMBHrestCaptionText}}
\label{fig:IMBHrest}
\end{figure*}
\begin{figure*} $
\ContinuedFloat
\begin{array}{c c}
\includegraphics[width=0.48\textwidth]{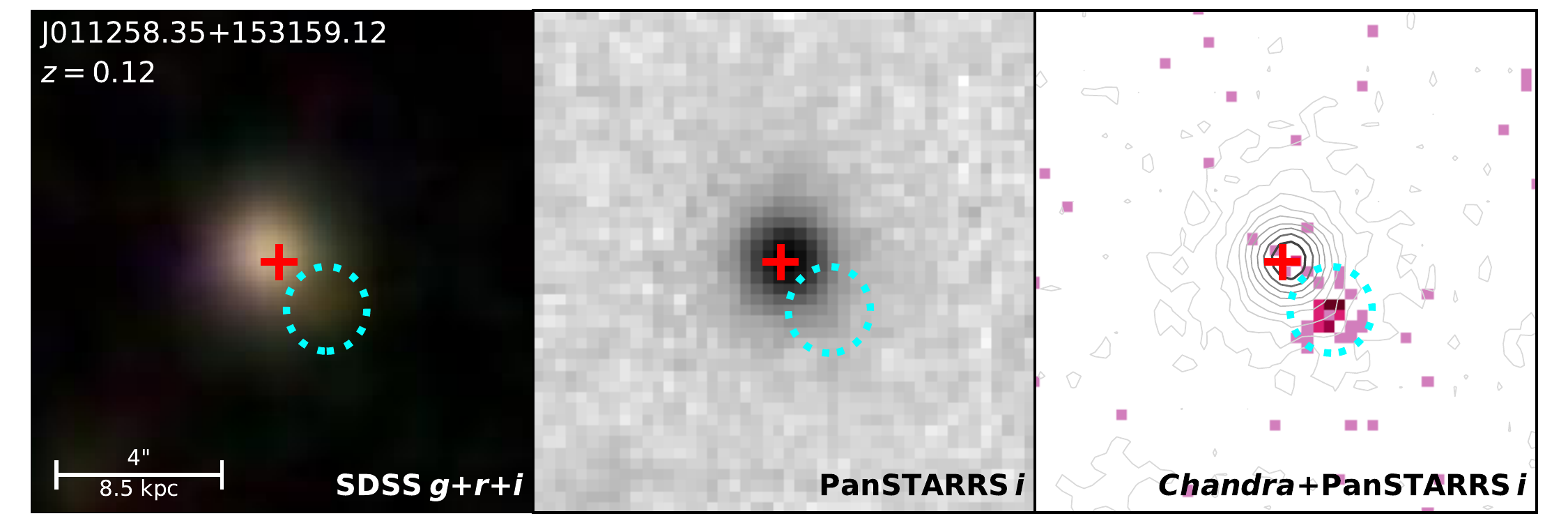} &
\includegraphics[width=0.48\textwidth]{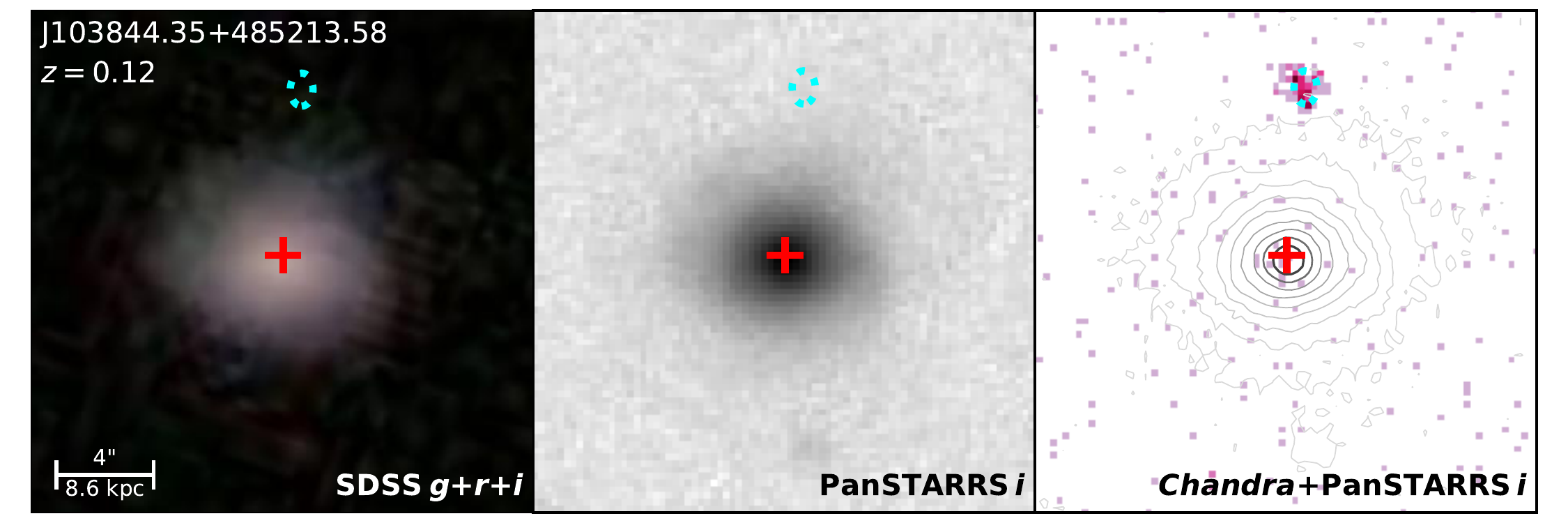} \\
\includegraphics[width=0.48\textwidth]{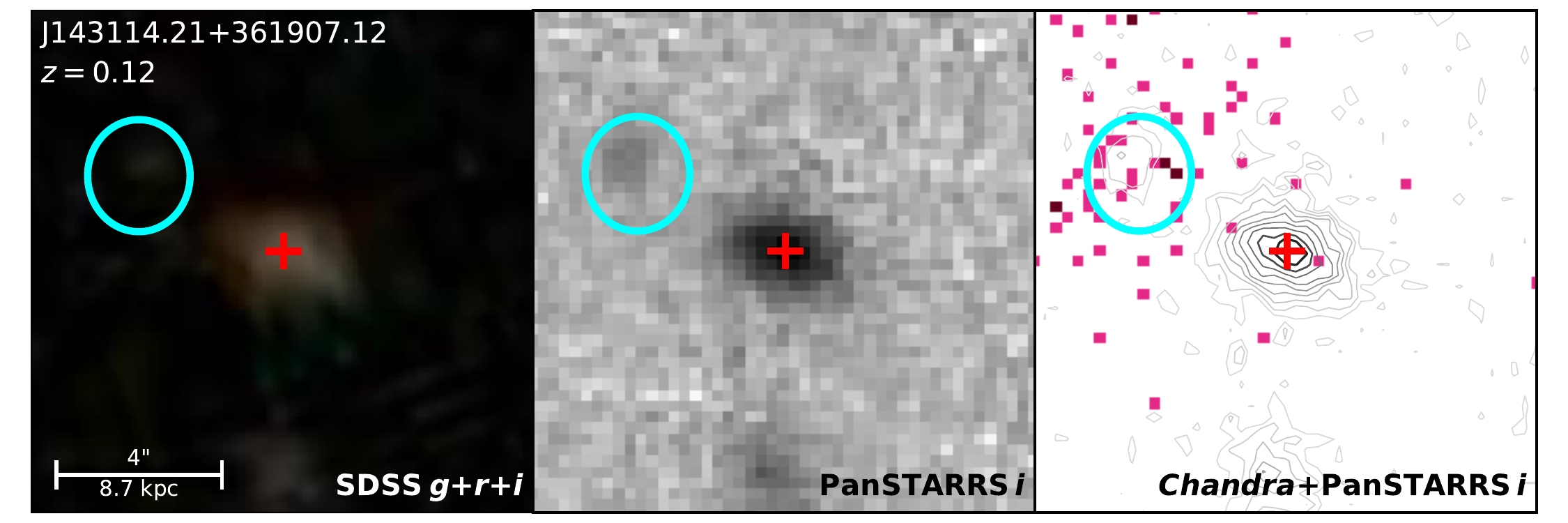} &
\includegraphics[width=0.48\textwidth]{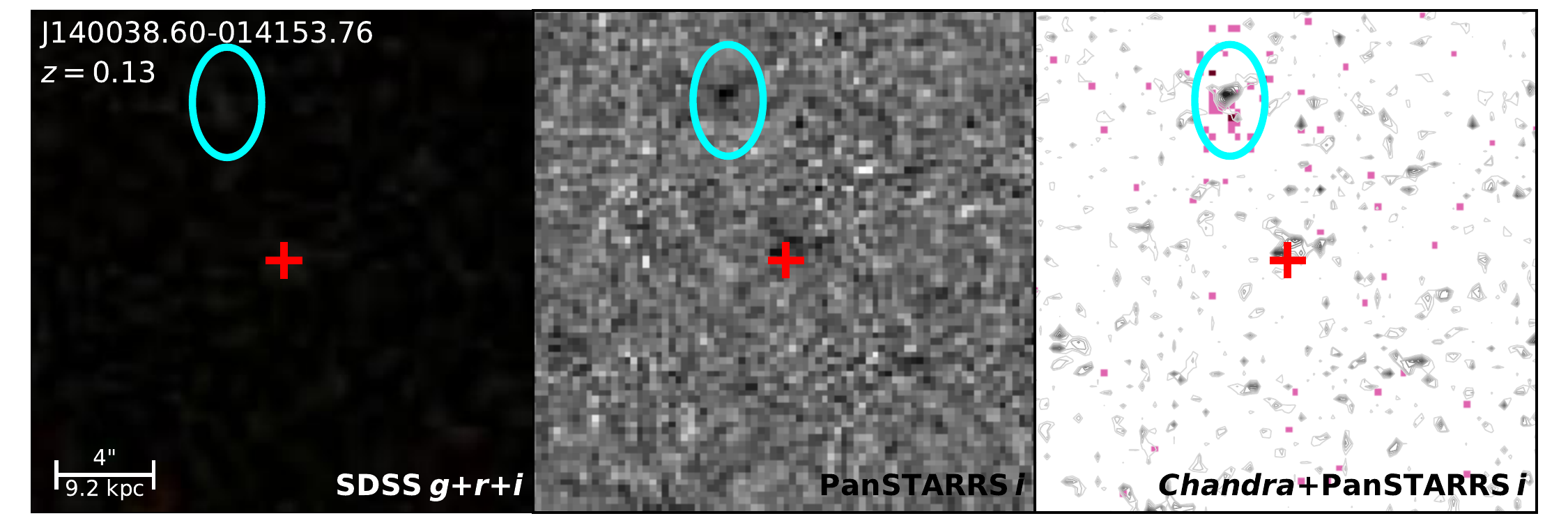} \\
\includegraphics[width=0.48\textwidth]{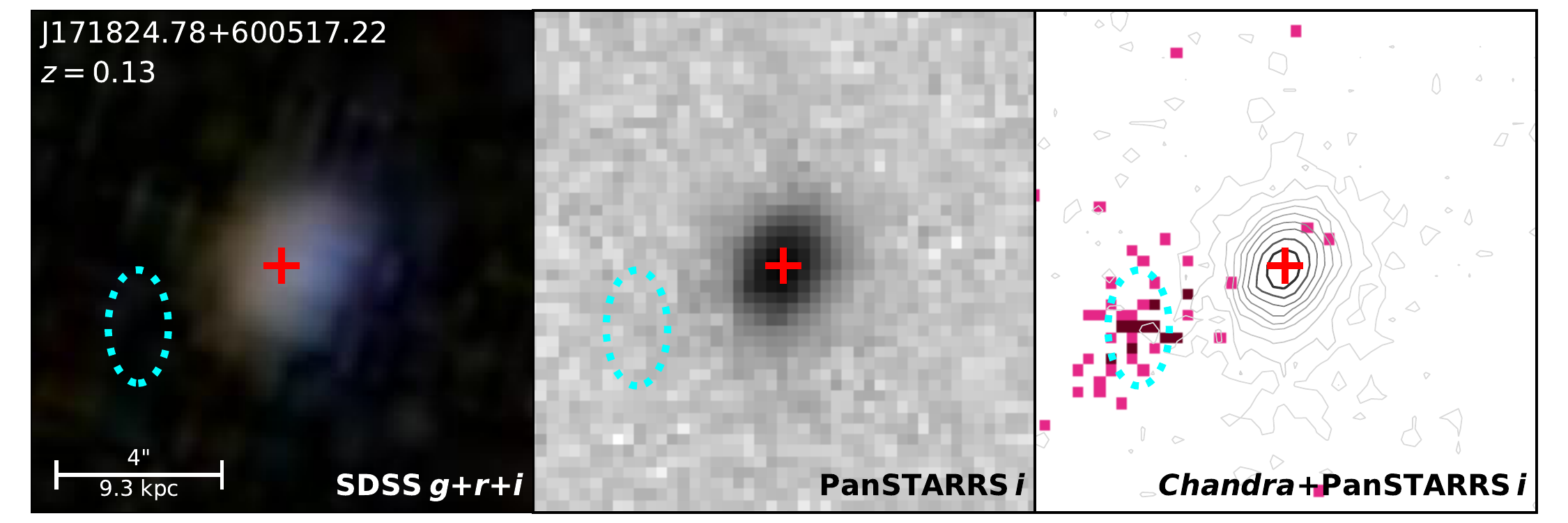} &
\includegraphics[width=0.48\textwidth]{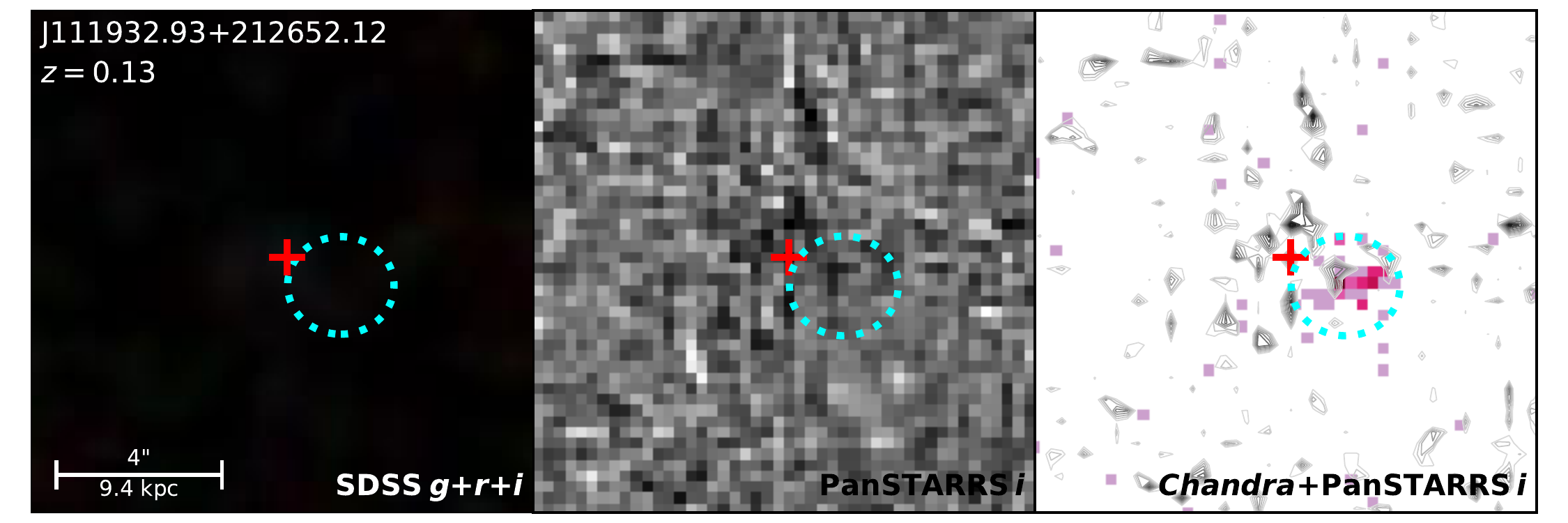} \\
\includegraphics[width=0.48\textwidth]{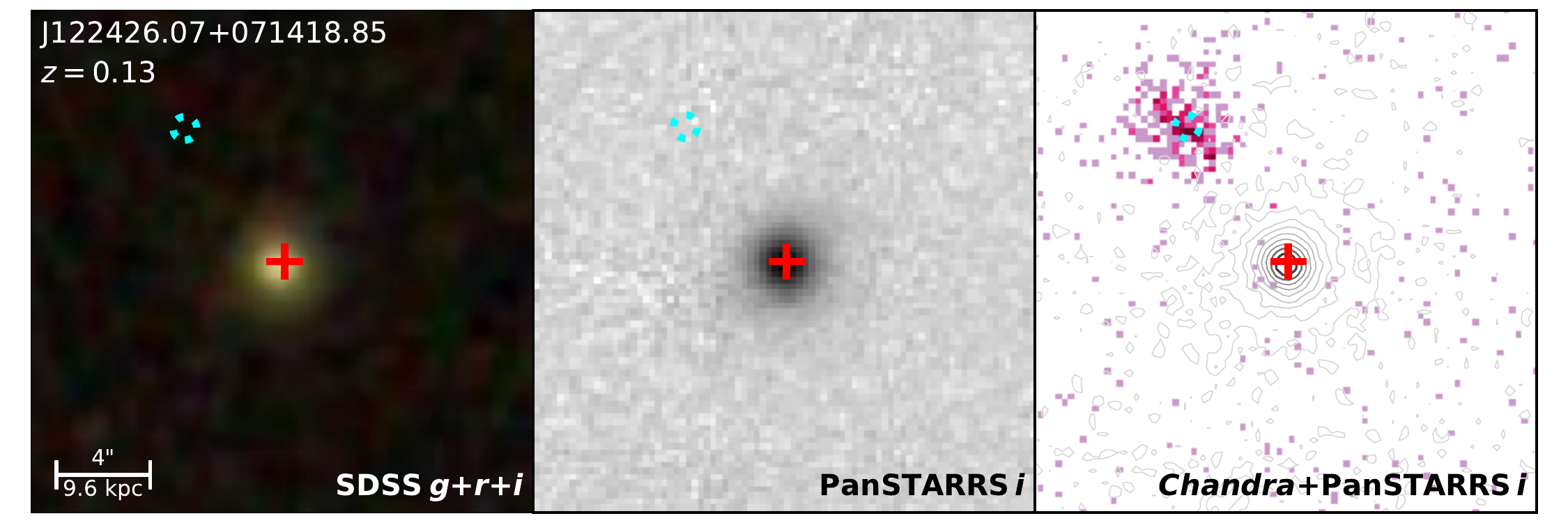} &
\includegraphics[width=0.48\textwidth]{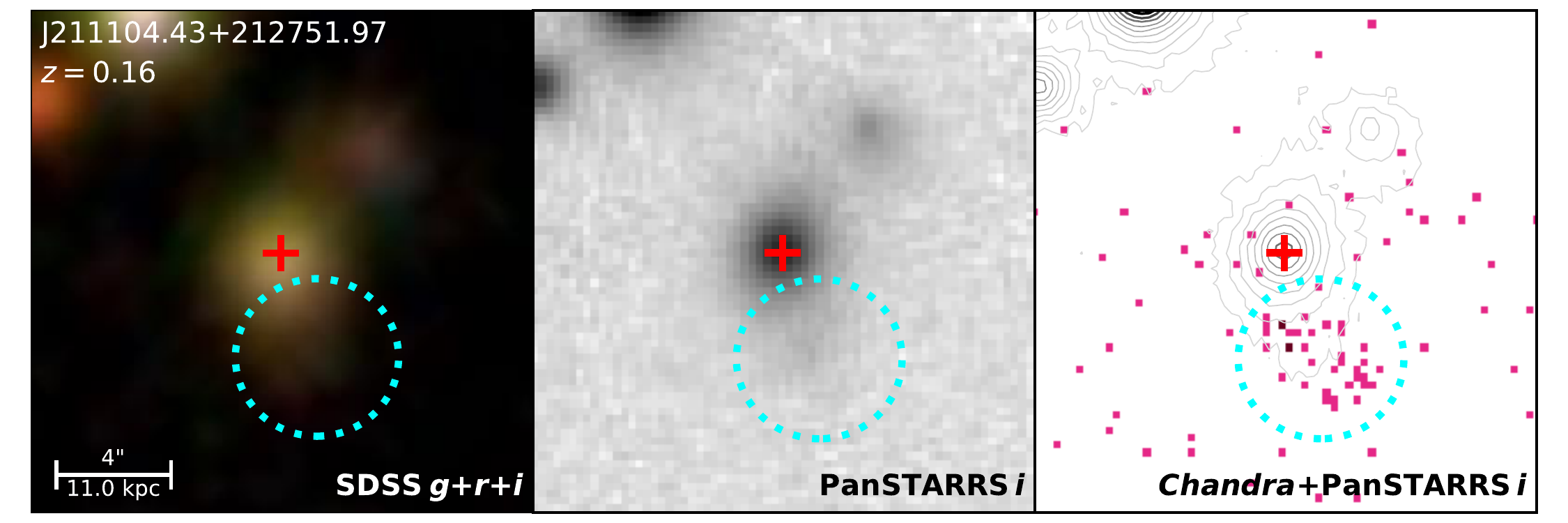} \\
\includegraphics[width=0.48\textwidth]{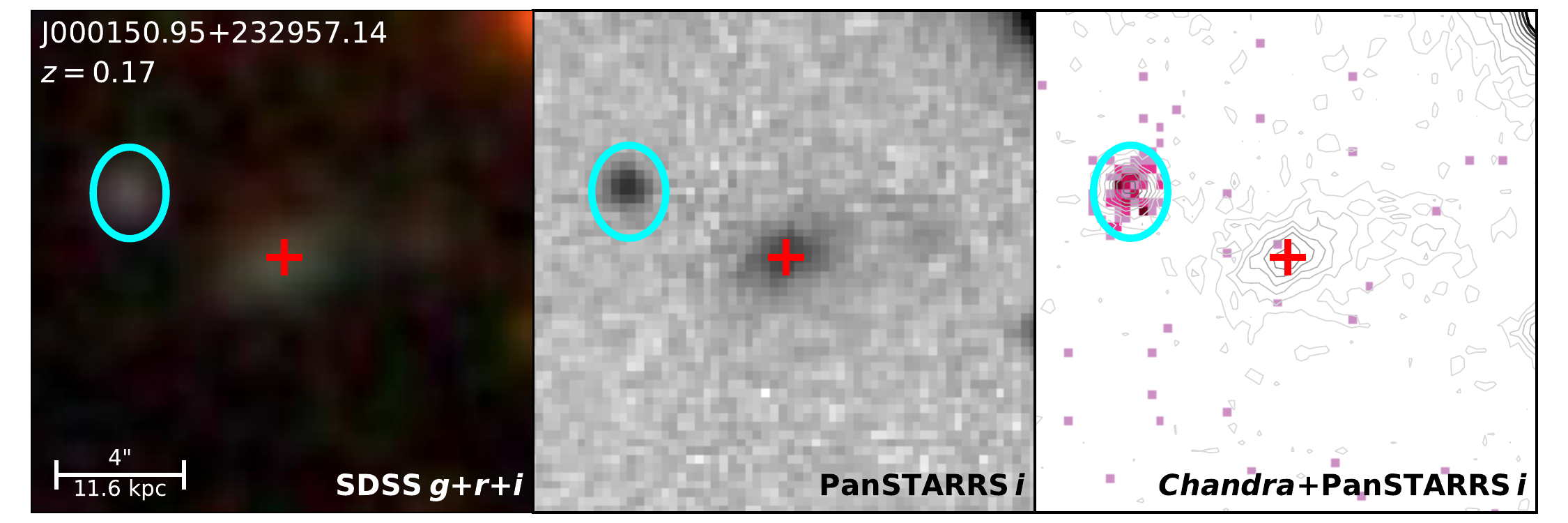} &
\includegraphics[width=0.48\textwidth]{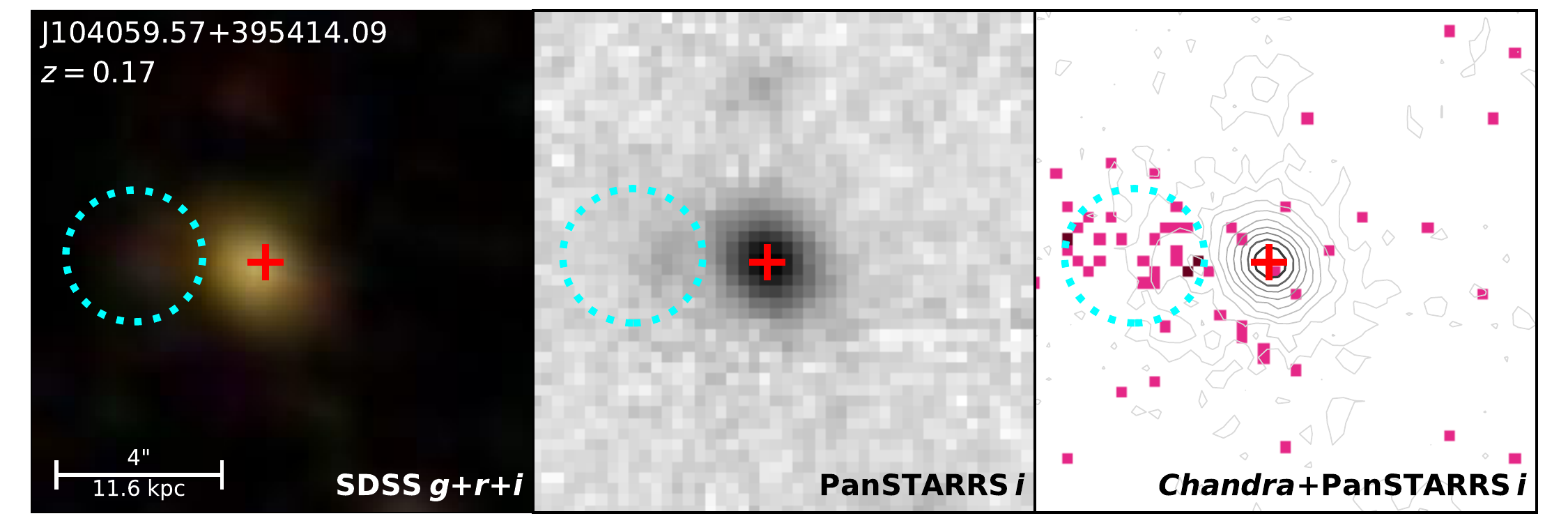} \\
\includegraphics[width=0.48\textwidth]{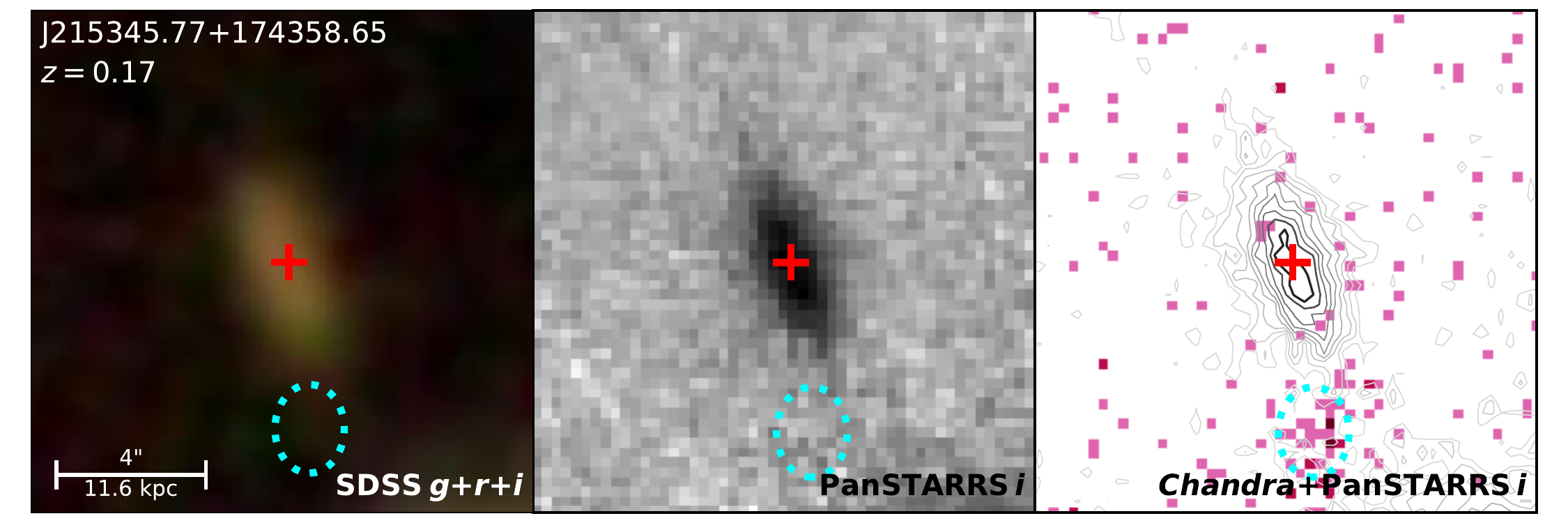} &
\includegraphics[width=0.48\textwidth]{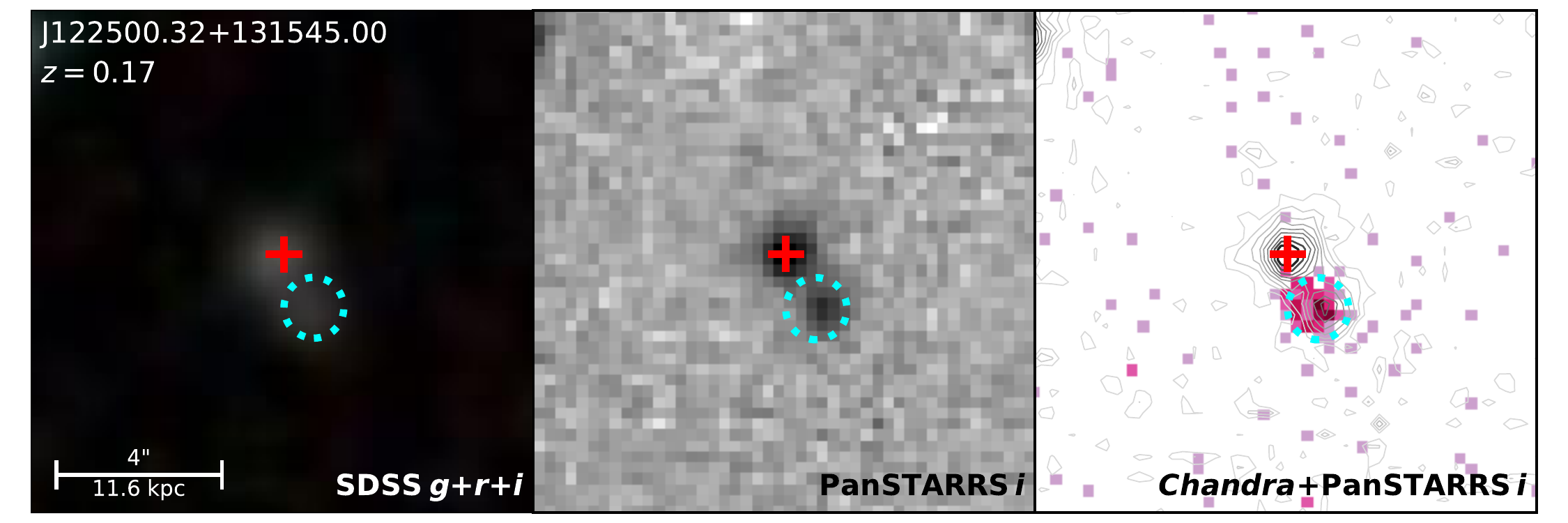} \\
\includegraphics[width=0.48\textwidth]{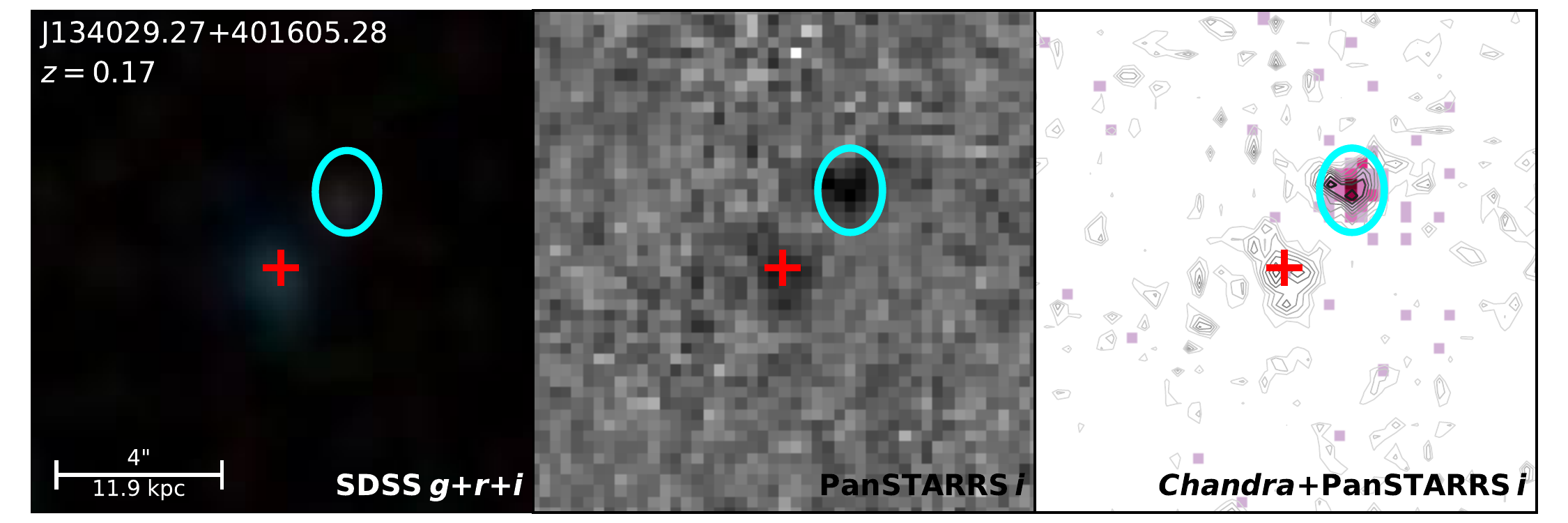} &
\includegraphics[width=0.48\textwidth]{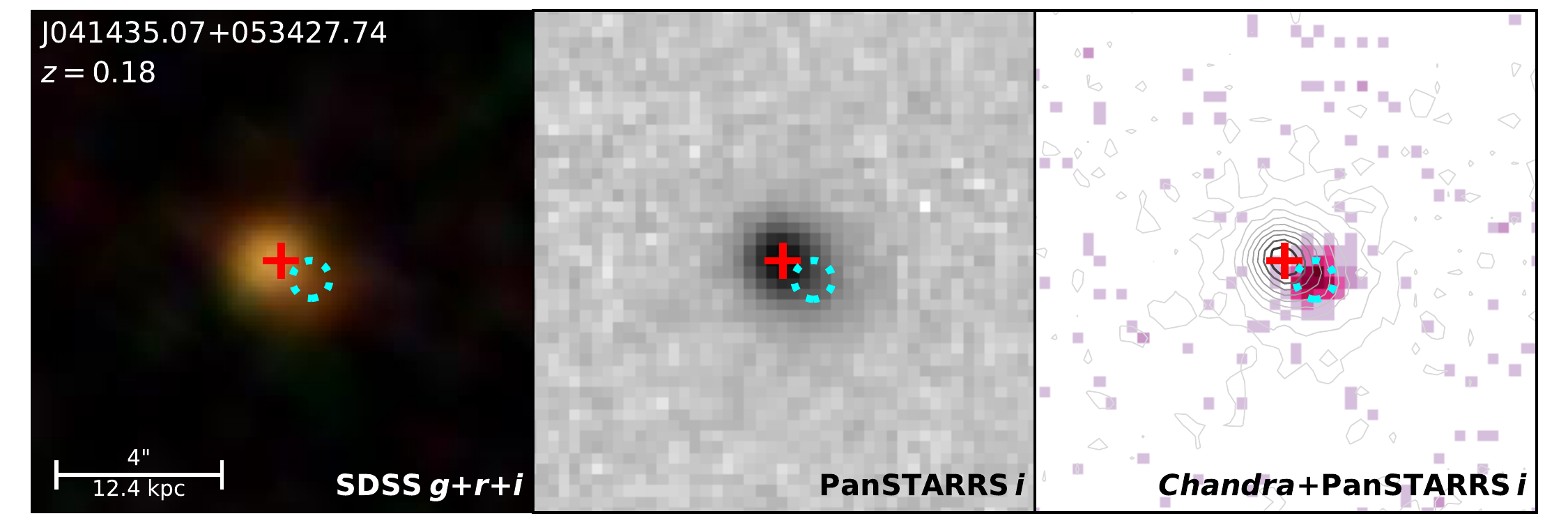} \\
\end{array} $
\caption{continued.}
\end{figure*}
\begin{figure*} $
\ContinuedFloat
\begin{array}{c c}
\includegraphics[width=0.48\textwidth]{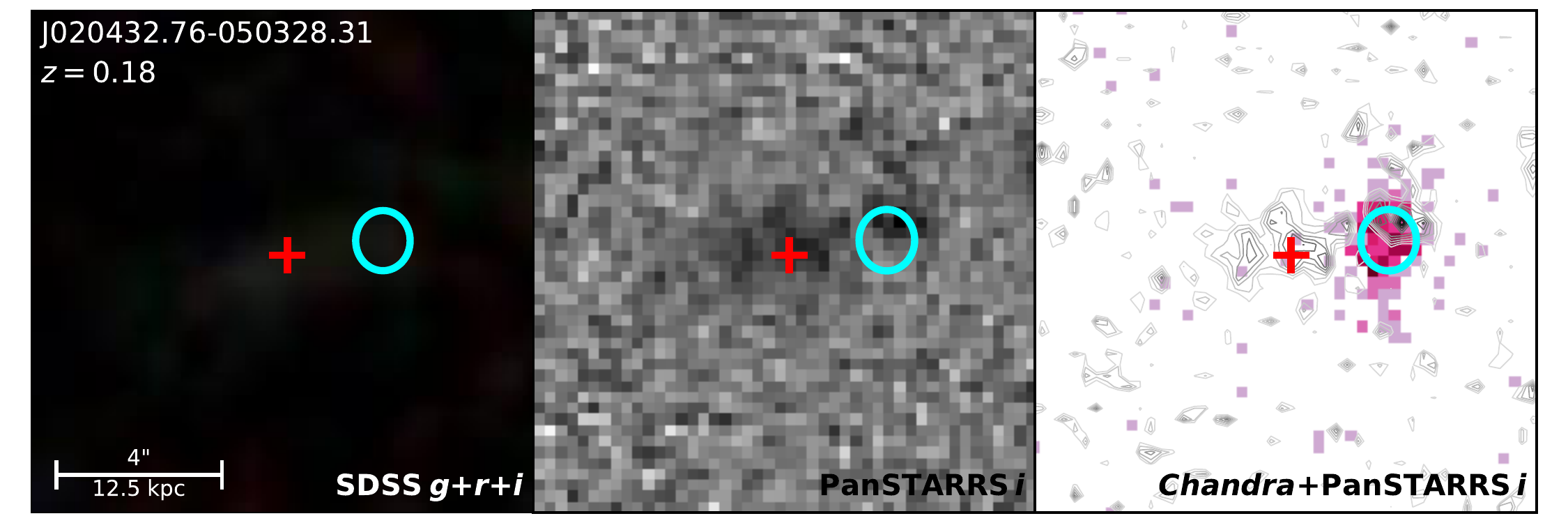} &
\includegraphics[width=0.48\textwidth]{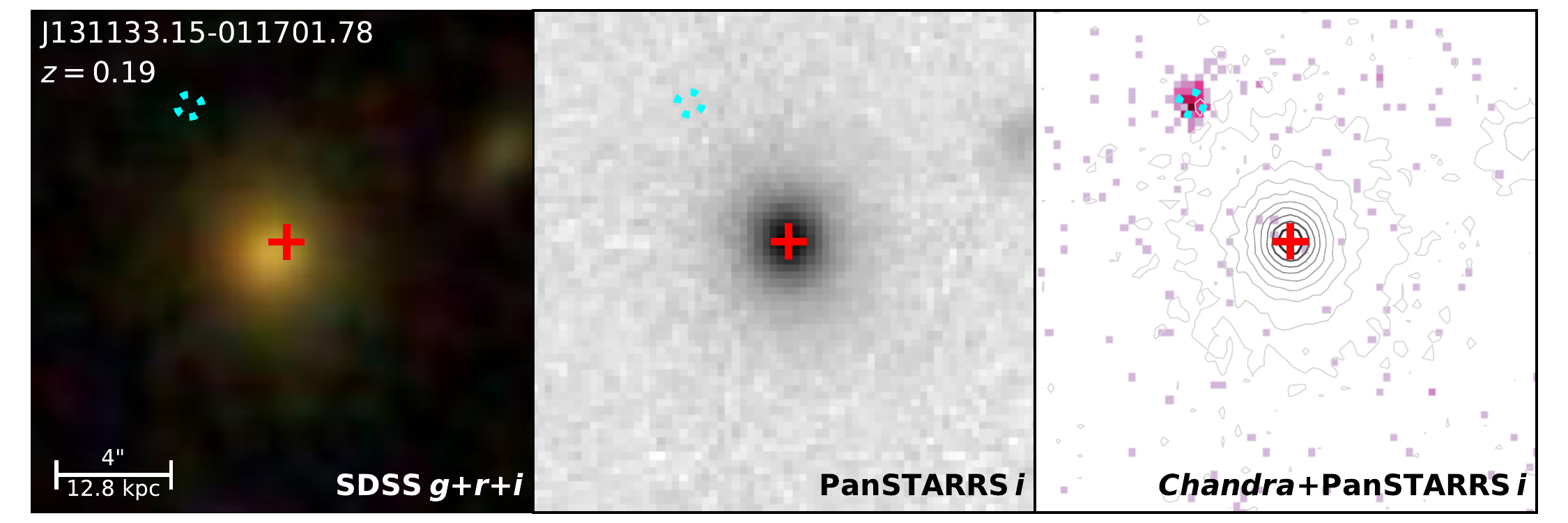} \\
\includegraphics[width=0.48\textwidth]{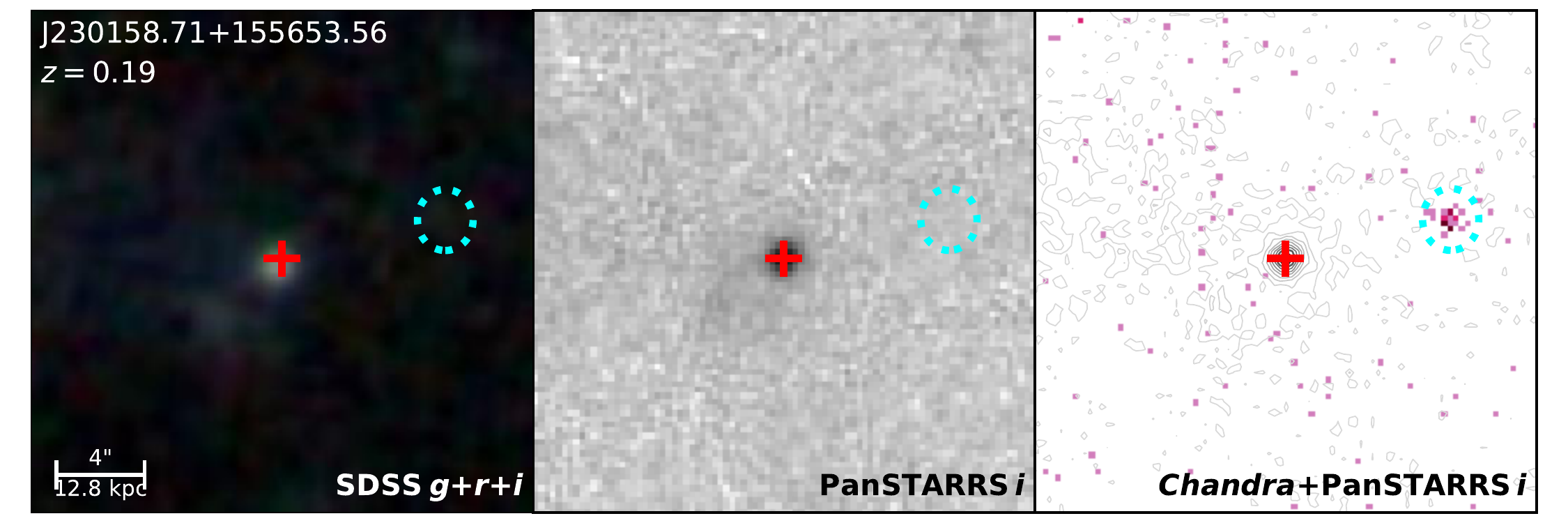} &
\includegraphics[width=0.48\textwidth]{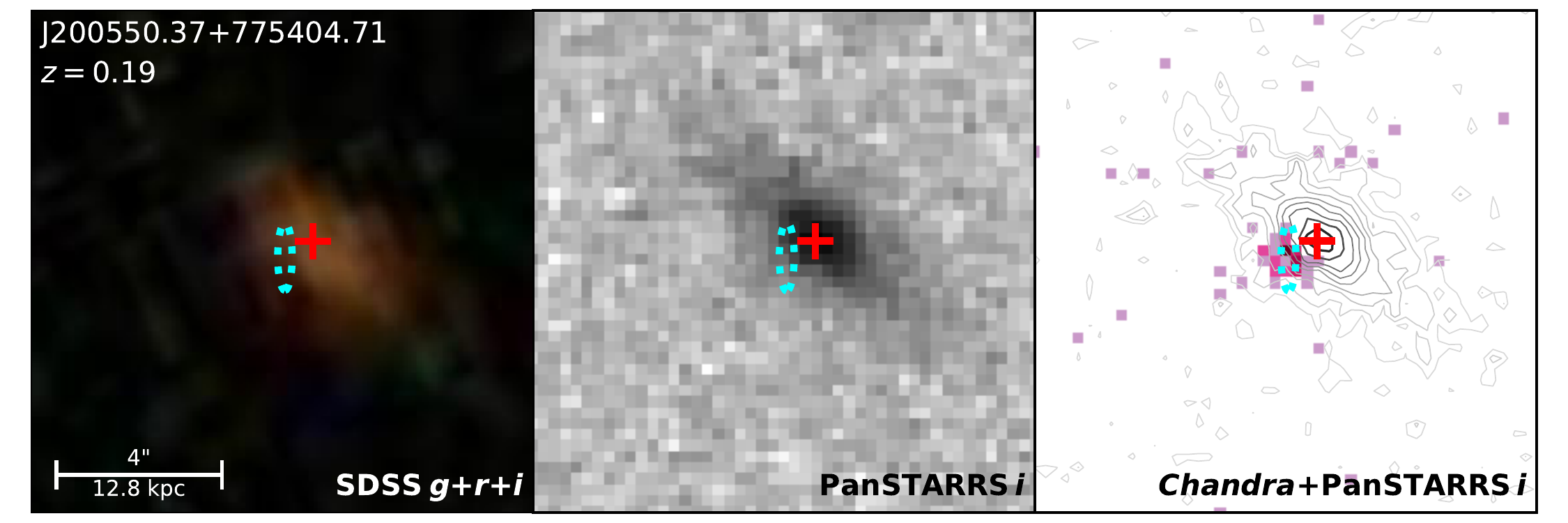} \\
\includegraphics[width=0.48\textwidth]{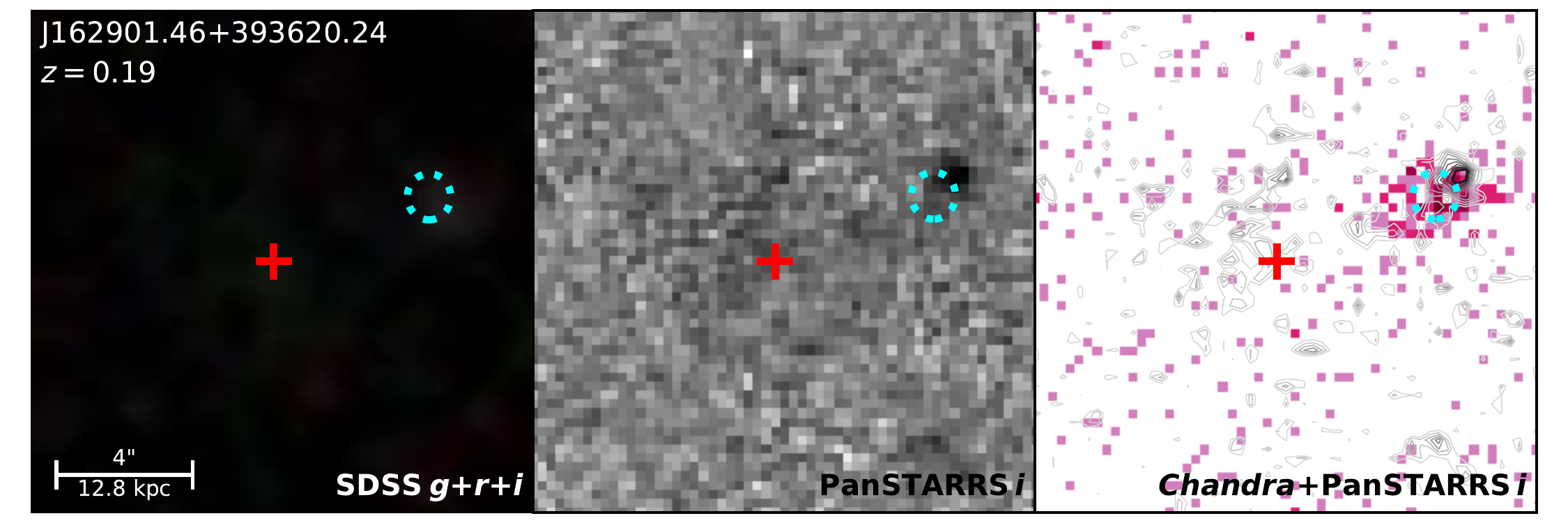} &
\includegraphics[width=0.48\textwidth]{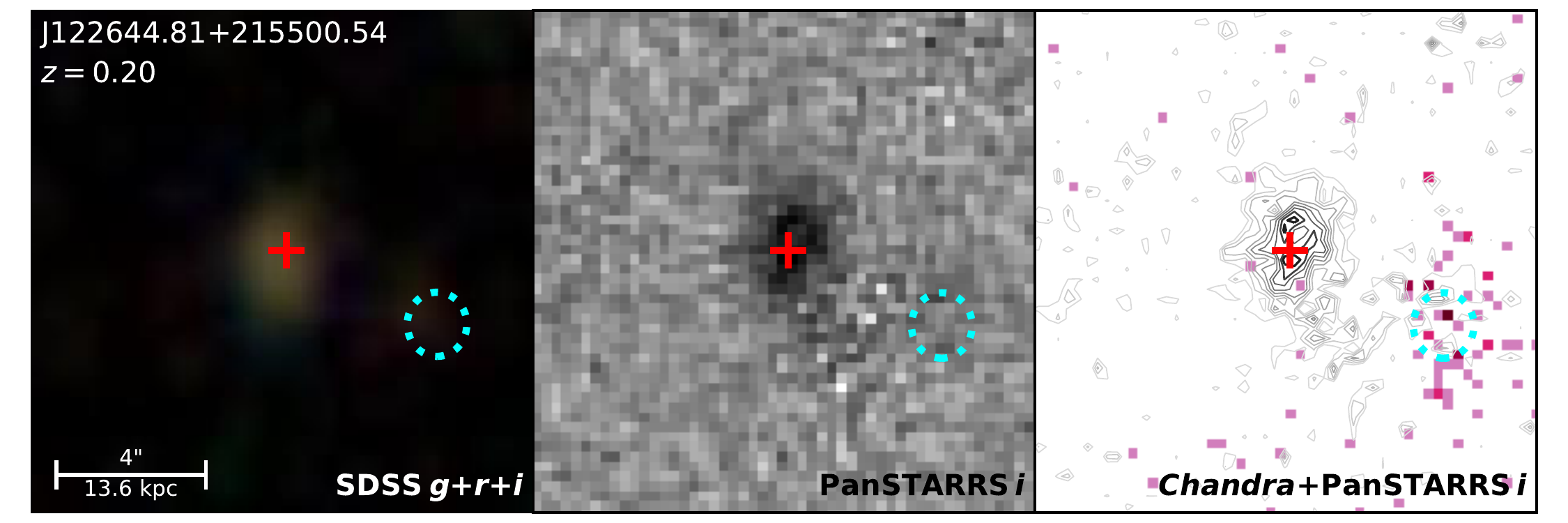} \\
\includegraphics[width=0.48\textwidth]{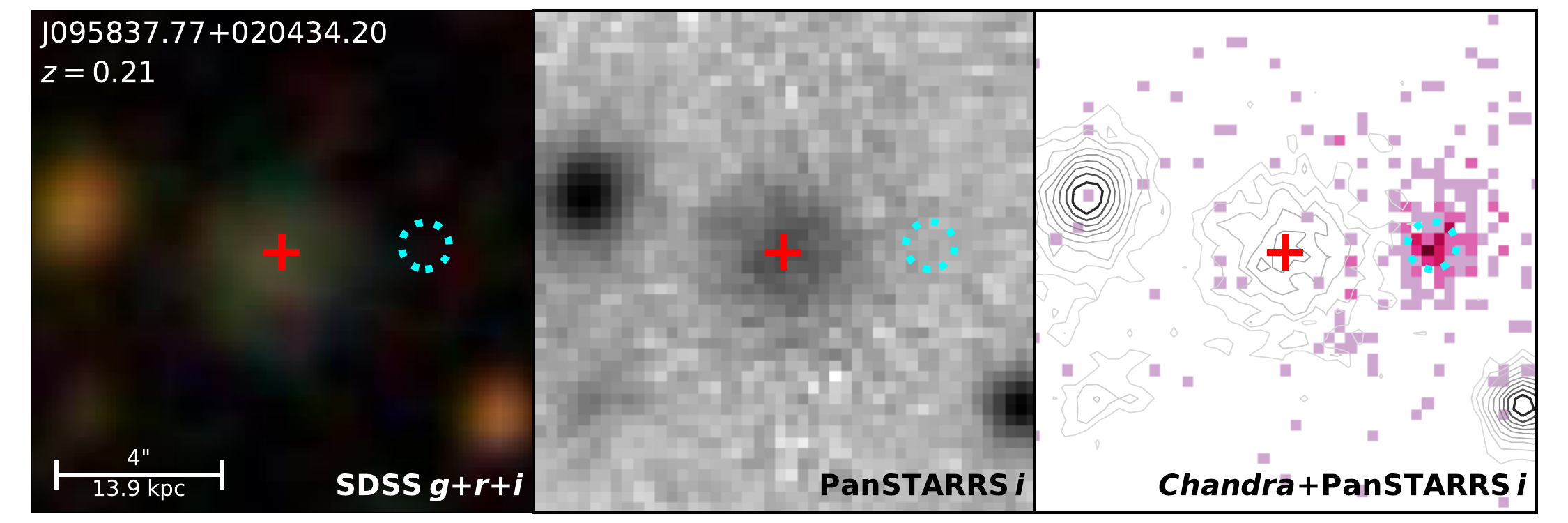} &
\includegraphics[width=0.48\textwidth]{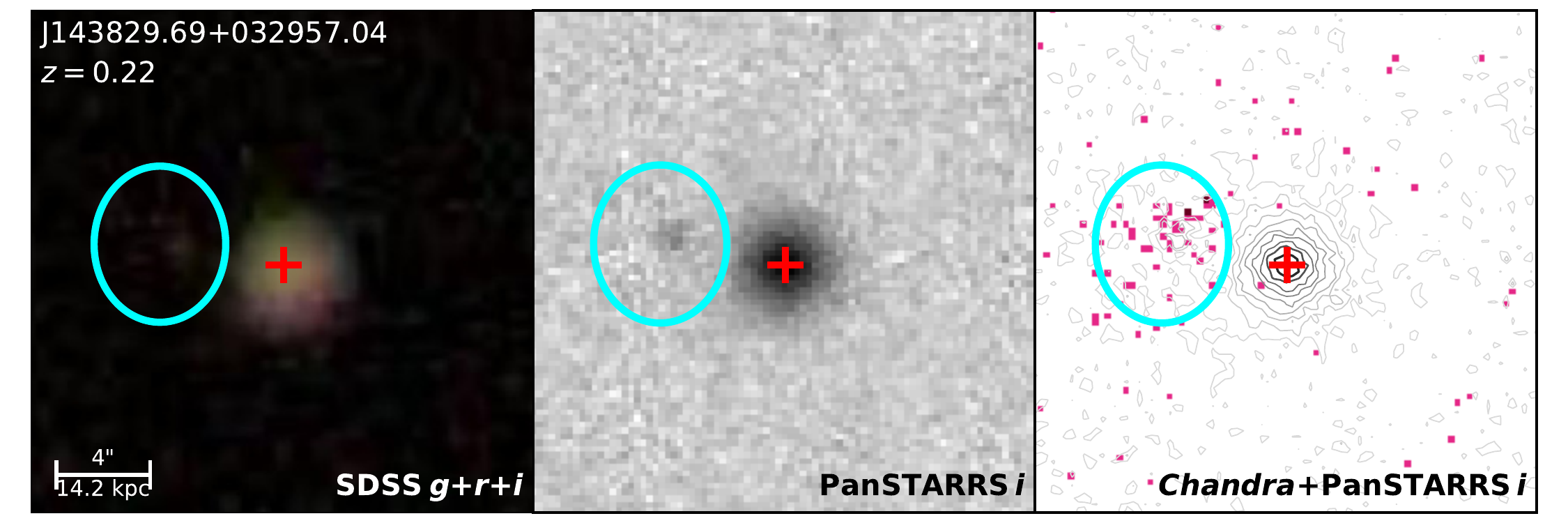} \\
\includegraphics[width=0.48\textwidth]{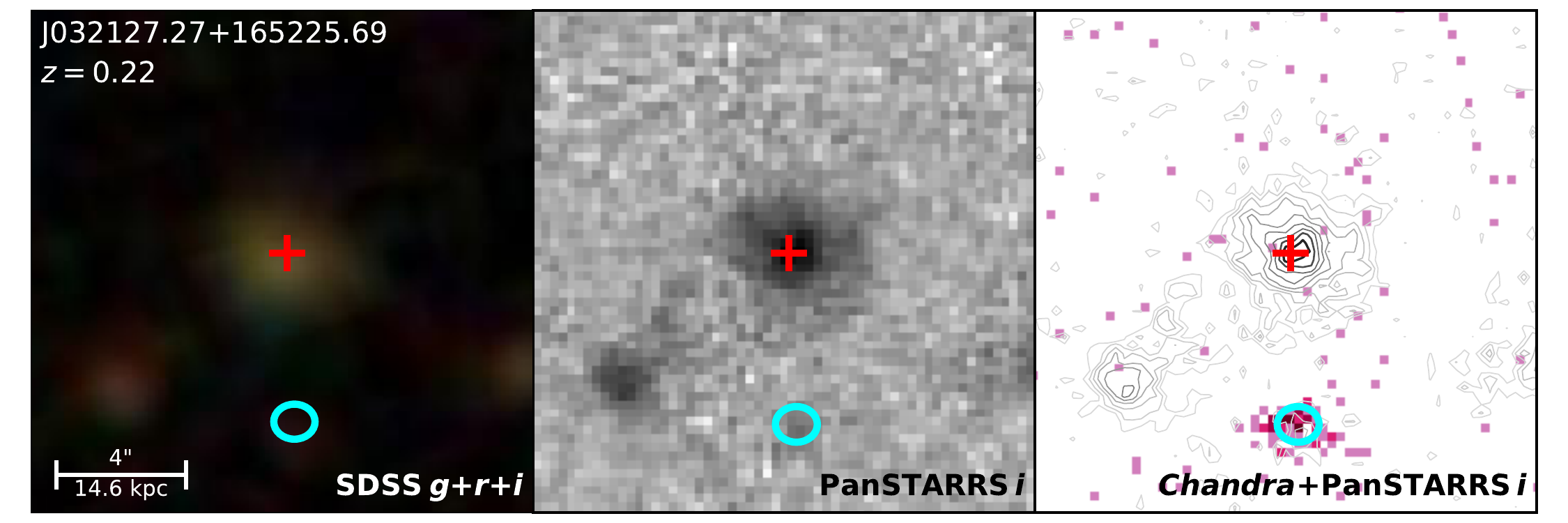} &
\includegraphics[width=0.48\textwidth]{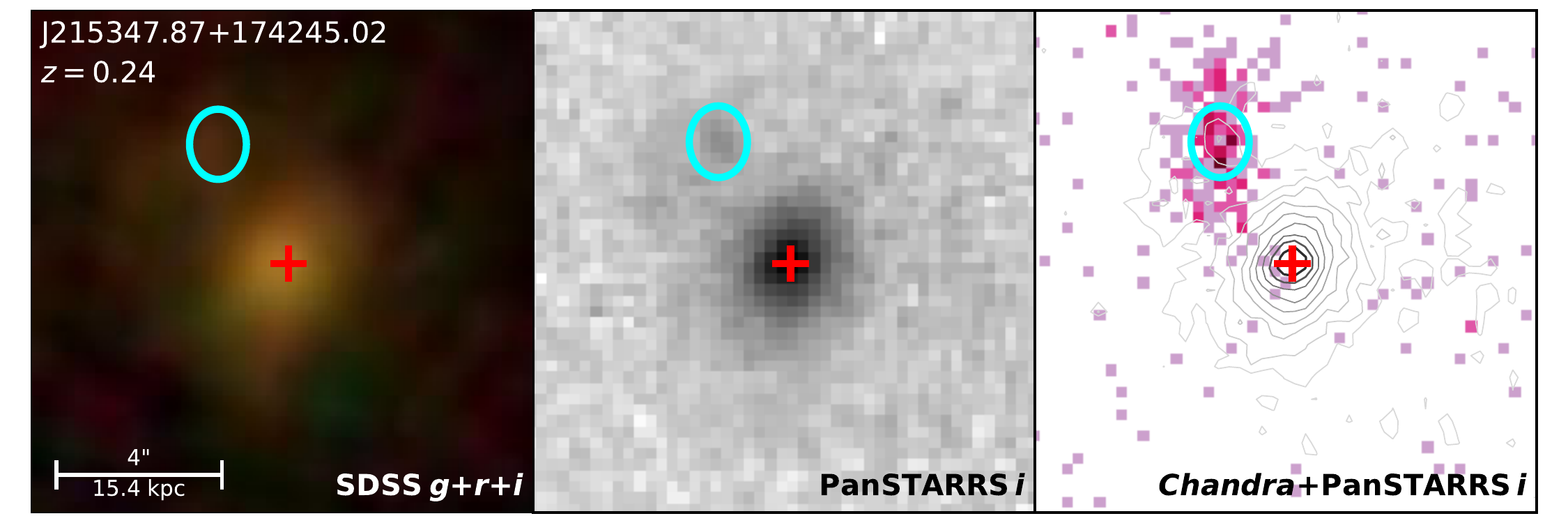} \\
\includegraphics[width=0.48\textwidth]{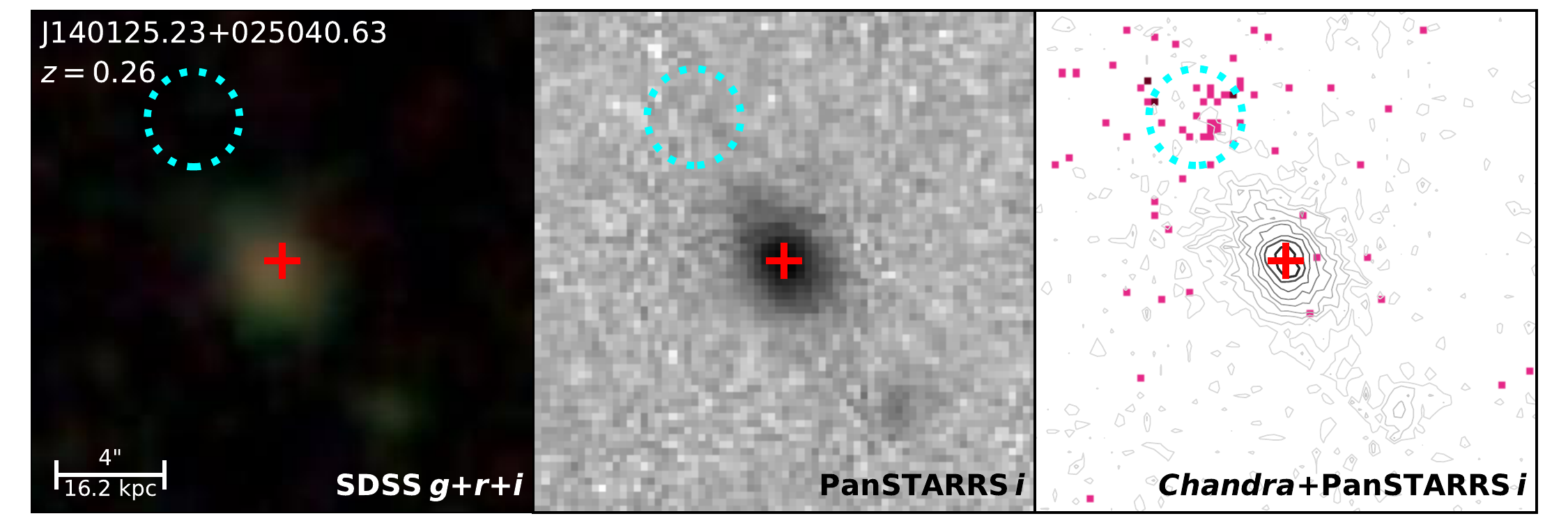} &
\includegraphics[width=0.48\textwidth]{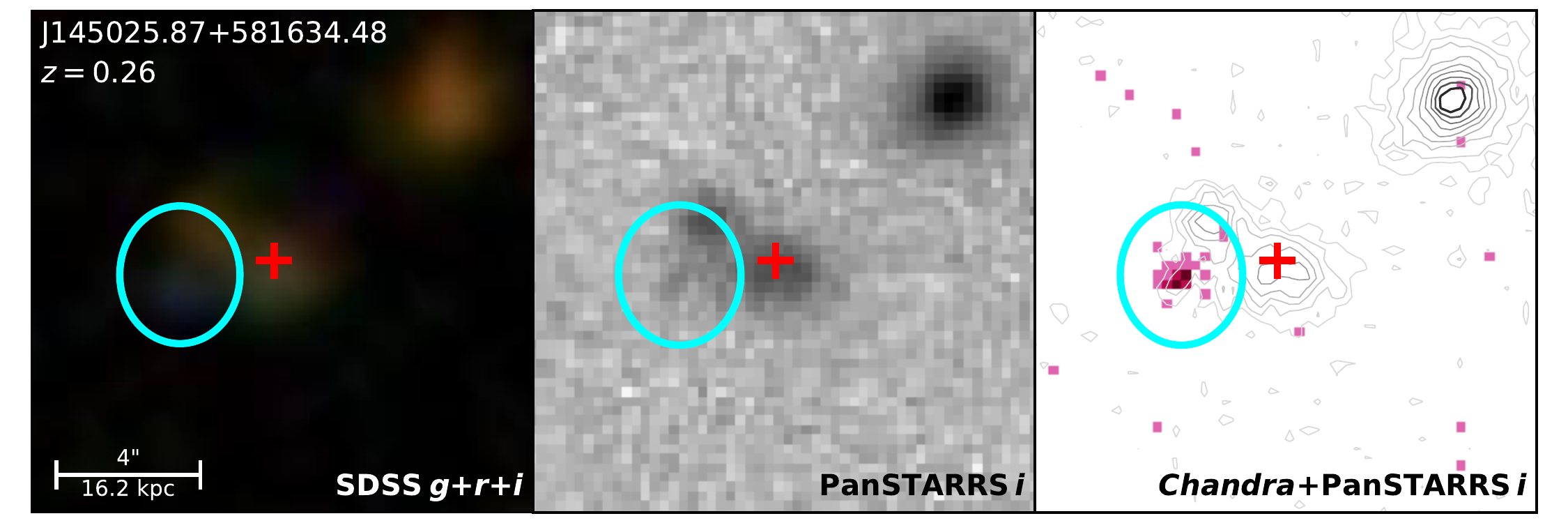} \\
\includegraphics[width=0.48\textwidth]{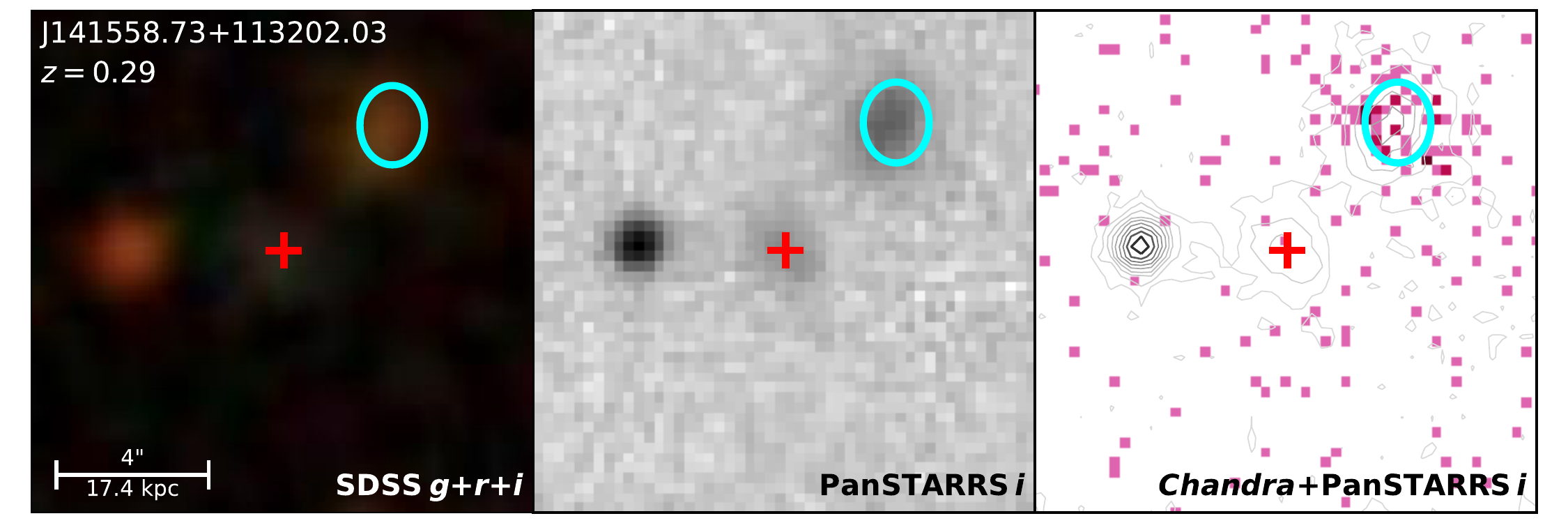} &
\includegraphics[width=0.48\textwidth]{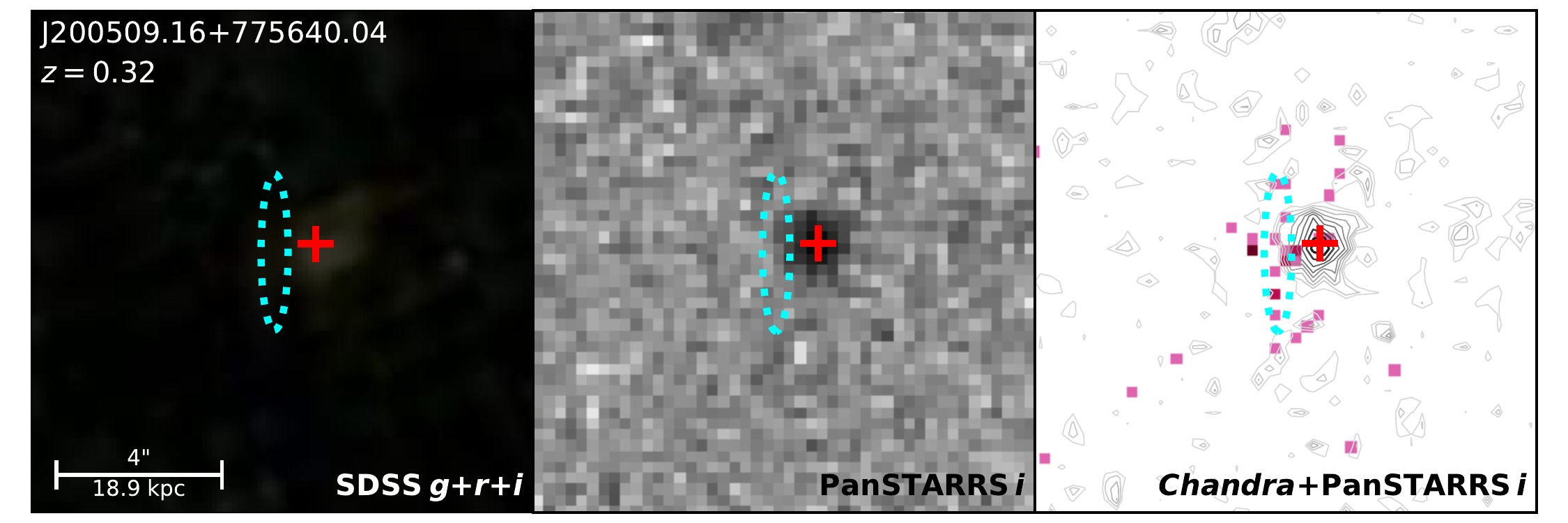} \\
\end{array} $
\caption{continued.}
\end{figure*}
\begin{figure*} $
\ContinuedFloat
\begin{array}{c c}
\includegraphics[width=0.48\textwidth]{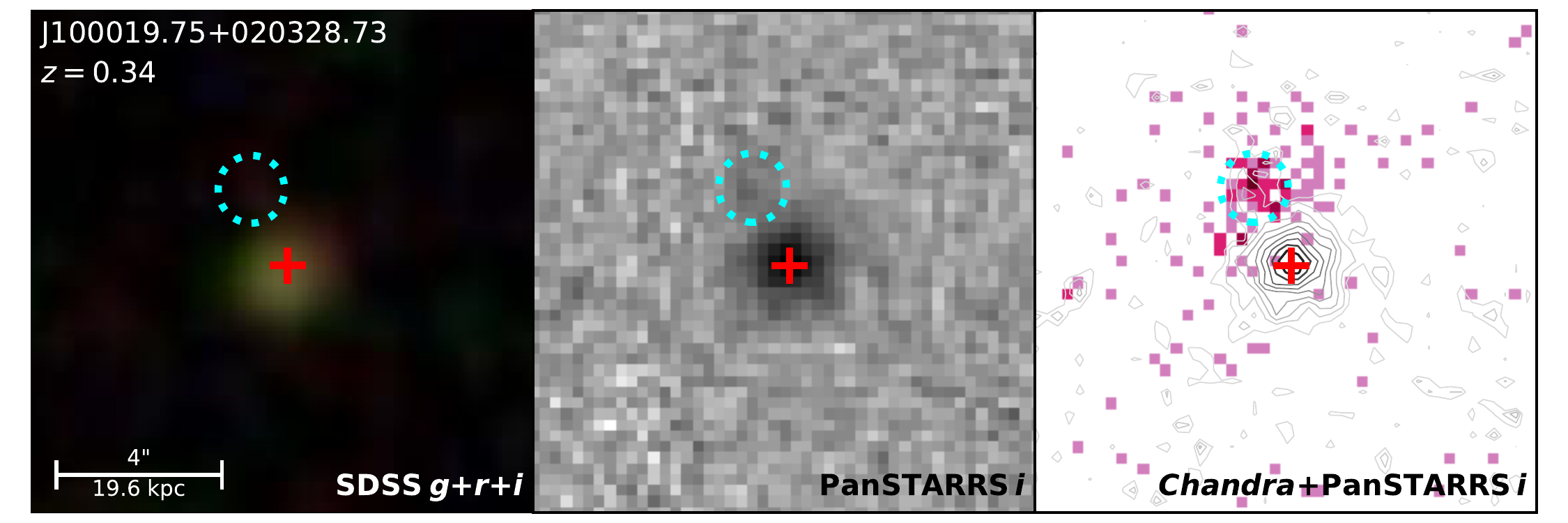} &
\includegraphics[width=0.48\textwidth]{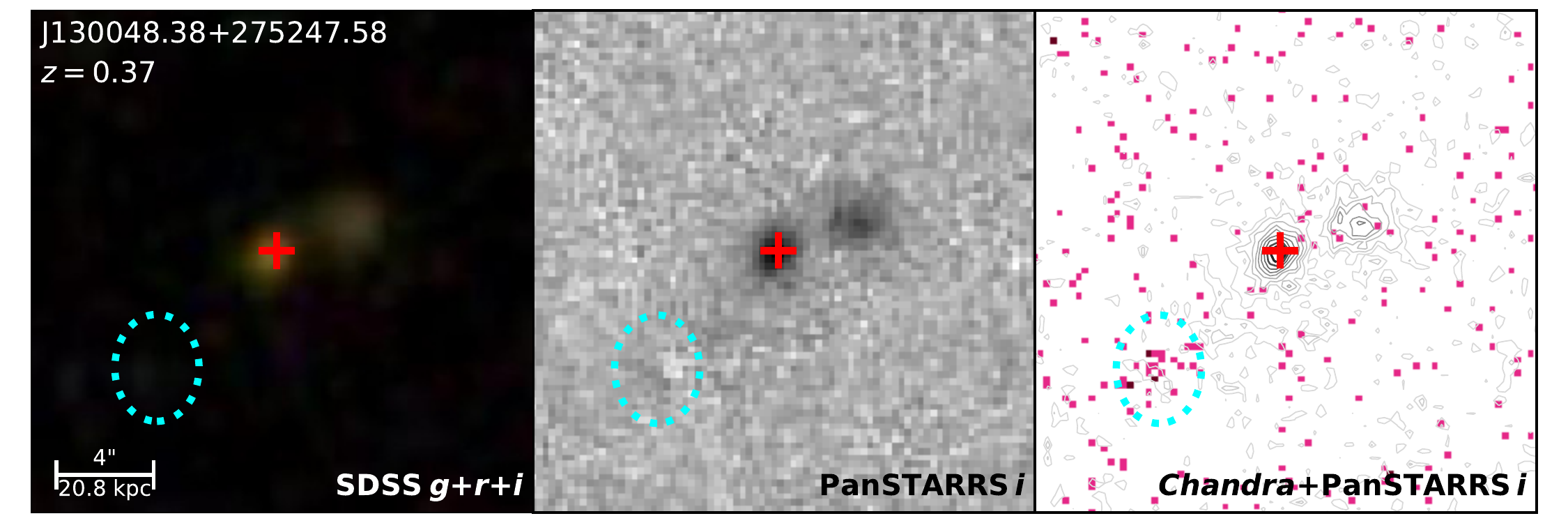} \\
\includegraphics[width=0.48\textwidth]{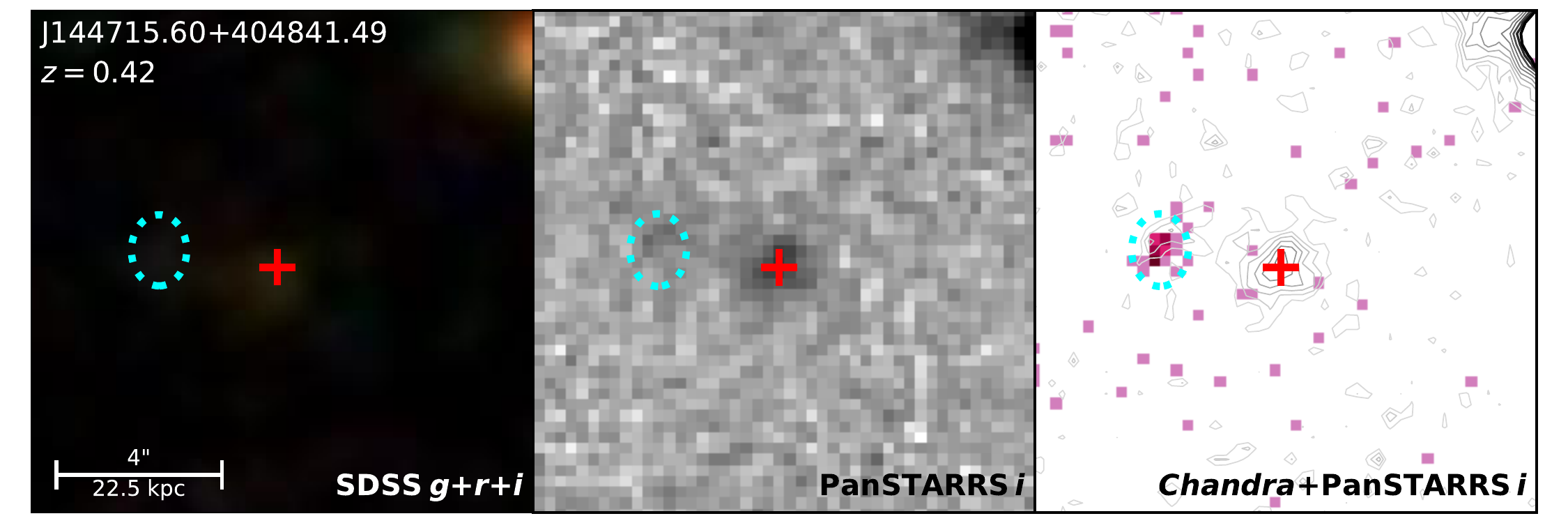} &
\includegraphics[width=0.48\textwidth]{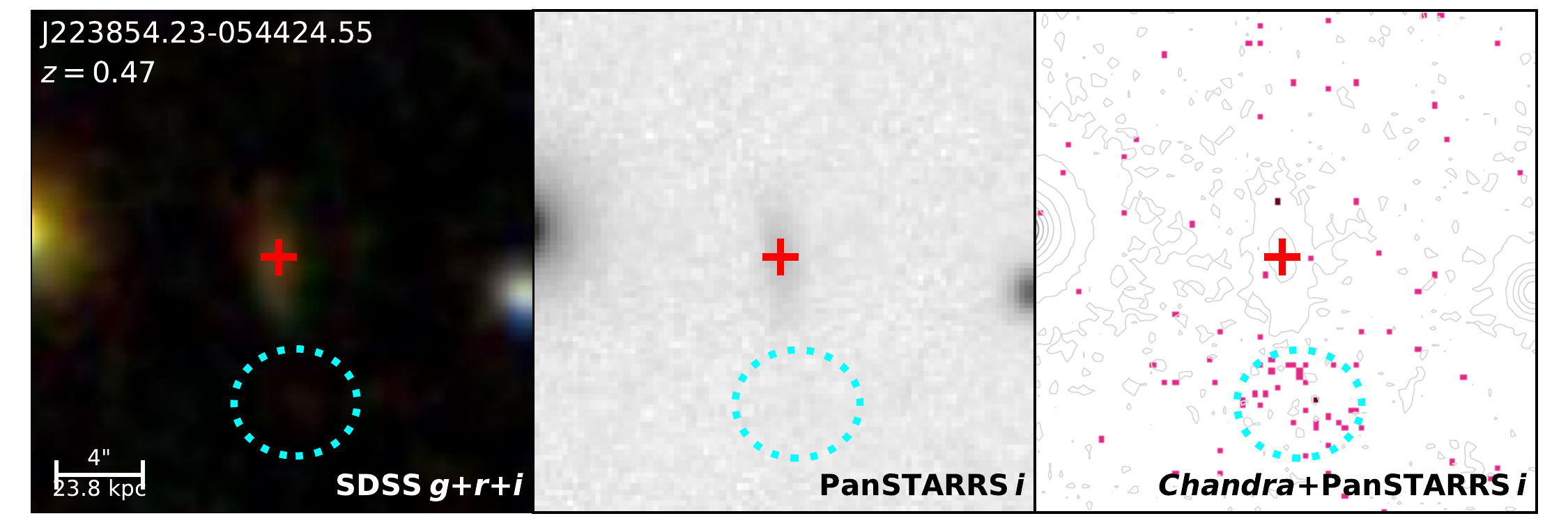} \\
\end{array} $
\caption{continued.}
\end{figure*}


\clearpage

\begin{thebibliography}{124}
\expandafter\ifx\csname natexlab\endcsname\relax\def\natexlab#1{#1}\fi

\bibitem[{{Abazajian} {et~al.}(2009){Abazajian}, {Adelman-McCarthy},
  {Ag{\"u}eros}, {Allam}, {Allende Prieto}, {An}, {Anderson}, {Anderson},
  {Annis}, {Bahcall}, \& et~al.}]{Abazajian09}
{Abazajian}, K.~N., {et~al.} 2009, \apjs, 182, 543

\bibitem[{Abolfathi {et~al.}(2018)Abolfathi, Aguado, Aguilar, Prieto, Almeida,
  Ananna, Anders, Anderson, Andrews, Anguiano, {et~al.}}]{Abolfathi:2018}
Abolfathi, B., {et~al.} 2018, The Astrophysical Journal Supplement Series, 235,
  42

\bibitem[{{Angl{\'e}s-Alc{\'a}zar} {et~al.}(2017){Angl{\'e}s-Alc{\'a}zar},
  {Faucher-Gigu{\`e}re}, {Quataert}, {Hopkins}, {Feldmann}, {Torrey}, {Wetzel},
  \& {Kere{\v s}}}]{Angles-Alcazar:2017}
{Angl{\'e}s-Alc{\'a}zar}, D., {Faucher-Gigu{\`e}re}, C.-A., {Quataert}, E.,
  {Hopkins}, P.~F., {Feldmann}, R., {Torrey}, P., {Wetzel}, A., \& {Kere{\v
  s}}, D. 2017, \mnras, 472, L109

\bibitem[{{Assef} {et~al.}(2018){Assef}, {Stern}, {Noirot}, {Jun}, {Cutri}, \&
  {Eisenhardt}}]{Assef:2018}
{Assef}, R.~J., {Stern}, D., {Noirot}, G., {Jun}, H.~D., {Cutri}, R.~M., \&
  {Eisenhardt}, P.~R.~M. 2018, \apjs, 234, 23

\bibitem[{{Assef} {et~al.}(2011){Assef}, {Kochanek}, {Ashby}, {Brodwin},
  {Brown}, {Cool}, {Forman}, {Gonzalez}, {Hickox}, {Jannuzi}, {Jones}, {Le
  Floc'h}, {Moustakas}, {Murray}, \& {Stern}}]{Assef:2011}
{Assef}, R.~J., {et~al.} 2011, \apj, 728, 56

\bibitem[{{Assef} {et~al.}(2013){Assef}, {Stern}, {Kochanek}, {Blain},
  {Brodwin}, {Brown}, {Donoso}, {Eisenhardt}, {Jannuzi}, {Jarrett}, {Stanford},
  {Tsai}, {Wu}, \& {Yan}}]{Assef:2013}
---. 2013, \apj, 772, 26

\bibitem[{{Astropy Collaboration} {et~al.}(2013){Astropy Collaboration},
  {Robitaille}, {Tollerud}, {Greenfield}, {Droettboom}, {Bray}, {Aldcroft},
  {Davis}, {Ginsburg}, {Price-Whelan}, {Kerzendorf}, {Conley}, {Crighton},
  {Barbary}, {Muna}, {Ferguson}, {Grollier}, {Parikh}, {Nair}, {Unther},
  {Deil}, {Woillez}, {Conseil}, {Kramer}, {Turner}, {Singer}, {Fox}, {Weaver},
  {Zabalza}, {Edwards}, {Azalee Bostroem}, {Burke}, {Casey}, {Crawford},
  {Dencheva}, {Ely}, {Jenness}, {Labrie}, {Lim}, {Pierfederici}, {Pontzen},
  {Ptak}, {Refsdal}, {Servillat}, \& {Streicher}}]{Astropy:2013}
{Astropy Collaboration} {et~al.} 2013, \aap, 558, A33

\bibitem[{{Astropy Collaboration} {et~al.}(2018){Astropy Collaboration},
  {Price-Whelan}, {Sip{\H o}cz}, {G{\"u}nther}, {Lim}, {Crawford}, {Conseil},
  {Shupe}, {Craig}, {Dencheva}, {Ginsburg}, {VanderPlas}, {Bradley},
  {P{\'e}rez-Su{\'a}rez}, {de Val-Borro}, {Aldcroft}, {Cruz}, {Robitaille},
  {Tollerud}, {Ardelean}, {Babej}, {Bach}, {Bachetti}, {Bakanov}, {Bamford},
  {Barentsen}, {Barmby}, {Baumbach}, {Berry}, {Biscani}, {Boquien}, {Bostroem},
  {Bouma}, {Brammer}, {Bray}, {Breytenbach}, {Buddelmeijer}, {Burke},
  {Calderone}, {Cano Rodr{\'{\i}}guez}, {Cara}, {Cardoso}, {Cheedella},
  {Copin}, {Corrales}, {Crichton}, {D'Avella}, {Deil}, {Depagne}, {Dietrich},
  {Donath}, {Droettboom}, {Earl}, {Erben}, {Fabbro}, {Ferreira}, {Finethy},
  {Fox}, {Garrison}, {Gibbons}, {Goldstein}, {Gommers}, {Greco}, {Greenfield},
  {Groener}, {Grollier}, {Hagen}, {Hirst}, {Homeier}, {Horton}, {Hosseinzadeh},
  {Hu}, {Hunkeler}, {Ivezi{\'c}}, {Jain}, {Jenness}, {Kanarek}, {Kendrew},
  {Kern}, {Kerzendorf}, {Khvalko}, {King}, {Kirkby}, {Kulkarni}, {Kumar},
  {Lee}, {Lenz}, {Littlefair}, {Ma}, {Macleod}, {Mastropietro}, {McCully},
  {Montagnac}, {Morris}, {Mueller}, {Mumford}, {Muna}, {Murphy}, {Nelson},
  {Nguyen}, {Ninan}, {N{\"o}the}, {Ogaz}, {Oh}, {Parejko}, {Parley}, {Pascual},
  {Patil}, {Patil}, {Plunkett}, {Prochaska}, {Rastogi}, {Reddy Janga},
  {Sabater}, {Sakurikar}, {Seifert}, {Sherbert}, {Sherwood-Taylor}, {Shih},
  {Sick}, {Silbiger}, {Singanamalla}, {Singer}, {Sladen}, {Sooley},
  {Sornarajah}, {Streicher}, {Teuben}, {Thomas}, {Tremblay}, {Turner},
  {Terr{\'o}n}, {van Kerkwijk}, {de la Vega}, {Watkins}, {Weaver}, {Whitmore},
  {Woillez}, {Zabalza}, \& {Astropy Contributors}}]{Astropy:2018}
---. 2018, \aj, 156, 123

\bibitem[{{Bachetti} {et~al.}(2014){Bachetti}, {Harrison}, {Walton},
  {Grefenstette}, {Chakrabarty}, {F{\"u}rst}, {Barret}, {Beloborodov}, {Boggs},
  {Christensen}, {Craig}, {Fabian}, {Hailey}, {Hornschemeier}, {Kaspi},
  {Kulkarni}, {Maccarone}, {Miller}, {Rana}, {Stern}, {Tendulkar}, {Tomsick},
  {Webb}, \& {Zhang}}]{Bachetti:2014}
{Bachetti}, M., {et~al.} 2014, \nat, 514, 202

\bibitem[{{Baldassare} {et~al.}(2015){Baldassare}, {Reines}, {Gallo}, \&
  {Greene}}]{Baldassare:2015}
{Baldassare}, V.~F., {Reines}, A.~E., {Gallo}, E., \& {Greene}, J.~E. 2015,
  \apj, 809, L14

\bibitem[{{Baldassare} {et~al.}(2017){Baldassare}, {Reines}, {Gallo}, \&
  {Greene}}]{Baldassare:2017}
---. 2017, \apj, 836, 20

\bibitem[{{Baldassare} {et~al.}(2016){Baldassare}, {Reines}, {Gallo}, {Greene},
  {Graur}, {Geha}, {Hainline}, {Carroll}, \& {Hickox}}]{Baldassare:2016}
{Baldassare}, V.~F., {et~al.} 2016, \apj, 829, 57

\bibitem[{{Barrows} {et~al.}(2016){Barrows}, {Comerford}, {Greene}, \&
  {Pooley}}]{Barrows:2016}
{Barrows}, R.~S., {Comerford}, J.~M., {Greene}, J.~E., \& {Pooley}, D. 2016,
  \apj, 829, 37

\bibitem[{{Barrows} {et~al.}(2017){Barrows}, {Comerford}, {Zakamska}, \&
  {Cooper}}]{Barrows:2017b}
{Barrows}, R.~S., {Comerford}, J.~M., {Zakamska}, N.~L., \& {Cooper}, M.~C.
  2017, \apj, 850, 27

\bibitem[{{Begelman}(2002)}]{Begelman:2002}
{Begelman}, M.~C. 2002, \apjl, 568, L97

\bibitem[{{Bell} {et~al.}(2003){Bell}, {McIntosh}, {Katz}, \&
  {Weinberg}}]{Bell:2003}
{Bell}, E.~F., {McIntosh}, D.~H., {Katz}, N., \& {Weinberg}, M.~D. 2003, \apjs,
  149, 289

\bibitem[{{Bellovary} {et~al.}(2019){Bellovary}, {Cleary}, {Munshi}, {Tremmel},
  {Christensen}, {Brooks}, \& {Quinn}}]{Bellovary:2019}
{Bellovary}, J.~M., {Cleary}, C.~E., {Munshi}, F., {Tremmel}, M.,
  {Christensen}, C.~R., {Brooks}, A., \& {Quinn}, T.~R. 2019, \mnras, 482, 2913

\bibitem[{{Bertin} \& {Arnouts}(1996)}]{Bertin:Arnouts:1996}
{Bertin}, E., \& {Arnouts}, S. 1996, \aaps, 117, 393

\bibitem[{Bian(2005)}]{Bian:2005}
Bian, W.-H. 2005, Chinese Journal of Astronomy and Astrophysics, 5, 289

\bibitem[{{Capelo} {et~al.}(2015){Capelo}, {Volonteri}, {Dotti}, {Bellovary},
  {Mayer}, \& {Governato}}]{Capelo:2015}
{Capelo}, P.~R., {Volonteri}, M., {Dotti}, M., {Bellovary}, J.~M., {Mayer}, L.,
  \& {Governato}, F. 2015, \mnras, 447, 2123

\bibitem[{{Chambers} {et~al.}(2016){Chambers}, {Magnier}, {Metcalfe},
  {Flewelling}, {Huber}, {Waters}, {Denneau}, {Draper}, {Farrow}, {Finkbeiner},
  {Holmberg}, {Koppenhoefer}, {Price}, {Saglia}, {Schlafly}, {Smartt},
  {Sweeney}, {Wainscoat}, {Burgett}, {Grav}, {Heasley}, {Hodapp}, {Jedicke},
  {Kaiser}, {Kudritzki}, {Luppino}, {Lupton}, {Monet}, {Morgan}, {Onaka},
  {Stubbs}, {Tonry}, {Banados}, {Bell}, {Bender}, {Bernard}, {Botticella},
  {Casertano}, {Chastel}, {Chen}, {Chen}, {Cole}, {Deacon}, {Frenk},
  {Fitzsimmons}, {Gezari}, {Goessl}, {Goggia}, {Goldman}, {Grebel}, {Hambly},
  {Hasinger}, {Heavens}, {Heckman}, {Henderson}, {Henning}, {Holman}, {Hopp},
  {Ip}, {Isani}, {Keyes}, {Koekemoer}, {Kotak}, {Long}, {Lucey}, {Liu},
  {Martin}, {McLean}, {Morganson}, {Murphy}, {Nieto-Santisteban}, {Norberg},
  {Peacock}, {Pier}, {Postman}, {Primak}, {Rae}, {Rest}, {Riess}, {Riffeser},
  {Rix}, {Roser}, {Schilbach}, {Schultz}, {Scolnic}, {Szalay}, {Seitz},
  {Shiao}, {Small}, {Smith}, {Soderblom}, {Taylor}, {Thakar}, {Thiel},
  {Thilker}, {Urata}, {Valenti}, {Walter}, {Watters}, {Werner}, {White},
  {Wood-Vasey}, \& {Wyse}}]{Chambers:2016}
{Chambers}, K.~C., {et~al.} 2016, ArXiv e-prints

\bibitem[{{Chilingarian} {et~al.}(2018){Chilingarian}, {Katkov}, {Zolotukhin},
  {Grishin}, {Beletsky}, {Boutsia}, \& {Osip}}]{Chilingarian:2018}
{Chilingarian}, I.~V., {Katkov}, I.~Y., {Zolotukhin}, I.~Y., {Grishin}, K.~A.,
  {Beletsky}, Y., {Boutsia}, K., \& {Osip}, D.~J. 2018, \apj, 863, 1

\bibitem[{{Cisternas} {et~al.}(2011){Cisternas}, {Jahnke}, {Inskip},
  {Kartaltepe}, {Koekemoer}, {Lisker}, {Robaina}, {Scodeggio}, {Sheth},
  {Trump}, {Andrae}, {Miyaji}, {Lusso}, {Brusa}, {Capak}, {Cappelluti},
  {Civano}, {Ilbert}, {Impey}, {Leauthaud}, {Lilly}, {Salvato}, {Scoville}, \&
  {Taniguchi}}]{Cisternas:2011}
{Cisternas}, M., {et~al.} 2011, \apj, 726, 57

\bibitem[{{Civano} {et~al.}(2012){Civano}, {Elvis}, {Brusa}, {Comastri},
  {Salvato}, {Zamorani}, {Aldcroft}, {Bongiorno}, {Capak}, {Cappelluti},
  {Cisternas}, {Fiore}, {Fruscione}, {Hao}, {Kartaltepe}, {Koekemoer}, {Gilli},
  {Impey}, {Lanzuisi}, {Lusso}, {Mainieri}, {Miyaji}, {Lilly}, {Masters},
  {Puccetti}, {Schawinski}, {Scoville}, {Silverman}, {Trump}, {Urry},
  {Vignali}, \& {Wright}}]{Civano:2012}
{Civano}, F., {et~al.} 2012, The Astrophysical Journal Supplement Series, 201,
  30

\bibitem[{{Civano} {et~al.}(2016){Civano}, {Marchesi}, {Comastri}, {Urry},
  {Elvis}, {Cappelluti}, {Puccetti}, {Brusa}, {Zamorani}, {Hasinger},
  {Aldcroft}, {Alexand er}, {Allevato}, {Brunner}, {Capak}, {Finoguenov},
  {Fiore}, {Fruscione}, {Gilli}, {Glotfelty}, {Griffiths}, {Hao}, {Harrison},
  {Jahnke}, {Kartaltepe}, {Karim}, {LaMassa}, {Lanzuisi}, {Miyaji}, {Ranalli},
  {Salvato}, {Sargent}, {Scoville}, {Schawinski}, {Schinnerer}, {Silverman},
  {Smolcic}, {Stern}, {Toft}, {Trakhtenbrot}, {Treister}, \&
  {Vignali}}]{Civano:2016}
---. 2016, \apj, 819, 62

\bibitem[{{Colbert} \& {Mushotzky}(1999)}]{Colbert:Mushotzky:1999}
{Colbert}, E.~J.~M., \& {Mushotzky}, R.~F. 1999, \apj, 519, 89

\bibitem[{{Comerford} {et~al.}(2015){Comerford}, {Pooley}, {Barrows}, {Greene},
  {Zakamska}, {Madejski}, \& {Cooper}}]{Comerford:2015}
{Comerford}, J.~M., {Pooley}, D., {Barrows}, R.~S., {Greene}, J.~E.,
  {Zakamska}, N.~L., {Madejski}, G.~M., \& {Cooper}, M.~C. 2015, \apj, 806, 219

\bibitem[{{Cunha} {et~al.}(2009){Cunha}, {Lima}, {Oyaizu}, {Frieman}, \&
  {Lin}}]{Cunha:2009}
{Cunha}, C.~E., {Lima}, M., {Oyaizu}, H., {Frieman}, J., \& {Lin}, H. 2009,
  \mnras, 396, 2379

\bibitem[{{Dadina} {et~al.}(2013){Dadina}, {Masetti}, {Cappi}, {Malaguti},
  {Miniutti}, {Ponti}, {Gandhi}, \& {De Marco}}]{Dadina:2013}
{Dadina}, M., {Masetti}, N., {Cappi}, M., {Malaguti}, G., {Miniutti}, G.,
  {Ponti}, G., {Gandhi}, P., \& {De Marco}, B. 2013, \aap, 559, A86

\bibitem[{{Davis} {et~al.}(2011){Davis}, {Narayan}, {Zhu}, {Barret}, {Farrell},
  {Godet}, {Servillat}, \& {Webb}}]{Davis:2011:HLX1}
{Davis}, S.~W., {Narayan}, R., {Zhu}, Y., {Barret}, D., {Farrell}, S.~A.,
  {Godet}, O., {Servillat}, M., \& {Webb}, N.~A. 2011, \apj, 734, 111

\bibitem[{{de Vaucouleurs} {et~al.}(1991){de Vaucouleurs}, {de Vaucouleurs},
  {Corwin}, {Buta}, {Paturel}, \& {Fouque}}]{deVaucouleurs:1991}
{de Vaucouleurs}, G., {de Vaucouleurs}, A., {Corwin}, Herold~G., J., {Buta},
  R.~J., {Paturel}, G., \& {Fouque}, P. 1991, {Third Reference Catalogue of
  Bright Galaxies}

\bibitem[{{Dong} {et~al.}(2012){Dong}, {Greene}, \& {Ho}}]{RDong:2012}
{Dong}, R., {Greene}, J.~E., \& {Ho}, L.~C. 2012, \apj, 761, 73

\bibitem[{{Earnshaw} {et~al.}(2018){Earnshaw}, {Roberts}, {Middleton},
  {Walton}, \& {Mateos}}]{Earnshaw:2018}
{Earnshaw}, H.~P., {Roberts}, T.~P., {Middleton}, M.~J., {Walton}, D.~J., \&
  {Mateos}, S. 2018, arXiv e-prints

\bibitem[{Evans {et~al.}(2010)Evans, Primini, Glotfelty, Anderson, Bonaventura,
  Chen, Davis, Doe, Evans, Fabbiano, {et~al.}}]{Evans:2010}
Evans, I.~N., {et~al.} 2010, The Astrophysical Journal Supplement Series, 189,
  37

\bibitem[{{Farrell} {et~al.}(2009){Farrell}, {Webb}, {Barret}, {Godet}, \&
  {Rodrigues}}]{Farrell:2009}
{Farrell}, S.~A., {Webb}, N.~A., {Barret}, D., {Godet}, O., \& {Rodrigues},
  J.~M. 2009, \nat, 460, 73

\bibitem[{Gaetz \& Jerius(2004)}]{Gaetz:2004}
Gaetz, T., \& Jerius, D. 2004

\bibitem[{{Gao} {et~al.}(2003){Gao}, {Wang}, {Appleton}, \& {Lucas}}]{Gao:2003}
{Gao}, Y., {Wang}, Q.~D., {Appleton}, P.~N., \& {Lucas}, R.~A. 2003, \apj, 596,
  L171

\bibitem[{{Ghosh} \& {White}(2001)}]{Ghosh:White:2001}
{Ghosh}, P., \& {White}, N.~E. 2001, \apjl, 559, L97

\bibitem[{{Gong} {et~al.}(2016){Gong}, {Liu}, \& {Maccarone}}]{Gong:2016}
{Gong}, H., {Liu}, J., \& {Maccarone}, T. 2016, \apjs, 222, 12

\bibitem[{{Graham} {et~al.}(2005){Graham}, {Driver}, {Petrosian}, {Conselice},
  {Bershady}, {Crawford}, \& {Goto}}]{Graham:2005b}
{Graham}, A.~W., {Driver}, S.~P., {Petrosian}, V., {Conselice}, C.~J.,
  {Bershady}, M.~A., {Crawford}, S.~M., \& {Goto}, T. 2005, \aj, 130, 1535

\bibitem[{{Greene}(2012)}]{Greene:2012}
{Greene}, J.~E. 2012, Nature Communications, 3, 1304

\bibitem[{{Greene} \& {Ho}(2004)}]{Greene:Ho:2004}
{Greene}, J.~E., \& {Ho}, L.~C. 2004, \apj, 610, 722

\bibitem[{{Greene} \& {Ho}(2007)}]{Greene:2007}
---. 2007, \apj, 670, 92

\bibitem[{{Heida} {et~al.}(2014){Heida}, {Jonker}, {Torres}, {Kool},
  {Servillat}, {Roberts}, {Groot}, {Walton}, {Moon}, \&
  {Harrison}}]{Heida:2014}
{Heida}, M., {et~al.} 2014, \mnras, 442, 1054

\bibitem[{{Hewitt} \& {Burbidge}(1993)}]{Hewitt:Burbidge:1993}
{Hewitt}, A., \& {Burbidge}, G. 1993, \apjs, 87, 451

\bibitem[{{Hickox} {et~al.}(2009){Hickox}, {Jones}, {Forman}, {Murray},
  {Kochanek}, {Eisenstein}, {Jannuzi}, {Dey}, {Brown}, {Stern}, {Eisenhardt},
  {Gorjian}, {Brodwin}, {Narayan}, {Cool}, {Kenter}, {Caldwell}, \&
  {Anderson}}]{Hickox:2009}
{Hickox}, R.~C., {et~al.} 2009, \apj, 696, 891

\bibitem[{{Hinshaw} {et~al.}(2013){Hinshaw}, {Larson}, {Komatsu}, {Spergel},
  {Bennett}, {Dunkley}, {Nolta}, {Halpern}, {Hill}, {Odegard}, {Page}, {Smith},
  {Weiland}, {Gold}, {Jarosik}, {Kogut}, {Limon}, {Meyer}, {Tucker}, {Wollack},
  \& {Wright}}]{Hinshaw:2013}
{Hinshaw}, G., {et~al.} 2013, The Astrophysical Journal Supplement Series, 208,
  19

\bibitem[{{Hopkins} {et~al.}(2003){Hopkins}, {Miller}, {Nichol}, {Connolly},
  {Bernardi}, {G{\'o}mez}, {Goto}, {Tremonti}, {Brinkmann}, {Ivezi{\'c}}, \&
  {Lamb}}]{Hopkins:2003}
{Hopkins}, A.~M., {et~al.} 2003, \apj, 599, 971

\bibitem[{Ishibashi \& Courvoisier(2010)}]{Ishibashi:2010}
Ishibashi, W., \& Courvoisier, T.-L. 2010, Astronomy \& Astrophysics, 512, A58

\bibitem[{Israel {et~al.}(2016)Israel, Papitto, Esposito, Stella, Zampieri,
  Belfiore, Rodr{\'\i}guez~Castillo, De~Luca, Tiengo, Haberl,
  {et~al.}}]{Israel:2016}
Israel, G., {et~al.} 2016, Monthly Notices of the Royal Astronomical Society:
  Letters, 466, L48

\bibitem[{{Jester} {et~al.}(2005){Jester}, {Schneider}, {Richards}, {Green},
  {Schmidt}, {Hall}, {Strauss}, {Vanden Berk}, {Stoughton}, {Gunn},
  {Brinkmann}, {Kent}, {Smith}, {Tucker}, \& {Yanny}}]{Jester05}
{Jester}, S., {et~al.} 2005, AJ, 130, 873

\bibitem[{{Jonker} {et~al.}(2010){Jonker}, {Torres}, {Fabian}, {Heida},
  {Miniutti}, \& {Pooley}}]{Jonker:2010}
{Jonker}, P.~G., {Torres}, M.~A.~P., {Fabian}, A.~C., {Heida}, M., {Miniutti},
  G., \& {Pooley}, D. 2010, \mnras, 407, 645

\bibitem[{{Kaaret} {et~al.}(2017){Kaaret}, {Feng}, \& {Roberts}}]{Kaaret:2017}
{Kaaret}, P., {Feng}, H., \& {Roberts}, T.~P. 2017, Annual Review of Astronomy
  and Astrophysics, 55, 303

\bibitem[{{Kim} {et~al.}(2015){Kim}, {Ho}, {Wang}, {Fabbiano}, {Bianchi},
  {Cappi}, {Dadina}, {Malaguti}, \& {Wang}}]{Kim:2015}
{Kim}, M., {et~al.} 2015, ArXiv e-prints

\bibitem[{{King} \& {Dehnen}(2005)}]{King:2005}
{King}, A.~R., \& {Dehnen}, W. 2005, \mnras, 357, 275

\bibitem[{{Lehmer} {et~al.}(2010){Lehmer}, {Alexander}, {Bauer}, {Brandt},
  {Goulding}, {Jenkins}, {Ptak}, \& {Roberts}}]{Lehmer:2010}
{Lehmer}, B.~D., {Alexander}, D.~M., {Bauer}, F.~E., {Brandt}, W.~N.,
  {Goulding}, A.~D., {Jenkins}, L.~P., {Ptak}, A., \& {Roberts}, T.~P. 2010,
  \apj, 724, 559

\bibitem[{{Lehmer} {et~al.}(2016){Lehmer}, {Basu-Zych}, {Mineo}, {Brandt},
  {Eufrasio}, {Fragos}, {Hornschemeier}, {Luo}, {Xue}, {Bauer}, {Gilfanov},
  {Ranalli}, {Schneider}, {Shemmer}, {Tozzi}, {Trump}, {Vignali}, {Wang},
  {Yukita}, \& {Zezas}}]{Lehmer:2016}
{Lehmer}, B.~D., {et~al.} 2016, \apj, 825, 7

\bibitem[{{Lemons} {et~al.}(2015){Lemons}, {Reines}, {Plotkin}, {Gallo}, \&
  {Greene}}]{Lemons:2015}
{Lemons}, S.~M., {Reines}, A.~E., {Plotkin}, R.~M., {Gallo}, E., \& {Greene},
  J.~E. 2015, \apj, 805, 12

\bibitem[{Levenberg(1944)}]{Levenberg:1944}
Levenberg, K. 1944, Quarterly of applied mathematics, 2, 164

\bibitem[{{Lin} {et~al.}(2012){Lin}, {Webb}, \& {Barret}}]{Lin:2012}
{Lin}, D., {Webb}, N.~A., \& {Barret}, D. 2012, \apj, 756, 27

\bibitem[{{Liu} \& {Bregman}(2005)}]{Liu:2005b}
{Liu}, J.-F., \& {Bregman}, J.~N. 2005, The Astrophysical Journal Supplement
  Series, 157, 59

\bibitem[{{Liu} \& {Mirabel}(2005)}]{Liu:2005a}
{Liu}, Q.~Z., \& {Mirabel}, I.~F. 2005, \aap, 429, 1125

\bibitem[{{Lodato} \& {Natarajan}(2006)}]{Lodato:2006}
{Lodato}, G., \& {Natarajan}, P. 2006, \mnras, 371, 1813

\bibitem[{{Lofthouse} {et~al.}(2018){Lofthouse}, {Kaviraj}, {Smith}, \&
  {Hardcastle}}]{Lofthouse:2018}
{Lofthouse}, E.~K., {Kaviraj}, S., {Smith}, D.~J.~B., \& {Hardcastle}, M.~J.
  2018, \mnras, 479, 807

\bibitem[{{L{\'o}pez-Corredoira} \& {Guti{\'e}rrez}(2006)}]{Corredoira:2006}
{L{\'o}pez-Corredoira}, M., \& {Guti{\'e}rrez}, C.~M. 2006, \aap, 454, 77

\bibitem[{{Lotz} {et~al.}(2011){Lotz}, {Jonsson}, {Cox}, {Croton}, {Primack},
  {Somerville}, \& {Stewart}}]{Lotz:2011}
{Lotz}, J.~M., {Jonsson}, P., {Cox}, T.~J., {Croton}, D., {Primack}, J.~R.,
  {Somerville}, R.~S., \& {Stewart}, K. 2011, \apj, 742, 103

\bibitem[{{Maccacaro} {et~al.}(1988){Maccacaro}, {Gioia}, {Wolter}, {Zamorani},
  \& {Stocke}}]{Maccacaro:1988}
{Maccacaro}, T., {Gioia}, I.~M., {Wolter}, A., {Zamorani}, G., \& {Stocke},
  J.~T. 1988, \apj, 326, 680

\bibitem[{{Madau} {et~al.}(1998){Madau}, {Pozzetti}, \&
  {Dickinson}}]{Madau:1998}
{Madau}, P., {Pozzetti}, L., \& {Dickinson}, M. 1998, \apj, 498, 106

\bibitem[{{Mapelli} {et~al.}(2012){Mapelli}, {Zampieri}, \&
  {Mayer}}]{Mapelli:2012}
{Mapelli}, M., {Zampieri}, L., \& {Mayer}, L. 2012, \mnras, 423, 1309

\bibitem[{{Marconi} {et~al.}(2004){Marconi}, {Risaliti}, {Gilli}, {Hunt},
  {Maiolino}, \& {Salvati}}]{Marconi04}
{Marconi}, A., {Risaliti}, G., {Gilli}, R., {Hunt}, L.~K., {Maiolino}, R., \&
  {Salvati}, M. 2004, MNRAS, 351, 169

\bibitem[{{Marleau} {et~al.}(2013){Marleau}, {Clancy}, \&
  {Bianconi}}]{Marleau:2013}
{Marleau}, F.~R., {Clancy}, D., \& {Bianconi}, M. 2013, \mnras, 435, 3085

\bibitem[{Marquardt(1963)}]{Marquardt:1963}
Marquardt, D.~W. 1963, Journal of the society for Industrial and Applied
  Mathematics, 11, 431

\bibitem[{{Martin} {et~al.}(2018){Martin}, {Kaviraj}, {Volonteri}, {Simmons},
  {Devriendt}, {Lintott}, {Smethurst}, {Dubois}, \& {Pichon}}]{Martin:2018}
{Martin}, G., {et~al.} 2018, \mnras, 476, 2801

\bibitem[{{Mart{\'\i}n-Navarro} \& {Mezcua}(2018)}]{Martin-Navarro:2018}
{Mart{\'\i}n-Navarro}, I., \& {Mezcua}, M. 2018, \apj, 855, L20

\bibitem[{{Matsumoto} {et~al.}(2001){Matsumoto}, {Tsuru}, {Koyama}, {Awaki},
  {Canizares}, {Kawai}, {Matsushita}, \& {Kawabe}}]{Matsumoto:2001}
{Matsumoto}, H., {Tsuru}, T.~G., {Koyama}, K., {Awaki}, H., {Canizares}, C.~R.,
  {Kawai}, N., {Matsushita}, S., \& {Kawabe}, R. 2001, \apjl, 547, L25

\bibitem[{{Mezcua}(2017)}]{Mezcua:2017}
{Mezcua}, M. 2017, International Journal of Modern Physics D, 26, 1730021

\bibitem[{{Mezcua}(2019)}]{Mezcua:2019}
---. 2019, Nature Astronomy, 3, 6

\bibitem[{{Mezcua} {et~al.}(2016){Mezcua}, {Civano}, {Fabbiano}, {Miyaji}, \&
  {Marchesi}}]{Mezcua:2016}
{Mezcua}, M., {Civano}, F., {Fabbiano}, G., {Miyaji}, T., \& {Marchesi}, S.
  2016, \apj, 817, 20

\bibitem[{{Mezcua} {et~al.}(2018{\natexlab{a}}){Mezcua}, {Civano}, {Marchesi},
  {Suh}, {Fabbiano}, \& {Volonteri}}]{Mezcua:2018}
{Mezcua}, M., {Civano}, F., {Marchesi}, S., {Suh}, H., {Fabbiano}, G., \&
  {Volonteri}, M. 2018{\natexlab{a}}, \mnras, 478, 2576

\bibitem[{{Mezcua} {et~al.}(2013{\natexlab{a}}){Mezcua}, {Farrell},
  {Gladstone}, \& {Lobanov}}]{Mezcua:2013a}
{Mezcua}, M., {Farrell}, S.~A., {Gladstone}, J.~C., \& {Lobanov}, A.~P.
  2013{\natexlab{a}}, \mnras, 436, 1546

\bibitem[{{Mezcua} {et~al.}(2018{\natexlab{b}}){Mezcua}, {Kim}, {Ho}, \&
  {Lonsdale}}]{Mezcua:2018b}
{Mezcua}, M., {Kim}, M., {Ho}, L.~C., \& {Lonsdale}, C.~J. 2018{\natexlab{b}},
  \mnras, 480, L74

\bibitem[{{Mezcua} {et~al.}(2015){Mezcua}, {Roberts}, {Lobanov}, \&
  {Sutton}}]{Mezcua:2015}
{Mezcua}, M., {Roberts}, T.~P., {Lobanov}, A.~P., \& {Sutton}, A.~D. 2015,
  \mnras, 448, 1893

\bibitem[{{Mezcua} {et~al.}(2013{\natexlab{b}}){Mezcua}, {Roberts}, {Sutton},
  \& {Lobanov}}]{Mezcua:2013c}
{Mezcua}, M., {Roberts}, T.~P., {Sutton}, A.~D., \& {Lobanov}, A.~P.
  2013{\natexlab{b}}, \mnras, 436, 3128

\bibitem[{{Middleton} {et~al.}(2008){Middleton}, {Done}, \&
  {Schurch}}]{Middleton:2008}
{Middleton}, M., {Done}, C., \& {Schurch}, N. 2008, \mnras, 383, 1501

\bibitem[{{Miller} {et~al.}(2003){Miller}, {Fabbiano}, {Miller}, \&
  {Fabian}}]{Miller:2003}
{Miller}, J.~M., {Fabbiano}, G., {Miller}, M.~C., \& {Fabian}, A.~C. 2003,
  \apjl, 585, L37

\bibitem[{{Miller} \& {Colbert}(2004)}]{Miller:Colbert:2004}
{Miller}, M.~C., \& {Colbert}, E.~J.~M. 2004, International Journal of Modern
  Physics D, 13, 1

\bibitem[{{Mineo} {et~al.}(2012){Mineo}, {Gilfanov}, \& {Sunyaev}}]{Mineo:2012}
{Mineo}, S., {Gilfanov}, M., \& {Sunyaev}, R. 2012, \mnras, 419, 2095

\bibitem[{{Moretti} {et~al.}(2003){Moretti}, {Campana}, {Lazzati}, \&
  {Tagliaferri}}]{Moretti:2003}
{Moretti}, A., {Campana}, S., {Lazzati}, D., \& {Tagliaferri}, G. 2003, \apj,
  588, 696

\bibitem[{{Nandra} \& {Pounds}(1994)}]{Nandra:1994}
{Nandra}, K., \& {Pounds}, K.~A. 1994, \mnras, 268, 405

\bibitem[{{Oser} {et~al.}(2012){Oser}, {Naab}, {Ostriker}, \&
  {Johansson}}]{Oser:2012}
{Oser}, L., {Naab}, T., {Ostriker}, J.~P., \& {Johansson}, P.~H. 2012, \apj,
  744, 63

\bibitem[{{Pacucci} {et~al.}(2018){Pacucci}, {Loeb}, {Mezcua}, \&
  {Mart{\'\i}n-Navarro}}]{Pacucci:2018}
{Pacucci}, F., {Loeb}, A., {Mezcua}, M., \& {Mart{\'\i}n-Navarro}, I. 2018,
  \apj, 864, L6

\bibitem[{{Park} {et~al.}(2006){Park}, {Kashyap}, {Siemiginowska}, {van Dyk},
  {Zezas}, {Heinke}, \& {Wargelin}}]{Park:2006}
{Park}, T., {Kashyap}, V.~L., {Siemiginowska}, A., {van Dyk}, D.~A., {Zezas},
  A., {Heinke}, C., \& {Wargelin}, B.~J. 2006, \apj, 652, 610

\bibitem[{{Piconcelli} {et~al.}(2005){Piconcelli}, {Jimenez-Bail{\'o}n},
  {Guainazzi}, {Schartel}, {Rodr{\'{\i}}guez-Pascual}, \&
  {Santos-Lle{\'o}}}]{Piconcelli:2005}
{Piconcelli}, E., {Jimenez-Bail{\'o}n}, E., {Guainazzi}, M., {Schartel}, N.,
  {Rodr{\'{\i}}guez-Pascual}, P.~M., \& {Santos-Lle{\'o}}, M. 2005, \aap, 432,
  15

\bibitem[{{Plotkin} {et~al.}(2016){Plotkin}, {Gallo}, {Haardt}, {Miller},
  {Wood}, {Reines}, {Wu}, \& {Greene}}]{Plotkin:2016}
{Plotkin}, R.~M., {Gallo}, E., {Haardt}, F., {Miller}, B.~P., {Wood}, C. J.~L.,
  {Reines}, A.~E., {Wu}, J., \& {Greene}, J.~E. 2016, \apj, 825, 139

\bibitem[{{Reeves} \& {Turner}(2000)}]{Reeves:2000}
{Reeves}, J.~N., \& {Turner}, M.~J.~L. 2000, \mnras, 316, 234

\bibitem[{{Reines} {et~al.}(2013){Reines}, {Greene}, \& {Geha}}]{Reines:2013}
{Reines}, A.~E., {Greene}, J.~E., \& {Geha}, M. 2013, \apj, 775, 116

\bibitem[{{Reines} \& {Volonteri}(2015)}]{Reines:2015}
{Reines}, A.~E., \& {Volonteri}, M. 2015, \apj, 813, 82

\bibitem[{{Remillard} \& {McClintock}(2006)}]{Remillard:McClintock:2006}
{Remillard}, R.~A., \& {McClintock}, J.~E. 2006, Annual Review of Astronomy and
  Astrophysics, 44, 49

\bibitem[{{Scargle} {et~al.}(2013){Scargle}, {Norris}, {Jackson}, \&
  {Chiang}}]{Scargle:2013}
{Scargle}, J.~D., {Norris}, J.~P., {Jackson}, B., \& {Chiang}, J. 2013, \apj,
  764, 167

\bibitem[{{Shemmer} {et~al.}(2006){Shemmer}, {Brandt}, {Netzer}, {Maiolino}, \&
  {Kaspi}}]{Shemmer:2006}
{Shemmer}, O., {Brandt}, W.~N., {Netzer}, H., {Maiolino}, R., \& {Kaspi}, S.
  2006, \apj, 646, L29

\bibitem[{{Soria} {et~al.}(2013){Soria}, {Hau}, \& {Pakull}}]{Soria:2013}
{Soria}, R., {Hau}, G. K.~T., \& {Pakull}, M.~W. 2013, \apj, 768, L22

\bibitem[{{Springel} \& {Hernquist}(2005)}]{Springel:Hernquist:2005}
{Springel}, V., \& {Hernquist}, L. 2005, \apjl, 622, L9

\bibitem[{{Stern}(2015)}]{Stern:2015}
{Stern}, D. 2015, \apj, 807, 129

\bibitem[{{Stern} {et~al.}(2012){Stern}, {Assef}, {Benford}, {Blain}, {Cutri},
  {Dey}, {Eisenhardt}, {Griffith}, {Jarrett}, {Lake}, {Masci}, {Petty},
  {Stanford}, {Tsai}, {Wright}, {Yan}, {Harrison}, \& {Madsen}}]{Stern:2012}
{Stern}, D., {et~al.} 2012, \apj, 753, 30

\bibitem[{{Stocke} {et~al.}(1991){Stocke}, {Morris}, {Gioia}, {Maccacaro},
  {Schild}, {Wolter}, {Fleming}, \& {Henry}}]{Stocke:1991}
{Stocke}, J.~T., {Morris}, S.~L., {Gioia}, I.~M., {Maccacaro}, T., {Schild},
  R., {Wolter}, A., {Fleming}, T.~A., \& {Henry}, J.~P. 1991, The Astrophysical
  Journal Supplement Series, 76, 813

\bibitem[{{Sutton} {et~al.}(2015){Sutton}, {Roberts}, {Gladstone}, \&
  {Walton}}]{Sutton:2015}
{Sutton}, A.~D., {Roberts}, T.~P., {Gladstone}, J.~C., \& {Walton}, D.~J. 2015,
  \mnras, 450, 787

\bibitem[{{Swartz} {et~al.}(2004){Swartz}, {Ghosh}, {Tennant}, \&
  {Wu}}]{Swartz:2004}
{Swartz}, D.~A., {Ghosh}, K.~K., {Tennant}, A.~F., \& {Wu}, K. 2004, \apjs,
  154, 519

\bibitem[{{Swartz} {et~al.}(2011){Swartz}, {Soria}, {Tennant}, \&
  {Yukita}}]{Swartz:2011}
{Swartz}, D.~A., {Soria}, R., {Tennant}, A.~F., \& {Yukita}, M. 2011, \apj,
  741, 49

\bibitem[{{Swartz} {et~al.}(2009){Swartz}, {Tennant}, \& {Soria}}]{Swartz:2009}
{Swartz}, D.~A., {Tennant}, A.~F., \& {Soria}, R. 2009, \apj, 703, 159

\bibitem[{{Tananbaum} {et~al.}(1979){Tananbaum}, {Avni}, {Branduardi}, {Elvis},
  {Fabbiano}, {Feigelson}, {Giacconi}, {Henry}, {Pye}, {Soltan}, \&
  {Zamorani}}]{Tananbaum:1979}
{Tananbaum}, H., {et~al.} 1979, \apj, 234, L9

\bibitem[{{Tao} {et~al.}(2011){Tao}, {Feng}, {Gris{\'e}}, \&
  {Kaaret}}]{Tao:2011}
{Tao}, L., {Feng}, H., {Gris{\'e}}, F., \& {Kaaret}, P. 2011, \apj, 737, 81

\bibitem[{{van Wassenhove} {et~al.}(2010){van Wassenhove}, {Volonteri},
  {Walker}, \& {Gair}}]{Van_Wassenhove:2010}
{van Wassenhove}, S., {Volonteri}, M., {Walker}, M.~G., \& {Gair}, J.~R. 2010,
  \mnras, 408, 1139

\bibitem[{{Volonteri}(2010)}]{Volonteri:2010}
{Volonteri}, M. 2010, \aapr, 18, 279

\bibitem[{{Volonteri} \& {Bellovary}(2012)}]{Volonteri:2012}
{Volonteri}, M., \& {Bellovary}, J. 2012, Reports on Progress in Physics, 75,
  124901

\bibitem[{{Volonteri} \& {Natarajan}(2009)}]{Volonteri:2009b}
{Volonteri}, M., \& {Natarajan}, P. 2009, \mnras, 400, 1911

\bibitem[{{Walton} {et~al.}(2011){Walton}, {Roberts}, {Mateos}, \&
  {Heard}}]{Walton:2011}
{Walton}, D.~J., {Roberts}, T.~P., {Mateos}, S., \& {Heard}, V. 2011, \mnras,
  416, 1844

\bibitem[{Webb {et~al.}(2017)Webb, Gu{\'e}rou, Ciambur, Detoeuf, Coriat, Godet,
  Barret, Combes, Contini, Graham, {et~al.}}]{Webb:2017}
Webb, N., {et~al.} 2017, Astronomy \& Astrophysics, 602, A103

\bibitem[{{Woods} {et~al.}(2018){Woods}, {Agarwal}, {Bromm}, {Bunker}, {Chen},
  {Chon}, {Ferrara}, {Glover}, {Haemmerle}, \& {Haiman}}]{Woods:2018}
{Woods}, T.~E., {et~al.} 2018, arXiv e-prints, arXiv:1810.12310

\bibitem[{{Wright} {et~al.}(2010){Wright}, {Eisenhardt}, {Mainzer}, {Ressler},
  {Cutri}, {Jarrett}, {Kirkpatrick}, {Padgett}, {McMillan}, {Skrutskie},
  {Stanford}, {Cohen}, {Walker}, {Mather}, {Leisawitz}, {Gautier}, {McLean},
  {Benford}, {Lonsdale}, {Blain}, {Mendez}, {Irace}, {Duval}, {Liu}, {Royer},
  {Heinrichsen}, {Howard}, {Shannon}, {Kendall}, {Walsh}, {Larsen}, {Cardon},
  {Schick}, {Schwalm}, {Abid}, {Fabinsky}, {Naes}, \& {Tsai}}]{Wright:2010}
{Wright}, E.~L., {et~al.} 2010, \aj, 140, 1868

\bibitem[{{York} {et~al.}(2000){York}, {Adelman}, {Anderson}, {Anderson},
  {Annis}, {Bahcall}, {Bakken}, {Barkhouser}, {Bastian}, {Berman}, {Boroski},
  {Bracker}, {Briegel}, {Briggs}, {Brinkmann}, {Brunner}, {Burles}, {Carey},
  {Carr}, {Castander}, {Chen}, {Colestock}, {Connolly}, {Crocker}, {Csabai},
  {Czarapata}, {Davis}, {Doi}, {Dombeck}, {Eisenstein}, {Ellman}, {Elms},
  {Evans}, {Fan}, {Federwitz}, {Fiscelli}, {Friedman}, {Frieman}, {Fukugita},
  {Gillespie}, {Gunn}, {Gurbani}, {de Haas}, {Haldeman}, {Harris}, {Hayes},
  {Heckman}, {Hennessy}, {Hindsley}, {Holm}, {Holmgren}, {Huang}, {Hull},
  {Husby}, {Ichikawa}, {Ichikawa}, {Ivezi{\'c}}, {Kent}, {Kim}, {Kinney},
  {Klaene}, {Kleinman}, {Kleinman}, {Knapp}, {Korienek}, {Kron}, {Kunszt},
  {Lamb}, {Lee}, {Leger}, {Limmongkol}, {Lindenmeyer}, {Long}, {Loomis},
  {Loveday}, {Lucinio}, {Lupton}, {MacKinnon}, {Mannery}, {Mantsch}, {Margon},
  {McGehee}, {McKay}, {Meiksin}, {Merelli}, {Monet}, {Munn}, {Narayanan},
  {Nash}, {Neilsen}, {Neswold}, {Newberg}, {Nichol}, {Nicinski}, {Nonino},
  {Okada}, {Okamura}, {Ostriker}, {Owen}, {Pauls}, {Peoples}, {Peterson},
  {Petravick}, {Pier}, {Pope}, {Pordes}, {Prosapio}, {Rechenmacher}, {Quinn},
  {Richards}, {Richmond}, {Rivetta}, {Rockosi}, {Ruthmansdorfer}, {Sandford},
  {Schlegel}, {Schneider}, {Sekiguchi}, {Sergey}, {Shimasaku}, {Siegmund},
  {Smee}, {Smith}, {Snedden}, {Stone}, {Stoughton}, {Strauss}, {Stubbs},
  {SubbaRao}, {Szalay}, {Szapudi}, {Szokoly}, {Thakar}, {Tremonti}, {Tucker},
  {Uomoto}, {Vanden Berk}, {Vogeley}, {Waddell}, {Wang}, {Watanabe},
  {Weinberg}, {Yanny}, {Yasuda}, \& {SDSS Collaboration}}]{York:2000}
{York}, D.~G., {et~al.} 2000, \aj, 120, 1579

\bibitem[{{Yuan} {et~al.}(1998){Yuan}, {Siebert}, \& {Brinkmann}}]{Yuan:1998}
{Yuan}, W., {Siebert}, J., \& {Brinkmann}, W. 1998, \aap, 334, 498

\bibitem[{{Yuan} {et~al.}(2014){Yuan}, {Zhou}, {Dou}, {Dong}, {Fan}, \&
  {Wang}}]{Yuan:2014}
{Yuan}, W., {Zhou}, H., {Dou}, L., {Dong}, X.~B., {Fan}, X., \& {Wang}, T.~G.
  2014, \apj, 782, 55

\bibitem[{{Zolotukhin} {et~al.}(2016){Zolotukhin}, {Webb}, {Godet}, {Bachetti},
  \& {Barret}}]{Zolotukhin:2016}
{Zolotukhin}, I., {Webb}, N.~A., {Godet}, O., {Bachetti}, M., \& {Barret}, D.
  2016, \apj, 817, 88

\bibitem[{Zwart \& McMillan(2002)}]{Zwart:2002}
Zwart, S. F.~P., \& McMillan, S.~L. 2002, The Astrophysical Journal, 576, 899

\end{thebibliography}
\end{document}